\documentclass{aa}  

\usepackage{graphicx}
\usepackage{cool}

\usepackage{subcaption}
\usepackage{amsmath}
\usepackage{booktabs}
\usepackage{units}
\usepackage{txfonts}
\usepackage{hyperref}
\hypersetup{colorlinks=true,citecolor=blue}

\begin{document}
\defcitealias{grillo13}{G13}

\title{Constraining the multi-scale dark-matter distribution in CASSOWARY 31 with strong gravitational lensing and stellar dynamics}

\author{H.~Wang\inst{1,2} \and R.~Ca\~nameras\inst{1} \and G.~B.~Caminha\inst{1} \and S.~H.~Suyu\inst{1,2,3} \and A.~Y{\i}ld{\i}r{\i}m\inst{1} \and G. Chiriv\`i\inst{1} \and L.~Christensen\inst{4} \and C.~Grillo\inst{5,6} \and S.~Schuldt\inst{1,2}}

\institute{Max-Planck-Institut f\"ur Astrophysik, Karl-Schwarzschild-Str. 1, 85748 Garching, Germany \\
{\tt e-mail: wanghan@mpa-garching.mpg.de}
\and
Technische Universit\"at M\"unchen, Physik Department, James-Franck Str. 1, 85748 Garching, Germany
\and
Academia Sinica Institute of Astronomy and Astrophysics (ASIAA), 11F of ASMAB, No. 1, Section 4, Roosevelt Road, Taipei 10617, Taiwan
\and
Cosmic Dawn Center, Niels Bohr Institute, Univ. of Copenhagen, Jagtvej 128, 2200 Copenhagen, Denmark
\and
Dipartimento di Fisica, Universit\`a degli Studi di Milano, via Celoria 16, I-20133 Milano, Italy
\and
INAF-IASF Milano, via A. Corti 12, I-20133 Milano, Italy
}

\titlerunning{Dark-matter distribution in CSWA\,31}

\authorrunning{H. Wang et al.}

\date{Received --; accepted --}

\abstract{We study the inner structure of the group-scale lens CASSOWARY 31 (CSWA\,31) by adopting both strong lensing and dynamical modeling. CSWA\,31 is a peculiar lens system. The brightest group galaxy (BGG) is an ultra-massive elliptical galaxy at $z=0.683$ with a weighted mean velocity dispersion of $\sigma = 432 \pm 31$\,km s$^{-1}$. It is surrounded by group members and several lensed arcs probing up to $\simeq$150\,kpc in projection. Our results significantly improve previous analyses of CSWA\,31 thanks to the new {\it HST} imaging and MUSE integral-field spectroscopy. From the secure identification of five sets of multiple images and measurements of the spatially-resolved stellar kinematics of the BGG, we conduct a detailed analysis of the multi-scale mass distribution using various modeling approaches, both in the single and multiple lens-plane scenarios. Our best-fit mass models reproduce the positions of multiple images and provide robust reconstructions for two background galaxies at $z=1.4869$ and $z=2.763$. Despite small variations related to the different sets of input constraints, the relative contributions from the BGG and group-scale halo are remarkably consistent in our three reference models, demonstrating the self-consistency between strong lensing analyses based on image position and extended image modeling. We find that the ultra-massive BGG dominates the projected total mass profiles within 20~kpc, while the group-scale halo dominates at larger radii. The total projected mass enclosed within $R_{\rm eff} = 27.2$\,kpc is $1.10_{-0.04}^{+0.02} \times 10^{13}$\,M$_\odot$. We find that CSWA\,31 is a peculiar fossil group, strongly dark-matter dominated towards the central region, and with a projected total mass profile similar to higher-mass cluster-scale halos. The total mass-density slope within the effective radius is shallower than isothermal, consistent with previous analyses of early-type galaxies in overdense environments.}


\maketitle

\section{Introduction}

In the $\Lambda$ cold dark-matter (CDM) paradigm, dark-matter halos evolve and grow hierarchically by accretion of smaller halos \citep[e.g.,][]{subramanian00}, and higher-density environments collapse and form galaxies earlier. The evolution history of the most massive galaxies with $M_*>10^{11}$~M$_{\odot}$ is well described by this hierarchical model. This population formed on average earlier than lower-mass galaxies, during short bursts of intense star formation maintained over a few 100~Myr, and followed by rapid quenching \citep[e.g.,][]{thomas05,pacifici16,tacchella21}. Large amounts of observations and simulations suggest that the main star formation episodes occured at $z>2$, and that an active galactic nuclei (AGNs) are primarily responsible for the quenching phase \citep[e.g.,][]{springel05,croton06}, leaving compact red quiescent galaxies at $z \sim 2$ \citep[e.g.][]{moster20} which subsequently undergo dry mergers \citep[e.g.,][]{naab07,remus13}, grow in size \citep[e.g.,][]{naab09,vanderwel14}, and form the massive elliptical galaxies in the local Universe. Each evolutionary process gives direct fingerprints on the properties of the descendants observed at low redshift. For instance, higher-mass ellipticals are thought to have more violent merger histories, which lowers their stellar angular momentum compared to less massive counterparts \citep{emsellem07}. The properties of galaxies in the highest-end of the mass distribution, with ultra-high stellar velocity dispersions ($\sigma \sim 500$~km s$^{-1}$), are particularly useful to further improve our understanding of this evolutionary sequence. At low redshifts $z \sim 0$ , these systems are larger and redder than equivalents at lower masses \citep[e.g.,][]{bernardi11}. They are also extremely rare, and their precise number density can be related to key properties, such as the redshift of their main growth phase \citep[e.g.,][]{loeb03}.

On larger scales, the mass distribution of cluster-scale dark-matter halos has been extensively studied with various techniques, including galaxy kinematics, X-rays and gravitational lensing. Detailed diagnostics have also been obtained for the baryonic and dark-matter content of massive ellipticals residing near the cores of such massive, dynamically-relaxed clusters \citep[e.g.,][]{newman13a,newman13b}. In addition, galaxy groups are the most common structures in the Universe and are expected to play a crucial role in the hierarchical formation of large-scale halos \citep[e.g.,][]{eke04,sommer06}. Despite the successful searches in wide-field surveys \citep[e.g.,][]{belokurov09,more12}, only small numbers of galaxy groups have precise mass distribution measurements \citep[][]{spiniello11,deason13,newman15}, which complicates the interpretation of the apparent diversity in their physical properties \citep[e.g.,][]{limousin09,munoz13}.

Standard CDM simulations predict universal mass-density profiles for dark-matter halos, independent of the total halo mass \citep{navarro96,navarro97}. In reality, baryonic physics also play a fundamental role, and the interplay between baryons and dark matter directly affects the total mass distibutions. Gas cooling can result in adiabatic contraction of the dark-matter halos \citep{blumenthal86} while, on the contrary, mergers and feedback from massive stars, supernovae, or AGNs, can expand the dark-matter distributions \citep[e.g.,][]{nipoti04,pontzen12}. Due to these competing processes, the inner mass-density slopes of dark-matter halos predicted by hydrodynamical simulations are either similar to the Navarro-Frenk-White (NFW) profile \citep[e.g.,][]{nfw1997, schaller15} or flatter \citep[e.g.,][]{martizzi12}, depending on the different prescriptions of feedback processes. Observational constraints on the halo properties can thus provide insights into the baryonic physics taking place during galaxy growth, and into their relative importance in setting the present-day dark-matter distributions.

Due to its sensitivity to all integrated mass along the line-of-sight, the strong gravitational lensing effect is very useful to measure robustly mass distributions and to explore the relation between dark and luminous mass \citep[e.g.,][]{spiniello11}. The positions of multiple images can be used to calculate the deflection angles, and to determine the lens mass within the separation between observed images (i.e; within the Einstein radius $\theta_{\rm E}$). Strong lensing studies have radically improved our understanding of the fundamental properties of galaxy and galaxy cluster dark-matter halos and subhalos \citep[][]{vegetti12}. Since this effect is sensitive to the total mass on $\theta_{\rm E}$ scales, generally of the order of the effective radius of the main foreground lens galaxy \citep[e.g.,][and references therein]{newman15}, strong lensing is often complemented with stellar dynamics. The combination of strong lensing and spatially-averaged stellar kinematics has tightly constrained the inner slope of the total mass-density profiles of isolated galaxies \citep[][]{koopmans03,treu04,auger09,sonnenfeld15}, and of massive ellipticals in overdense environments \citep[e.g.,][]{newman13b} acting as deflectors. These probes did not reveal significant dark-matter contraction \cite[e.g.,][]{dutton14,newman15}. For group and cluster-scale lenses, these joint analyses have helped improve the mass model accuracies towards the central regions \citep[e.g.,][]{sand08}.

Combining strong lensing with spatially-resolved stellar dynamics from integral-field-unit spectroscopy is particularly helpful to break the mass-sheet degeneracy \citep{falco85} by constraining the amount of mass sheet related to the main lens \citep[e.g.,][]{yildirim21}. Combining these fine observational constraints with stellar population synthesis analyses is generally sufficient to disentangle the dark-matter and baryonic contributions to the total mass-density profiles. This leads to reliable measurements of the radial slopes of the dark-matter density profiles to search for deviations from the standard NFW profile. Joint modeling of strong lensing and 2D stellar dynamics is relatively easier for galaxy groups or clusters in hydrostatic equilibrium, which are typically well deblended and have lower source contamination to the stellar kinematics of the foreground lens.

\begin{figure*}
  \centering
  \includegraphics[width=0.9\linewidth]{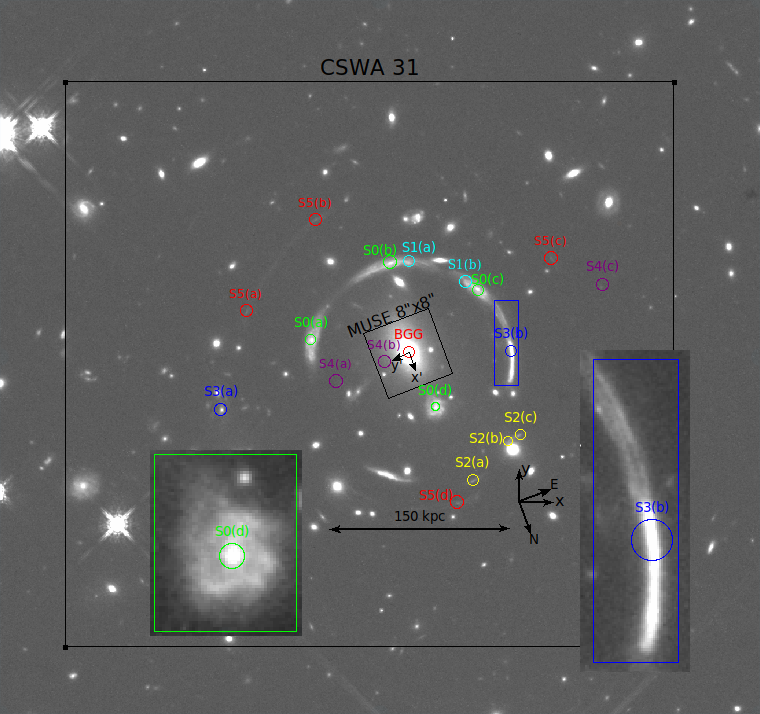}
  \caption{{\it HST}/WFC3 image of CSWA\,31 and its surrounding environment in the F160W band. The orientation is marked by the black arrows on the bottom-right of the figure, and the ruler indicates a physical scale of 150~kpc in the main lens plane at $z=0.683$. The large black box marks the total 1\arcmin\,$\times$\,1\arcmin\ coverage of MUSE observations. The small 8\arcsec\,$\times$\,8\arcsec\ black box in the center shows the field-of-view considered for the dynamical analysis of the main lens galaxy, and the $x'$ and $y'$ arrows show the orientation adopted for the kinematic maps (see Sect.~\ref{sec:glad_model}). The sets of spectroscopically-confirmed multiple images used in our analysis are marked with circles, using a different color for each set. The image sets S0 and S1 are bright knots from the same spiral galaxy at $z = 1.4869$, with image S0(d) zoomed-in on the left inset. The right inset shows the bright arc from source galaxy S3 at $z=2.763$. The multiple images of source 4 (S4) are identified with MUSE and undetected in this {\it HST} image.}
  \label{fig:CSWA31}
\end{figure*}

In this work, we present a detailed analysis of the inner mass structure of the group-scale lens CASSOWARY\,31 \citep[CSWA\,31, see also,][]{belokurov09,brewer11,stark13}. This strong gravitational lens opens interesting perspectives to constrain simultaneously the total and dark-matter mass distributions within a galaxy group at $z=0.683$ and its ultra-massive central elliptical galaxy. CSWA\,31 has a large Einstein radius of 70\,kpc and shows several multiply-imaged sources at various projected separations from the lens centroid \citep[][and Fig.~\ref{fig:CSWA31}]{grillo13}. In contrast to most group-scale lenses having small image separations $\leq$5\arcsec\ \citep[e.g.,][]{auger08,limousin09b,newman15}, the peculiar configuration of CSWA\,31 makes it well-suited to characterize the mass density slope of an extended group-scale halo beyond 100\,kpc. In addition, combining strong lensing and stellar dynamics modeling of CSWA\,31 has the potential to constrain the inner slope of the total mass-density profile and to robustly distinguish the relative contributions from the central galaxy and group-scale halo within 10\,kpc \citep[e.g.,][]{vandewen10,barnabe12}. This system can therefore be used as a testbed to unveil the inner structure of massive ellipticals and their host galaxy groups at intermediate redshift. We use new high-quality imaging from the Hubble Space Telescope ({\it HST}) and integral-field spectroscopy from the Multi-Unit Spectroscopic Explorer \citep[MUSE,][]{bacon14} in order to extend previous studies of CSWA\,31 with more detailed parametrizations of the lens mass distribution, and to develop novel methods to separate the multi-scale components.

The paper is organised as follows. In Sect.~\ref{sec:data}, we briefly summarize previous analyses of CSWA\,31, and we present the new imaging and spectroscopic observations, as well as redshift measurements. In Sect.~\ref{sec:method}, we overview the formalism of our strong lensing and stellar dynamics modeling, as well as the numerical methods to infer the best-fit parameters of the mass models. We present the lensing-only models of CSWA\,31 in Sect.~\ref{sec:sl_model}, and the joint lensing and dynamics modeling in Sect.~\ref{sec:glad_model}. In Sect.~\ref{sec:model_summary}, we compare the best-fit mass models from previous sections, and we put the results in context with other group- and cluster-scale lenses in the literature. In Sect.~\ref{sec:summary}, we summarize our results and give an outlook on general studies of galaxy groups. Throughout this work, we assume $H_{\rm 0} = 70$\,km s$^{-1}$ Mpc$^{-1}$, $\Omega_{\rm m} = 0.3$ and $\Omega_{\Lambda} = 0.7$. Hence, 1\arcsec\ corresponds to 7.08\,kpc at the lens redshift of $z = 0.683$.

\section{Observations}
\label{sec:data}

The CSWA\,31 lens system was discovered in SDSS imaging as part of the Cambridge And Sloan Survey Of Wide ARcs in the skY \citep[CASSOWARY,][]{belokurov09}, and was analysed by \citet{brewer11}, \citet{stark13}, and \citet[][hereafter \citetalias{grillo13}]{grillo13}. The main lens galaxy of CSWA\,31 is located in the group center and will be referred to as the brightest group galaxy (BGG). This is an early-type galaxy at $z=0.683$ with a very high stellar mass of about $3 \times 10^{12}$\,M$_{\odot}$ and an aperture-averaged stellar velocity dispersion of $450 \pm 80$\,km s$^{-1}$ from SDSS \citepalias{grillo13}, which suggests it is among the rarest ultra-massive elliptical galaxies at these redshifts \citep[with comoving number densities $\lesssim$ $10^{-8}$~Mpc$^{-3}$,][]{loeb03}. The BGG is surrounded by group members and by giant lensed arcs formed by a face-on spiral at $z=1.487$. \citetalias{grillo13} used Gemini/GMOS imaging and VLT/X-Shooter spectroscopy to characterize the lens total and stellar-to-total mass profiles, based on the four multiple images of this main background source\footnote{Other blue arcs were observed with VLT/X-shooter as part of programs 091.A-0852(A) (PI: Christensen) and 094.A-0684(A) (PI: Grillo) but did not provide secure identifications of additional multiple image families.}. They measured a total mass of $4 \times 10^{13}$~M$_\odot$ projected within the Einstein radius and found that CSWA\,31 is strongly dark-matter dominated in its center. Subsequently, \citet{leethochawalit16} took advantage of the lensing magnifications to characterize the spatially-resolved kinematics and gas-phase metallicity gradients within the background spiral galaxy based-on Keck/OSIRIS integral-field spectroscopy. We present here the additional observables inferred from our new {\it HST} and MUSE data set.

\begin{table*}
  \centering
  \caption{Position, spectroscopic redshift, and (F438W$-$F160W) color for the BGG and multiple images used in the modeling.}
  \begin{tabular}{ccccccc}
    \hline
    \hline
    ID & RA & dec & $z_{\rm spec}$ & color & $a$ [$\arcsec$] & $b$ [$\arcsec$]\\
    \hline
    BGG   & 9:21:25.738 & 18:10:17.70 & 0.6828 & $\dots$  & $\dots$ & $\dots$ \\
    S0(a) & 9:21:25.040 & 18:10:12.27 & 1.4869 & 4.69$\pm$0.90 & 0.14& 0.07 \\
    S0(b) & 9:21:25.858 & 18:10:07.24 & 1.4869 & 5.53$\pm$2.13 & 0.14& 0.07 \\
    S0(c) & 9:21:26.439 & 18:10:13.60 & 1.4869 & 4.35$\pm$1.67 & 0.14 & 0.07 \\
    S0(d) & 9:21:25.781 & 18:10:24.64 & 1.4869 & 4.19$\pm$0.53 & 0.14 & $\dots$ \\
    S1(a) & 9:21:25.987 & 18:10:07.84 & 1.4869 & 1.37$\pm$0.08 & 0.14& 0.07 \\
    S1(b) & 9:21:26.376 & 18:10:12.58 & 1.4869 & 1.32$\pm$0.08 & 0.14 & 0.07 \\
    S2(a) & 9:21:25.851 & 18:10:34.05 & 1.4874 & $\dots$ & 0.14 & 0.07 \\
    S2(b) & 9:21:26.252 & 18:10:31.13 & 1.4874 & $\dots$ & 0.14 & 0.07 \\
    S2(c) & 9:21:26.323 & 18:10:31.04 & 1.4874 & $\dots$ & 0.14 & 0.07 \\
    S3(a) & 9:21:24.159 & 18:10:16.03 & 2.763 & 3.13$\pm$0.24 & 0.14 & 0.07 \\
    S3(b) & 9:21:26.517 & 18:10:21.73 & 2.763 & 3.37$\pm$0.17 & 0.28 & 0.07 \\
    S4(a) & 9:21:25.105 & 18:10:17.76 & 3.4280 & $\dots$ & 0.21 & $\dots$ \\
    S4(b) & 9:21:25.531 & 18:10:17.62 & 3.4280 & $\dots$ & 0.21 & $\dots$ \\
    S4(c) & 9:21:27.395 & 18:10:18.65 & 3.4280 & $\dots$ & 0.21 & $\dots$ \\
    S5(a) & 9:21:24.634 & 18:10:06.47 & 4.205 & $\dots$ & 0.27 & 0.14 \\
    S5(b) & 9:21:25.418 & 18:09:59.46 & 4.205 & 2.33$\pm$0.69 & 0.27& 0.14 \\
    S5(c) & 9:21:27.085 & 18:10:13.39 & 4.205 & 2.23$\pm$0.85 & 0.27& 0.14 \\
    S5(d) & 9:21:25.670 & 18:10:35.81 & 4.205 & $\dots$ & 0.27 & 0.14 \\
    \hline
  \end{tabular}
  \label{tab:multima}
  \tablefoot{The right ascension and declination are measured with SExtractor in our {\it HST} F160W image, except for images of source 4 (S4) which are only detected by MUSE. All spectroscopic redshifts are secure and measured with MUSE. The (F438W$-$F160W) colors of multiple images in each set are consistent with each other given the 1-$\sigma$ uncertainties. Multiple images without color values are either falling into the WFC3 UVIS chip gaps, or undetected in both bands (for S4), and they are all confirmed by MUSE. The last two columns show the positional uncertainties along the elliptical major $a$ and minor $b$ axes that we used for the image position modeling (Sec.~\ref{sec:sl_model}). Images with a single listed value have circular positional uncertainties.}
\end{table*}

\subsection{{\it HST}/WFC3 imaging}
\label{ssec:hst}

We used high-resolution {\it HST} optical and near-infrared imaging in filters F438W and F160W over a field of view of $\sim$ 2\arcmin\,$\times$\,2\arcmin\ to resolve galaxies and lensed sources, and perform lens mass modeling. One-orbit exposures of $\sim$2400~sec were taken in each of the two filters with the Wide Field Camera 3 in November 2018 (program GO-15253; PI: Ca\~nameras). We redrizzled the individual exposures with the {\tt DrizzlePac} software package \citep{fruchter10} with optimal sampling, using ${\rm final\_pixfrac = 0.6}$ and 0.033\arcsec\ pix$^{-1}$ in F438W, and ${\rm final\_pixfrac = 0.8}$ and 0.066\arcsec\ pix$^{-1}$ in F160W. We used flat fields from the WFC3/IR monitoring campaigns to correct the small regions with decreased sensitivity in the F160W image. After correcting the astrometry, both images were saved on the same grid using Scamp and SWarp \citep{bertin06,bertin10}. We built point spread function (PSF) models by stacking four bright, unsaturated stars in the field, and measured PSF full width at half maximum (FWHM) of 0.09\arcsec\ and 0.19\arcsec\ in the reduced F438W and F160W images, respectively, about five times lower than ancillary ground-based imaging.

The WFC3 F160W image shown in Fig.~\ref{fig:CSWA31} not only probes the bulk of the old stellar populations in the BGG, group members and other galaxies in the field, but also provides a sharp, high signal-to-noise ratio (S/N) image of the main lensed arc formed by the face-on spiral galaxy at $z=1.487$, and detections of several additional faint arcs. The F438W band provides color information to help determine whether or not observed images are associated with the same background source. Sets of multiple images and selection of group members are then robustly confirmed with spectroscopy. Positions adopted in the modeling in Sect.~\ref{sec:sl_model} and \ref{sec:glad_model} are given in pixel units, using a reference coordinate at ${\rm RA = 9}$:21:19.349 and ${\rm dec = +}$18:10:52.50 (bottom-left corner of Fig.~\ref{fig:CSWA31}), and orientation following the $(x,y)$ arrows in Fig.~\ref{fig:CSWA31}.

\begin{figure*}
  \centering
  \includegraphics[width=0.85\linewidth]{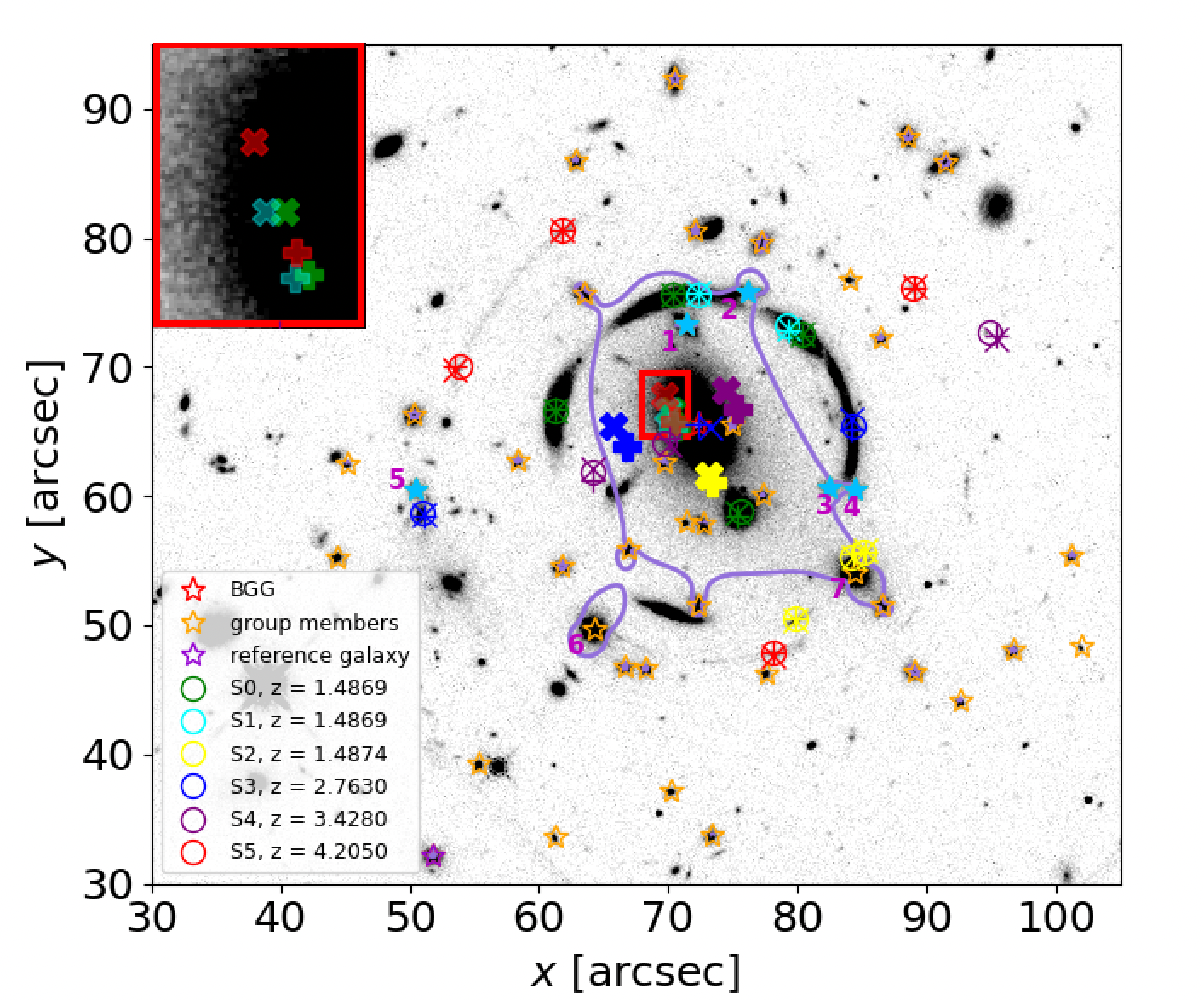}
  \caption{{\it HST} F160W image showing the BGG (red star), group members (orange stars), the reference galaxy used in the scaling relations (purple star), and sets of multiple images (circles). We modeled the emission of group members 1, 2, 3, 4, and 5 (cyan filled stars) together with the BGG in order to minimize light contamination on the nearby lensed arcs. Given the significant contribution of group members 1, 2, 6, and 7 to the light deflection angles, their Einstein radius was optimized separately. The image positions predicted by models Img-SP (L) and Img-MP (L) are marked by plus and cross symbols, respectively. Similarly, the mean weighted source positions reconstructed by models Img-SP (L) and Img-MP (L) are indicated with bold plus and cross symbols, respectively. The solid purple lines shows the critical curve of model Img-SP (L) regarding to S0. The upper-left inset zooms in on the 3.6\arcsec\,$\times$\,4.9\arcsec\ rectangle to highlight the intrinsic positions of S0, S1, and S5 in both models. The purple solid line shows the critical curve of model Img-MP(L) for $z=1.487$, the redshift of S0. The critical curves of other reference models are shown in Appendix \ref{sec:Critical curves of three reference lens models}. }
  \label{fig:input}
\end{figure*}

\subsection{VLT/MUSE spectroscopy}
\label{ssec:muse}

This work uses MUSE data from program 0104.A-0830(A) (PI: Ca\~nameras) in order to resolve the stellar velocity dispersion profile of the primary lens up to $\sim$10 kpc, and to secure the identification and redshift measurements of group members and multiple image candidates. Observations were carried out in December 2019 and January 2020, with seeing $\leq$1\arcsec, clear sky conditions, and airmass $<$1.6, with the MUSE wide-field mode corresponding to 1\arcmin\,$\times$\,1\arcmin\ field-of-view and 0.2\arcsec\,pix$^{-1}$ spatial sampling. We divided the observations into five individual OBs, applying a dithering pattern and 90$^{\circ}$ rotations between each OB, and obtained a total on-source exposure time of 5~hours.

The data were reduced with the standard MUSE pipeline \citep{weilbacher14} following the procedure described in \citet{caminha19}. After correcting raw exposures for the bias, flat fields and illumination frames, we applied wavelength and flux calibrations. The individual exposures were then combined into a stacked data cube. We optimized the sky subtraction with the {\tt ZAP} software \citep{soto16}, and defined the astrometry with respect to the HST F160W image. The reduced MUSE data cube has a PSF FWHM of 0.69\arcsec. The instrument spectral resolution ranging from 1770 at 4800\,\AA\ to 3590 at 9300\,\AA, and the constant 1.25~\AA\,pix$^{-1}$ sampling, make these data well-suited to reliably measure the lens stellar dynamics. In addition, the MUSE pointing shown in Fig.~\ref{fig:CSWA31} encloses all arcs and multiple image candidates detected with HST, as well as the majority of galaxies within the foreground group covering about 450~kpc.

To measure source redshifts, we followed \citet{caminha17} and \citet{caminha19} by analyzing the MUSE data cube using two different methods. First, we extracted the spectra at the positions of sources detected in the {\it HST} F160W image, using circular or optimized apertures depending on the source morphology. Second, we adapted the line search for sources with faint stellar continuum, undetected with {\it HST}, by extracting sources from a continuum-subtracted MUSE data cube. We then inspected both the spatially-averaged spectra and the spectra extracted along two perpendicular directions, in order to identify emission lines, absorption lines, or continuum breaks and measure redshifts. We also fitted templates to sources with bright stellar continuum to help infer their redshift. Finally, we assigned a quality flag to each object, by ranking the reliability of redshift measurements as QF = 3 (secure), 2 (likely), 1 (not reliable). Redshifts inferred from single, but unambiguous line identifications such as the [OII] doublet are considered secure. We obtained a total of 121 spectroscopic redshifts with QF $\geq$ 2 (see Appendix~\ref{sec:muse_multim}).

\subsection{Identification of multiple images}
\label{ssec:ima_fam}

From the MUSE redshift catalog, we identified a total of five sets of multiple images listed in Table~\ref{tab:multima} and shown in Fig.~\ref{fig:CSWA31}. The face-on spiral galaxy at $z=1.4869$ forming the bright arc is labelled as S0. Its four counter-images were the only ones spectroscopically-confirmed prior to our analysis, with S0(b) and S0(d) secured by Keck/DEIMOS and MMT spectroscopy in \citet{stark13}, and S0(a) and S0(c) confirmed by \citetalias{grillo13}. In addition to S0 centered on the galaxy bulge, we also find a doubly-imaged blue star-forming clump within the disk of this spiral galaxy, which we mark as S1. The second brightest arc labelled S3(b) has a single counter-image S3(a), both at $z=2.763$. The proposed counter images of S3(b), i.e., the radial image located between S0(a) and S0(d) and the tangentially elongated object in the southwest to S5(d) by \citetalias{grillo13} have spectrocopic redshifts of $z =0.3296$ (QF = 3) and $z =1.359$ (QF = 3), respectively, and are therefore ruled out by our MUSE data. Moreover, the two merging images S5(a) and S5(b) of the faint, diffuse external arc have reliable redshifts of $z=4.205$, as well as their counter-image S5(c). The redshift of S5(d) from a blind template fitting is $z=4.203$ (QF = 3), but with higher uncertainties than for S5(a-c) due to the lower spectrum S/N (see Fig.~\ref{fig:spec}). In addition, its colour is consistent with images S5(a-c), and its observed position and morphology match the expectations from simple lens models based on image families S1-S3. We therefore included S5(d) as a fourth counter-image and fixed its redshift to $z=4.205$. We checked that the (F438W$-$F160W) aperture colors measured with SExtractor \citep{bertin96} are within 1-$\sigma$ uncertainties for all multiple images in these sets (Table~\ref{tab:multima}).

Lastly, S4 is identified exclusively with MUSE, without stellar continuum counterpart in {\it HST}. We obtained three multiple image candidates with single emission lines interpreted as Ly$\alpha$ at $z=3.4280$. Since the positions of S4(a-c) are consistent with the predictions of our strong lensing models based on the other image families, we also included this set in the analysis.

These five multiple-image families in the field of CSWA\,31 provide a number of secure strong lensing constraints comparable with studies of galaxy clusters based on similarly deep {\it HST} images \cite[shallower than for the Frontier Field program, e.g.,][]{jauzac19,mahler19,caminha19,rescigno20,richard21}. This rich data set allows us to adopt a parametrized composite mass model to investigate the inner structure of this complex lens system in more details than previous analyses. Note that we measured the positions of multiple images from our highest resolution F160W image with SExtractor, except for S4 which is only detected with MUSE and thus has a larger positional uncertainty. The MUSE 1D spectra are shown in the Appendix~\ref{sec:muse_multim} together with the spectral features used to identify multiple image families.

\subsection{Properties of the foreground galaxy group}

 To select group members we relied on our best redshift estimate of $z=0.6828$ for the BGG. We then selected all galaxies from the MUSE spectroscopic catalogue located within $\pm 3000$~km s$^{-1}$ with respect to the rest-frame of the BGG and we obtained a total of 46 group members (see Fig.~\ref{fig:input}). We measured robust redshifts for members down to K(AB)$\sim$23~mag, about 3~mag fainter than L$^*$ galaxies at $z \sim 0.7$ \citep[see, e.g.,][]{fassbender11}, and we checked that their number is not sensitive to the exact velocity thresholds. The lack of spectroscopic coverage towards the group outer regions prevents us from inferring the total halo mass, $M_{\rm 200}$, based on the line-of-sight velocities of satellite galaxies at 1-3~Mpc from the lens center \citep[see e.g.,][]{munari13,deason13}. The MUSE data nonetheless cover the entire core region and provides valuable constraints to include the group members in our composite lens mass models.
 
 The number of members within the central 1\arcmin\,$\times$ 1\,\arcmin\ and their corrected isophotal magnitudes measured in F160W with SExtractor further suggest that CSWA\,31 is a rich galaxy group. Sparse clusters in the Abell catalogue have a minimum of 30 members within a magnitude range between m3 and m3+2, where m3 is the magnitude of the third brightest member, while confirmed members of CSWA\,31 span $\simeq$5.6~mag, and only 19 are between m3 and m3+2. This suggests that CSWA\,31 is not rich enough to meet the criteria from Abell. While additional members could lie outside the MUSE field-of-view and within the 2.5~Mpc radius corresponding to the compactness criteria from the Abell catalog, our {\it HST} photometry indicates that their number is much lower than towards the core.
 
CSWA\,31 has been identified as one of the most distant candidate fossil systems by \citetalias{grillo13} and \citet{johnson18}, in which case the extreme brightness of the BGG (21~mag~arcsec$^{-2}$ in $r$-band) would result from the past, slow accretion of all surrounding group members of intermediate masses \citep[e.g.,][]{khosroshahi07}. Fossil systems are considered as the final evolutionary stages of galaxy groups and \citet{johnson18} showed that, due to their elevated halo concentration, they are more efficient gravitational lenses than standard groups. The positions and brightnesses of group members spectroscopically-confirmed with our new observations further suggest that CSWA\,31 meets the fossil criteria from \citet{jones03}. However, additional data are still needed to confirm that CSWA\,31 is a progenitor of fossil groups seen in the local Universe. In particular, weak lensing constraints would help measure $R_{\rm 200}$ and, while \citetalias{grillo13} reported a non-detection in X-rays from the {\it ROSAT} All-Sky Survey (RASS), deeper X-ray observations would help characterize the hot gas and dynamical state of CSWA\,31.

We estimated the stellar mass of the BGG by modeling its spectral energy distribution (SED) with the Code Investigating GAlaxy Emission \citep[CIGALE,][]{burgarella05,noll09,boquien19}. We used a grid of models based on \citet{bruzual03} single stellar population templates, assuming delayed star formation histories with ages between 0.5 and 8~Gyr. We used the modified \citet{charlot00} extinction law, a \citet{salpeter55} stellar initial mass function, and a solar metallicity \citep[see also][]{conroy13,gallazzi14} to fit the PanSTARRS and {\it HST} photometry. Varying the dust extinction during the fit results in elevated $A_{\rm V} \gtrsim 3$~mag, essentially due to the lack of data points at infrared wavelengths. Assuming low dust extinction, as expected for massive ellipticals, gives a comparable fit while changing the stellar mass by 0.1~dex, and results in $M_* = (1.6 \pm 0.4) \times 10^{12}$~M$_{\odot}$ (see Fig.~\ref{fig:cigale}). This estimate is lower but consistent with the value reported in \citetalias{grillo13}.

\section{Methodology}
\label{sec:method}

In this work, we used the Gravitational Lens Efficient Explorer \citep[{\tt GLEE},][]{suyu10,suyu12} software to model the mass components and surface brightness of galaxies by adopting parametrized mass and light profiles. We also used Gravitational Lensing and Dynamics \citep[{\tt GLaD},][]{chirivi20,yildirim20}, an extension of {\tt GLEE} to perform a joint lensing and stellar dynamics modeling. {\tt GLaD} adopts the projected second-order velocity moment along the line-of-sight $\overline{v^2_{\rm LOS}}$ as inputs for Jeans anisotropic modeling \citep[JAM,][]{cappellari08} to estimate the dynamical parameters of galaxies. In Sect. \ref{ssec:sl}, we present the strong lensing formalism in the single- and multi-plane scenarios \citep[see also,][]{blandford86,schneider06}. In Sect.~\ref{ssec:dyn}, we have a short overview of the dynamical modeling approach \citep[see also,][]{binney87,cappellari08,barnabe12,yildirim20}. In Sect.~\ref{ssec:bestfit}, we summarize the sampling methods to infer the best-fitting parameters.

\subsection{Strong lensing}
\label{ssec:sl}
In the general relativity paradigm, a light ray can be deflected differentially due to the deformation of space-time along the line-of-sight induced by massive clumps with potential $\psi$
\begin{equation}
  \psi(\boldsymbol \theta) = \frac{1}{\pi} \int \rm d^2 \boldsymbol \theta^{'} \kappa (\boldsymbol \theta^{'})\ln  \left| \boldsymbol \theta - \boldsymbol \theta^{'} \right|
  \label{kappa related to potential}
\end{equation}
where $\kappa$ is the dimensionless surface mass density, so-called convergence, and $\boldsymbol \theta$ is the lensed source position. The potential connects to the scaled deflection angle $\boldsymbol \alpha$ via $\boldsymbol \alpha =\nabla \psi$. This leads to the following relation between $\kappa$ and $\boldsymbol \alpha$
\begin{equation}
  2\kappa = \nabla \cdot \boldsymbol \alpha.
  \label{linkKappa}
\end{equation}
The mass distributions of the deflectors are modeled using different parametrizations of the convergence $\kappa$. Firstly, we used the softened power-law elliptical mass distribution \cite[SPEMD,][]{barkana98} with $\kappa$ defined in Cartesian coordinates $(x_1, x_2)$ as follows, for a source at redshift $z_\text{s} = \Infinity$,
\begin{equation}
  \kappa (x_1,x_2) \mid_{z_s = \infty} = E \left(x_1^2+\frac{x_2^2}{q^2} +\frac{4r_{\rm core}^2}{(1+q)^2}\right)^{-\gamma},
  \label{eq:spemd}
\end{equation}
where the amplitude $E$ is related to the Einstein radius $\theta_\text{E}$ via
\begin{equation}
   E = \frac{2(1-\gamma)}{1 + q} \frac{\theta_{\rm E}^2}{\left( \theta_{\rm E}^2+\frac{4r_{\rm core}^2}{(1+q)^2}\right)^{1-\gamma} -\left( \frac{4r_{\rm core}^2}{(1+q)^2}\right)^{1-\gamma} }
   \label{E_scale}
\end{equation}
The first two terms in Eq.~\ref{eq:spemd} depict the elliptical shape of the mass distribution, $q$ is the axis ratio, $r_{\rm core}$ is the core radius, and $\gamma$ is the power-law slope which is related to the 3D density slope $\gamma'$ as $\gamma' = 2\gamma + 1$, where $\rho (r) \propto r^{-\gamma'}$. This corresponds to an isothermal mass profile for $\gamma = 0.5$ and $\gamma' = 2$. Secondly, we also use the truncated dual pseudo-isothermal elliptical mass distribution \citep[dPIE,][]{eliasdottir07,suyu10} defined as
\begin{equation}
  \kappa (x_1,x_2) \mid_{z_s = \infty}  =\frac{\theta_{\rm E}}{2} \frac{r_{\rm tr}^2}{r_{\rm tr}^2 - r_{\rm core}^2} \left( \frac{1}{\sqrt{R_{\rm em}^2 + r_{\rm core}^2}}-\frac{1}{\sqrt{R_{\rm em}^2 + r_{\rm tr}^2}} \right),
  \label{eq:dpie}
\end{equation}
where $r_{\rm tr}$ is the truncation radius, $r_{\rm core}$ is the core radius, $\theta_{\rm E}$ is the Einstein Radius, given $r_{\rm tr} = \Infinity$ and $r_{\rm core} = 0 $ and $R_{\rm em}$ is the elliptical mass radius related to the ellipticity $e = (1+q)/(1+q)$, where $q$ is the axis ratio, defined as:
\begin{equation}
  R_{\rm em}(x_1,x_2) = \sqrt{\frac{x_1^2}{(1+e)^2}+\frac{x_2^2}{(1-e)^2}}. 
\end{equation}
The corresponding 3D density distribution is proportional to $r^{-2}$ for $ r_{\rm core} \leq r < r_{\rm tr}$, and drops as $r^{-4}$ for $r > r_{\rm tr}$. This truncation can represent the tidal stripping of galaxy halos in dense group or cluster environments \citep[e.g.,][]{limousin09,suyu10}. Both SPEMD and dPIE can be rotated by a position angle $\theta_\text{PA}$ to account for the orientation of the mass distribution. 

 Thirdly, we introduce a constant external shear to account for the tidal stretching from neighbour galaxies. The lens potential for an external shear is parametrized in polar coordinates as
\begin{equation}
  \psi_{\rm ext} (r,\phi) = \frac{1}{2} \gamma_{\rm ext} r^2 \cos (2\phi - 2\theta_{\rm ext}),
  \label{eq:ext_shear}
\end{equation}
where $\gamma_{\rm ext}$ represents the strength of the external shear, and the shear angle $\theta_{\rm ext}$ represents the stretching orientation of the images. The shear center can be selected arbitrarily because it corresponds to an unobservable constant shift in the source plane. The external shear does not contribute to the surface mass density of the lens due to the vanishing external convergence $\kappa_{\rm ext}$ from $\kappa_{\rm ext} = \frac{1}{2} \nabla^2 \psi_{\rm ext}$, but it affects the shape of the observed images. For shear position angles $\theta_{\rm ext} = 0^{\circ}$ and $\theta_{\rm ext} = 90^{\circ}$, lensed image configurations are elongated along the horizontal and vertical axes, respectively.

We adopted the dPIE profile to model the total mass of individual lens galaxies such as the BGG and group members, and the SPEMD profile to model the dark matter component in the group-halo and BGG. Further detail on the mass parameters are given in Sect.~\ref{sec:sl_model} and \ref{ssec:glee_glad}.

To optimize the unknown parameters $\boldsymbol \eta$, i.e., the parameters introduced in mass profiles, in the adopted $\kappa$, we use the observed multiple images as constraints via Eq.~\ref{linkKappa} and the geometry relation between the background source position $\boldsymbol \beta$, and the observed image positions $\boldsymbol \theta$. In the single plane, the light rays from the background source are deflected by a single deflector (lens), yielding the lens equation,
\begin{equation}
  \boldsymbol \beta = \boldsymbol \theta - \boldsymbol \alpha (\boldsymbol \eta),
  \label{lens_sp_simple}
\end{equation}
with the scaled deflection angle $\boldsymbol \alpha$ expressed as
\begin{equation}
  \boldsymbol \alpha = \frac{D_{\rm ds}}{D_{\rm s}} \boldsymbol{\hat{\alpha}}(D_{\rm d} \boldsymbol \theta),
\end{equation}
where $D_{\rm ds}$, $D_{\rm s}$, $D_{\rm d}$ are the angular diameter distances between lens and source, between observer and source, and between observer and lens, respectively. In the multiplane scenario, the lens equation Eq.~\ref{lens_sp_simple} can be modified as follows to account for multiple deflections produced by a sequence of $n-1$ lenses distributed at different redshifts along the line-of-sight
\begin{equation}
  \boldsymbol \beta = \boldsymbol \theta_{n}(\boldsymbol \theta_1) = \boldsymbol \theta_{\rm 1} - \sum_{i=1}^{n-1} \frac{D_{in}}{D_n} \boldsymbol{\hat{\alpha}} (\boldsymbol \theta_i, \boldsymbol \eta).
  \label{lens_eq_mp}
\end{equation}
In this equation, $\boldsymbol \theta_n$ represents the position of the light ray in the $n$th plane, namely the source plane, with respect to the position of the light ray in the first observed lens plane $\boldsymbol \theta_{\rm 1}$. $\boldsymbol \theta_i$ is the image of the lensed source in the $i$th plane, $\boldsymbol{\hat{\alpha}} (\boldsymbol \theta_i, \boldsymbol \eta)$ is the deflection angle on the $i$th plane, $D_{in}$ is the angular diameter distance between the $i$th plane and $n$th plane, and $D_n$ is the angular diameter distance between the observer and $n$th plane. The case $n = 2$ corresponds to a single lens plane and Eq.~\ref{lens_eq_mp} reduces to Eq.~\ref{lens_sp_simple}.

For each lens model, we can calculate the lens surface mass density $\Sigma$ via $\kappa$ with a single lens plane, for a given source at redshift $z_{\rm s}$,
\begin{equation}
  \Sigma = \Sigma_{\rm crit} \times \kappa_{\rm z = \infty} \frac{D_{\rm ds}}{D_{\rm s}},
\end{equation}
with the critical surface mass density $\Sigma_{\rm crit}$ defined as
\begin{equation}
  \Sigma_{\rm crit} = \frac{c^2 D_{\rm s}}{4\pi G D_{\rm d} D_{\rm ds}}.
\end{equation}
This results in a definition of $\Sigma$ depending only on $D_{\rm d}$
\begin{equation}
  \Sigma = \frac{c^2}{4\pi G D_{\rm d}} \kappa_{\rm z = \infty}.
\end{equation}
We can then deduce the total mass enclosed within a radial distance $R$ from the defined lens center with
\begin{equation}
  M(<R) =  \int_{0}^{R} \Sigma(R^{'})2\pi R^{'}\,\text{d} R^{'}.
  \label{eq:total_mass}
\end{equation}
For multiple lens planes, $\Sigma$ becomes an effective surface mass density, corresponding to the gradient of the total deflection angle via Eq.~\ref{eq:spemd} that includes the contributions from all lenses along the line of sight. In that case, while the quantity inferred from Eq.~\ref{eq:total_mass} is not physical, it remains a good approximation for the enclosed mass.

\subsection{Stellar dynamics}
\label{ssec:dyn}

Stellar dynamics can capture the inner mass distribution within the effective radius $R_{\rm eff}$ by connecting the line-of-sight velocity $v$ and the velocity dispersion $\sigma$ to the mass potential. This probe is widely used in complement to strong gravitational lensing \citep[e.g.,][]{vandewen10,barnabe12,yildirim20}.

The motion of a group of stars within a galaxy can be characterised by the Collisionless Boltzmann Equation (CBE) with phase-space density $f(\boldsymbol x, \boldsymbol v)$ at the position $\boldsymbol x$ with velocity $\boldsymbol v$,
\begin{equation}
  \frac{\partial f}{\partial t} +  \sum_{i=1}^{3} v_i \frac{\partial f}{\partial x_i} - \frac{\partial \psi_{\rm D}}{\partial x_i} \frac{\partial f}{\partial v_i} = 0,
  \label{eq:cbe}
\end{equation}
which describes stars embedded in a gravitational field $\psi_{\rm D}$ following phase-space density conservation. The phase-space density is not accessible for distant galaxies, and we can only extract the velocities along the line-of-sight $v$ and velocity dispersions $\sigma$. To properly adopt the CBE, we therefore multiply velocities $v_R, v_z, v_\phi $ with Eq.~\ref{eq:cbe} and integrate over all velocity space in cylindrical coordinates ($R$, $z$, $\phi$). Assuming the alignment of an axisymmetric velocity ellipsoid with the cylindrical coordinate system, we obtain the following two equations, so-called axisymmetric Jeans equation,
\begin{equation}
  \frac{\beta_z \lambda \overline{v_z^2}-\lambda \overline{v_\phi^2}}{R} + \frac{\partial \left( \beta_z \lambda \overline{v_z^2}\right)} {\partial R} = -\lambda \frac{\partial \psi_{\rm D}}{\partial R},
  \label{eq11}
\end{equation}
\begin{equation}
  \frac{\partial \left(\lambda \overline{v_{\rm z}^2}\right)}{\partial z} = -\lambda \frac{\partial \psi_{\rm D}}{\partial z},
  \label{eq22}
\end{equation}
where the orbital anisotropy parameter, $\beta_z = 1-\overline{v_z^2}/ \overline{v_R^2}$, presents a flattening in the meridional plane. We use a simplified notation of the second-order velocity moments $\lambda \overline{v_k v_j}$
\begin{equation}
  \lambda \overline{v_k v_j} =  \int v_k v_j f \text{d}^3 v,  
\end{equation}
where $\lambda$ is the intrinsic luminosity density of galaxies. The second-order velocity moments are projected onto the plane of sky with an inclination $i$, to obtain $\overline{v^2_{\rm LOS}}$ as Eq.~\ref{eq:vlos}, which can be related with the measurable quantities $v$ and $\sigma$ as follows
\begin{flalign}
    \overline{v^2_{\rm LOS}} = \frac{1}{\mu(x',y')}   \int_{-\infty}^{\infty} \lambda [(\overline{v_R^2}   \text{sin}^2\phi + \overline{v_{\phi}^2} \text{cos}^2 \phi)~\text{sin}^2i + \\ \overline{v_z^2} \text{cos}^2i]~\text{dz}'
    = v^2 + \sigma^2 = V^2_{\rm rms}.
    \label{eq:vlos}
\end{flalign}
In this equation, $\overline{v^2_{\rm LOS}}$ is expressed in cartesian coordinates $(x',y',z')$, where $x'$ and $y'$ presents the coordinates on the plane of sky and $z'$ describes the direction along the line of sight. The observed surface brightness of galaxies $\mu$ is the intrinsic surface brightness $\lambda$ projected onto the plane of sky with an inclination $i$. Both the surface brightnesses ($\lambda$, $\mu$) and the potential $\psi_{\rm D}$ can be modeled with multiple two-dimensional Gaussians. We can thus connect these components via the stellar mass-to-light luminosity ratio $\Gamma = M_{*}/L$. The terms $\lambda$ and $\psi_{\rm D}$ can be incorporated into the Jeans equations and Eq.~\ref{eq:vlos} to infer the dynamical parameters of galaxies ($\beta_z$, $i$, $\Gamma$) given the measured second-order moments $V_{\rm rms}$.

\subsection{Best-fitting parameters}
\label{ssec:bestfit}

Both {\tt GLEE} and {\tt GLaD} use parametrized profiles and the Bayesian method to infer the values of the free model parameters. The posterior distributions of parameters $P(\boldsymbol \eta| \boldsymbol d)$ describe the probability of obtaining a set of parameters $\boldsymbol \eta$ given the data set $\boldsymbol d$ as
\begin{equation}
P(\boldsymbol \eta| \boldsymbol d) = \frac{\mathcal{L}(\boldsymbol d|\boldsymbol \eta)P(\boldsymbol \eta)}{P(\boldsymbol d)}, 
\label{eq:bayes}
\end{equation}
where $P(\boldsymbol \eta)$ is the prior on the parameters, which we assumed to be uniform, except for the centroids of mass components which have Gaussian priors. $P(\boldsymbol d)$ is the evidence, and $\mathcal{L}(\boldsymbol d| \boldsymbol \eta)$ is the likelihood presenting the probability that the measurements $\boldsymbol d$ are produced by $\boldsymbol \eta$. It is related to the squared residuals $\chi^2$ between the modeled and observed data,
\begin{equation}
\mathcal{L}(\boldsymbol d|\boldsymbol \eta) \propto \text{exp} \left \{ -\frac{\chi^2}{2}\right \}.    
\end{equation}
and the maximal likelihood therefore corresponds to the minimal $\chi^2$. Note that $\boldsymbol \eta$ refers to parameters in the mass and light profiles in our lensing-only models, and to the mass, light, and dynamical parameters in our joint lensing and dynamical analysis. The input data $\boldsymbol d$ depend on the configuration of each model and will be introduced specifically in Sects.~\ref{sec:sl_model} and \ref{sec:glad_model}. The posterior distributions of parameters $P(\boldsymbol \eta| \boldsymbol d)$ , given the data $\boldsymbol d$ are sampled with Markov chain Monte Carlo (MCMC) methods, using the {\tt emcee} ensemble sampling algorithm by \citet{foreman13}. We ensured that the MCMC chains converge to steady state distributions, which indicates that the posterior probability distribution functions (PDFs) of the free parameters are well sampled and can be used to find the maximum of $P(\boldsymbol \eta|\boldsymbol d)$.

In practice, we computed the models in two steps. First of all, we assumed realistic measurement errors on the observed quantities which are, depending on the model, either the centroid positions of multiple images, the position of pixels along the extended arcs, the lens stellar kinematics per spatial bin, or a combination. We tested various mass parametrizations, ran the parameter optimization for each model, and compared the output reduced $\chi^2$ per degrees-of-freedom (d.o.f.). The concrete $\chi^2$ form depending on the model will be introduced in Sects.~\ref{sec:sl_model} and \ref{sec:glad_model}. Secondly, a range of good mass models with low $\chi^2/{\rm d.o.f.}$ were selected and we ran MCMC chains to sample the posterior PDFs. For this step, we slightly increased the measurement errors in order to obtain a rescaled $\chi^2/{\rm d.o.f.} = 1$ and to infer the realistic parameter uncertainties listed together with the best-fit and marginalized values in Sect.~\ref{sec:sl_model} and~\ref{sec:glad_model}. This approach allowed for a direct comparison of models based on different sets of constraints or different mass parametrizations. Rescaling the $\chi^2$ also accounts for mass perturbations not represented by the parametric descriptions of the lens potentials, such as line-of-sight components, or asymmetries in the main lens plane.

\section{Mass modeling with strong lensing}
\label{sec:sl_model}

To begin, we modeled the total mass of the deflectors via the strong lensing technique. In Sect.~\ref{ssec:light_model}, we present the modeling of the lens light distribution which serves as an input for the lens mass modeling. In Sect.~\ref{ssec:pos_model}, we constrained the total mass of the deflectors using the observed centroid positions of multiple images, both in the single-plane and multiplane scenarios. To describe the dense environment of CSWA\,31, we first adopted composite mass models with a single lens plane at redshift $z=0.6828$ to account for the BGG, group members, the extended group-scale dark-matter halo, and a constant external shear. Given that light rays from the most distant lensed source can be deflected by other mass components near the line-of-sight, we then conducted multiplane modeling with a secondary lens at redshift $z=1.4869$. In Sect.~\ref{ssec:sb_model}, we present the modeling of the extended lensed source emission. Since the surface brightness is conserved between the source and observed images, we exploited the information from the pixel light intensities of extended lensed arcs in order to refine the best-fit mass models obtained from image positions. 

\begin{figure}
  \includegraphics[width=0.45\textwidth]{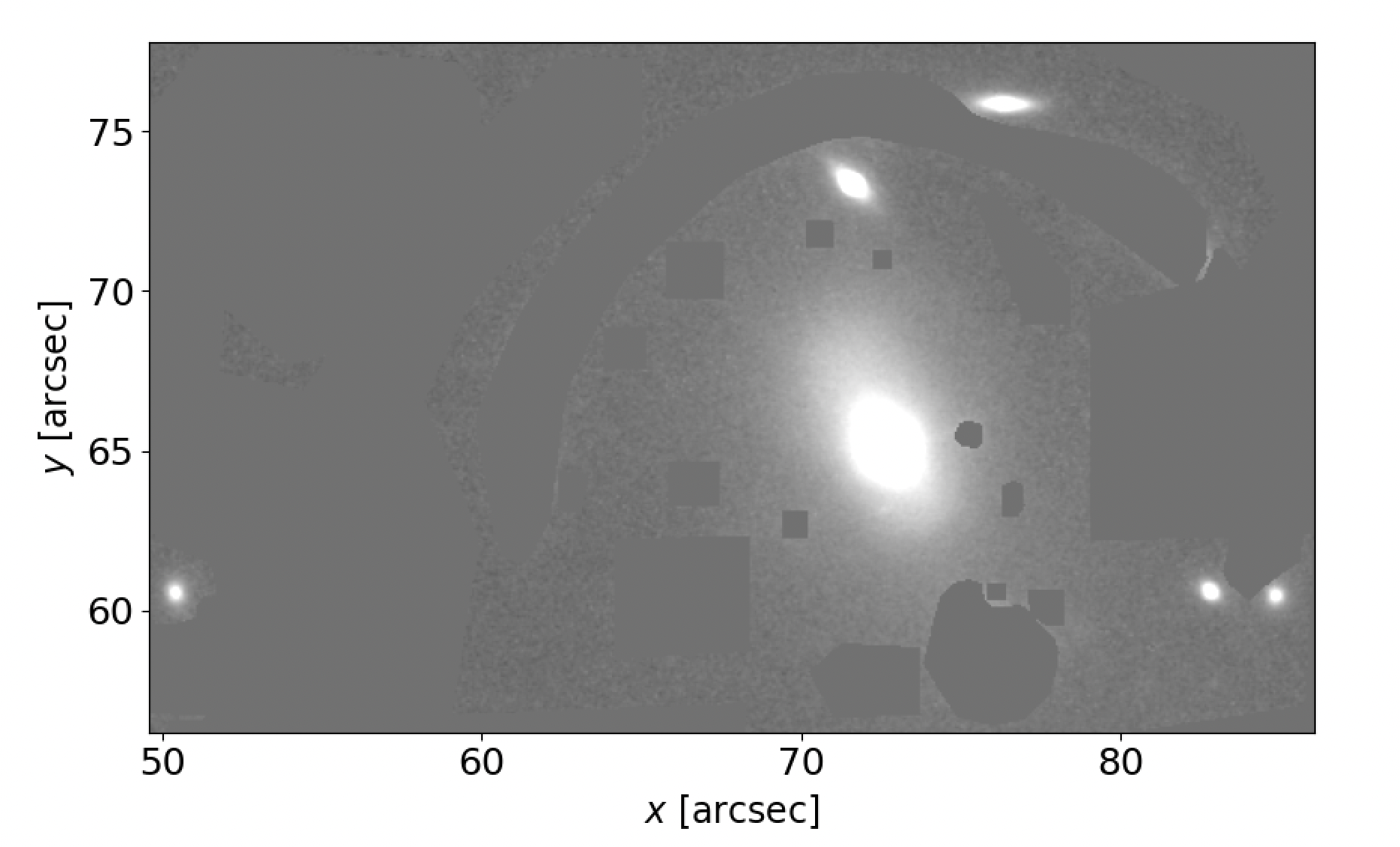}
  \includegraphics[width=0.45\textwidth]{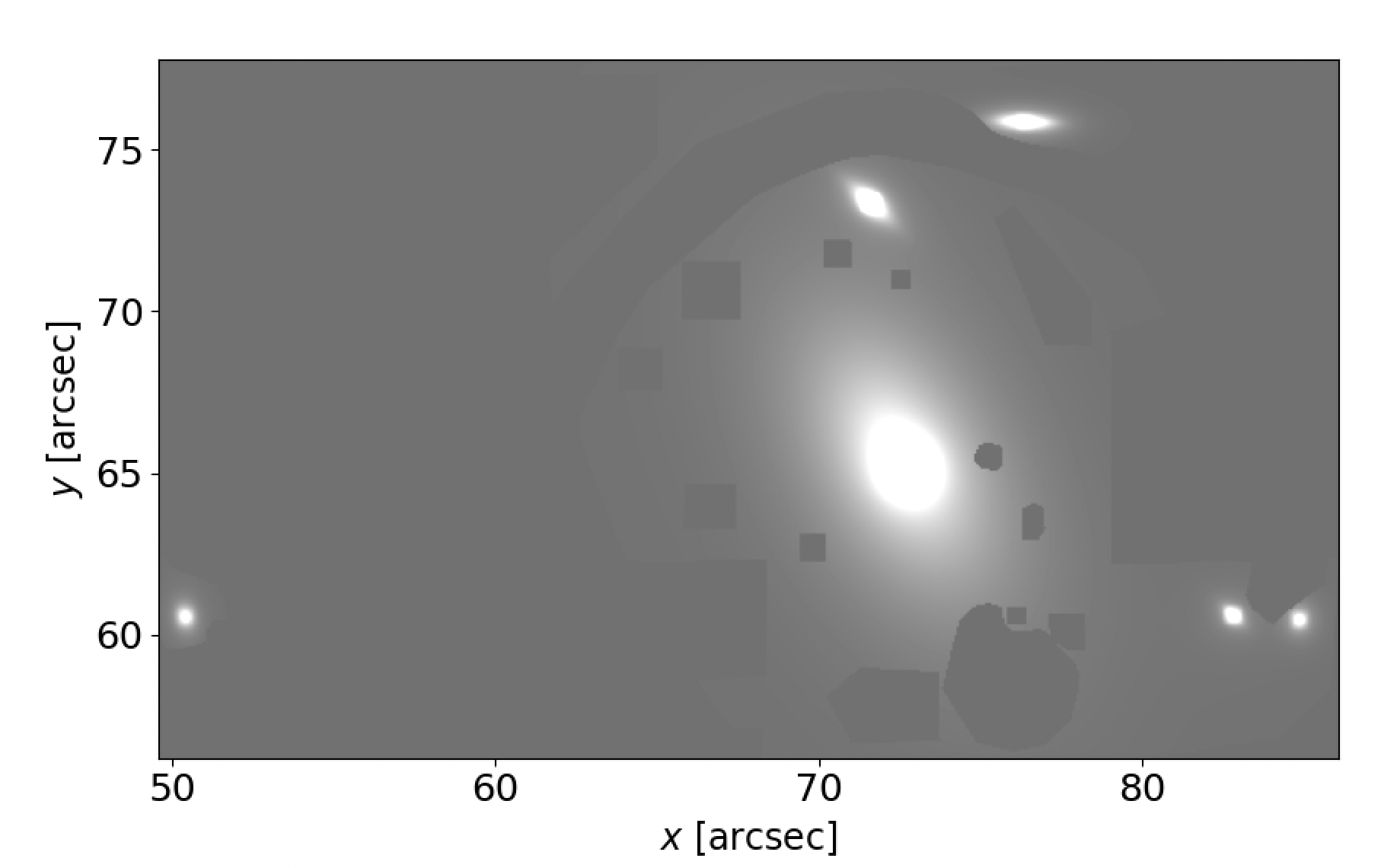}
  \includegraphics[width=0.45\textwidth]{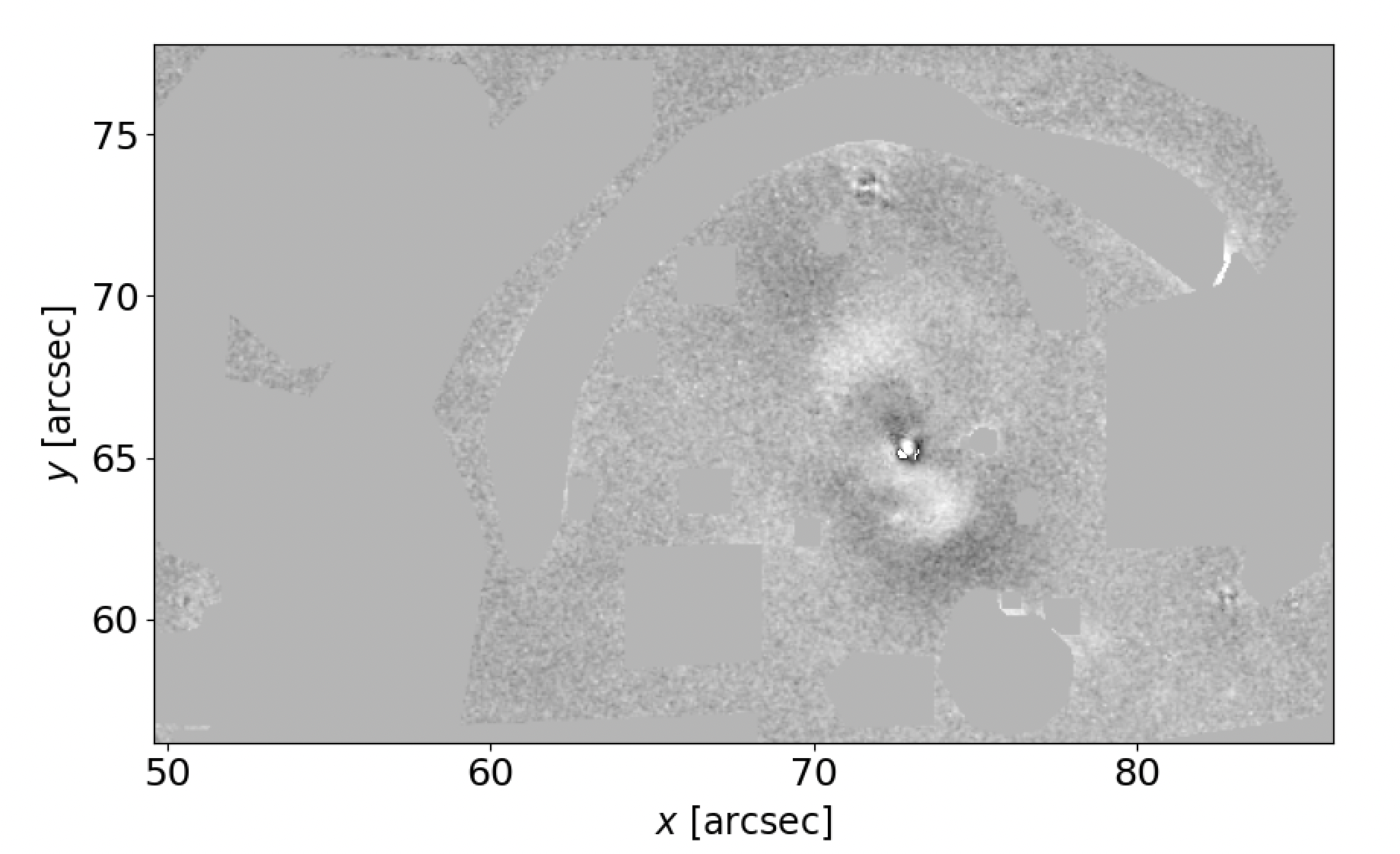}
  \caption{Best-fit model for the light distribution of the BGG and five nearby group members. {\it From top to bottom}: the observed {\it HST} F160W band image, the best-fit model, and the normalized residuals in a range between $-7\sigma$ to $7\sigma$. In all three images, the constant areas are the masks covering the lensed arcs and other objects in the field. The panels cover a field-of-view of 36\arcsec\,$\times$\,22\arcsec\ and are oriented in the same way as in Fig.~\ref{fig:CSWA31} with north, approximately downward pointing.
    }
  \label{fig:light_mod_six}
\end{figure}

\subsection{Lens light modeling}
\label{ssec:light_model}

We characterized the stellar continuum emission of the brightest foreground galaxies using the high-S/N F160W image. Modeling the light distribution of the BGG and other individual lens galaxies helps us define priors for several parameters of our lens models such as the centroids, axis ratios, and position angles of the galaxy-scale mass components. In addition, some group members are in close proximity to the lensed arcs, and we need to subtract their light contamination before conducting the extended image modeling in Sect.~\ref{ssec:dm_bar}. Lastly, light modeling is particularly useful to distinguish the baryonic mass component of the BGG from dark matter in Sect.~\ref{sec:glad_model}. 

For each galaxy, we adopted one or multiple light profiles to model the surface brightness of each galaxy $I_j$ in pixel $j$, by convolving with our PSF model in F160W band described in Sect.~\ref{ssec:hst}. We minimised the sum of the offsets between the predicted $I_j \otimes \text{PSF}$, and the observed $I_{\text{obs},j}$ intensities in $N_\text{p}$ pixels with the following $\chi^2$
\begin{equation}
  \chi_{\text{light}}^2 = \sum^{N_\text{p}}_{j = 1} \frac{\left(I_{\text{obs},j}-I_j \otimes \text{PSF} \right)^2}{\sigma_{\text{total,} j}^2},
  \label{Eq:noise}
\end{equation}
where the term $\sigma_{\rm total}$ includes background noise $\sigma_{\rm bg}$ and Poisson noise $\sigma_{\rm Poisson}$ as
\begin{equation}
\sigma_{\text{total,} j}^2 =  \sigma_{\text{bg,}j}^2 + \sigma_{\text{Poisson,}j}^2.   
\end{equation}

We simultaneously fitted the light distribution of the BGG and of five group members indicated in Fig.~\ref{fig:input} that are in proximity to the lensed images. Such early-type, elliptical galaxies are well described by S\'ersic profiles \citep{sersic63}. Fig.~\ref{fig:input} shows that the light emission of the BGG has a prominent diffuse component towards the South, and we adopted two S\'ersic profiles with different central positions to account for this asymmetry. To describe the surface brightness of the other five group members with symmetrical isophotes, we used a S\'ersic and a Gaussian profile with linked centroids. During the fitting process, we masked out the flux from lensed arcs and other neighboring galaxies over the field.

The best-fit parameters 1-$\sigma$ (68 \% CI) uncertainties for the light distribution of the BGG shown in the Table~\ref{tab:light_BGG} are the most probable values inferred from the peak of the joint posterior PDFs. We also listed the median and the 16th and 84th percentiles of the one-dimensional marginalized posterior PDFs. The best-fit model and its normalized residual are shown in Fig.~\ref{fig:light_mod_six}. We obtained $ \chi_{\rm light}^2 = 1.76 \times 10^5$ which corresponds to a reduced $\chi_{\rm light}^2/\text{d.o.f} = 2.4$. The BGG dominates the contributions to the global $\chi_{\rm light}^2$ for the six galaxies, due to the strong central residual and slight underfitting towards the South, and fitting only the BGG while masking other group members in the field leads to $ \chi_{\rm light, BGG}^2/\text{d.o.f} = 2.3$. We note that the central residuals are restricted to $\lesssim$~0.5\arcsec\ and do not alter the extended image modeling. The underfitting of the BGG towards the South highlights the limitations of our parametrization with symmetrical Sérsic profiles, and does not significantly impact the surface brightness distribution of the lensed arcs. The emission from group members shows lower residuals (Fig.~\ref{fig:light_mod_six}).

The first Sérsic profile in Table~\ref{tab:light_BGG} accounts for the majority of the light emission from the BGG and we used its effective radius, $R_{\rm eff}=27.2$~kpc at $z = 0.6828$, as measurement of the BGG size. This value is consistent with the effective radius fitted with a single S\'ersic component, with the estimate from \citetalias{grillo13}, and matches the size of BGGs at lower-redshift \citep[$z \sim 0.2$--0.45,][]{newman15}. We also used the best-fit position, axis ratio $q$ and position angle $\theta_\text{PA}$ of the first Sérsic profile to constrain the BGG mass potential in our lens models. As explained below, other group members in the field are described with simple spherical potentials fixed at the positions from our light modeling.

\begin{table*}
  \renewcommand\arraystretch{1.5}
  \centering
  \caption{Best-fit and marginalised values with 1-$\sigma$ uncertainties for the near-infrared light modeling of the BGG in CSWA\,31.}
  \begin{tabular}{ccccc}
    \hline
    \hline
    &\multicolumn{2}{c}{Sérsic profile 1 (BGG)}
    &
    \multicolumn{2}{c}{Sérsic profile 2 (BGG)} \\\cmidrule(r){2-3}\cmidrule(l){4-5}
    Parameter & Best-fit value & Marginalised value & Best-fit value & Marginalised value \\
    \hline
    $x$ [arcsec] & 72.342 & $72.341_{-0.001}^{+0.001}$ & 72.561 & $72.561_{-0.003}^{+0.003}$ \\
    $y$ [arcsec] & 65.505 & $65.508_{-0.002}^{+0.002}$ & 65.157 & $65.156_{-0.003}^{+0.003}$ \\
    $q$ & 0.570 & $0.570_{-0.001}^{+0.001}$ & 0.935 & $0.935_{-0.003}^{+0.003}$ \\
    $\theta_{\rm PA}$ [$^\circ$] & $116.204$ & $116.204_{-0.06}^{+0.06}$ & $-23.190$ & $-23.190_{-0.001}^{+0.001}$ \\
    $A$ [$\rm mag~arcsec^{-2}$] & 0.117 & $0.117_{-0.001}^{+0.001}$ & 1.281 & $1.280_{-0.004}^{+0.004}$ \\
    $R_{\rm eff}$ [arcsec] & 3.850 & $3.858_{-0.005}^{+0.005}$ & 0.471 & $0.471_{-0.002}^{+0.001}$ \\
    $n$ & 1.092 & $1.091_{-0.003}^{+0.003}$ & 1.297 & $1.298_{-0.004}^{+0.004}$ \\
    \hline
  \end{tabular}
  \tablefoot{The centroid coordinates $x$ and $y$ follow the orientation displayed in the Fig.~\ref{fig:CSWA31}. $q$ is the axis ratio, $A$ is the amplitude, $\theta_\text{PA}$ is the position angle measured counterclockwise from the x-axis, $R_{\rm eff}$ is the effective radius, and $n$ is the Sérsic index. We adopted the value $R_{\rm eff}=3.85$\arcsec\ of the first Sérsic profile as the effective radius of the BGG, since this first profile dominates the overall light distribution. This corresponds to $R_{\rm eff}=27.2$~kpc.}  
  \label{tab:light_BGG}
\end{table*}

\subsection{Image position modeling}
\label{ssec:pos_model}

Strong gravitational lensing provides tight constraints on the lens mass distribution over radial ranges where the lensed arcs emerge. In this section, we present the lens mass modeling inferred from the peak positions of multiple images belonging to the six families presented in Sect.~\ref{ssec:ima_fam} and plotted in Fig.~\ref{fig:input}. These sets provided us with 36 constraints to build composite models of the lens mass distribution, and to fit a large number of parameters in the convergence $\kappa$. The cumulative one-dimensional lens mass profiles were then computed via Eq.~\ref{eq:total_mass}. After deducing the best-fit lens mass parameters, the intrinsic source positions retraced from each multiple image are expected to agree with each other. In practice, since residuals in our mass modeling introduces small shifts, we defined the intrinsic source position as the mean source position weighted by the magnification $\mu$ of the observed images.

\subsubsection{Lens mass parametrization}
\label{ssec:Lens_mass_parametrization}
To characterize the structure of CSWA\,31 in detail and to separate the lens mass components on various scales, we used composite models including the BGG, group members, the dark-matter group halo, and a constant external shear for the dense environment around CSWA\,31. Strong gravitational lensing and dynamical modeling of early-type galaxies have shown that their total mass density profiles are well described by nearly-isothermal power-law models \citep{koopmans06,gavazzi07,barnabe09,sonnenfeld13,tortora14,cappellari15}. Consequentely, we modeled the total mass of the BGG and group members with dPIE profiles. Due to the limited number of multiple images, we scaled the total mass of the group members to a member selected arbitrarily, the so-called reference galaxy, as commonly done in galaxy-cluster lens modeling \citep[e.g.,][]{richard10,grillo16,limousin16,caminha19,chirivi18} and we assumed circular symmetry. Given that such galaxies lie on the fundamental plane, we assumed that their Einstein radius and truncation radius scale with their F160W band luminosities. We followed \citet{grillo15} and \citet{chirivi18} in setting the following scaling relations to ensure that total mass-to-light ratios follow the tilt of the fundamental plane
\begin{equation}
  \theta_{\text{E},i} = \theta_{\rm E,ref}\left(\frac{L_i}{L_{\rm ref}}\right)^{0.7}, \hspace{2cm} r_{\text{tr},i} = r_{\rm tr,ref}\left(\frac{L_i}{L_{\rm ref}}\right)^{0.5},
  \label{eq:scaling_rel}
\end{equation}
where $L_{\rm i}$ and $L_{\rm ref}$ are the F160W band luminosities of group member $i$ and reference galaxy, respectively. The mass of an arbitrary group member $i$ in CSWA 31 can be characterised by $\theta_{\rm E,ref}$ and $r_{\rm tr,ref}$, once we determined the light ratio $L_i/L_\text{ref}$ between the group member $i$ and the reference galaxy and plugged it into the scaling relation \ref{eq:scaling_rel}. Thus, the mass
of an individual group member is regarded as a multiple of the reference galaxy. Instead of varying the individual mass profiles, we optimized the reference galaxy mass in terms of $\theta_{\rm E,ref}$ and $r_{\rm tr,ref}$ to reduce the number of optimized parameters. 

\begin{table*}
  \renewcommand\arraystretch{1.5}
  \centering
  \caption{Best-fit and marginalized values with 1-$\sigma$ uncertainties for the parameters of our lensing-only models based on image positions, in the single-plane (Img-SP (L)) and multiplane scenarios (Img-MP (L)).}
  \begin{tabular*}{\textwidth}{c@{\extracolsep{\fill}}*{6}{c}}
    \hline
    \hline
    & & \multicolumn{2}{c}{Img-SP (L)}
    &
    \multicolumn{2}{c}{Img-MP (L)*} \\\cmidrule(r){3-4}\cmidrule(r){5-6}
    Component & Parameter & Best-fit & Marginalised & Best-fit & Marginalised \\
    \hline
    BGG &$r_\text{tr,BGG}$ [\arcsec] & 17.60 & $17.6_{-1.8}^{+1.6}$ & 16.28 & $16.3_{-1.4}^{+1.1}$ \\
    (dPIE) & $\theta_\text{E,BGG}$ [\arcsec] & 7.79 & $7.5_{-0.8}^{+0.8}$ & 9.07 & $8.9_{-1.1}^{+1.0}$ \\
    \hline
    & $x_\text{GH}$ [\arcsec] & 71.86 & $72.1_{-0.3}^{+0.2}$ & 72.79 & $72.7_{-0.3}^{+0.3}$ \\
    & $y_\text{GH}$ [\arcsec] & 64.34 & $64.4_{-0.3}^{+0.3}$ & 63.24 & $63.3_{-0.5}^{+0.5}$ \\
    Group Halo & $q_\text{GH}$ & 0.96 & $0.9_{-0.03}^{+0.05}$ & 0.833 & $0.8_{-0.04}^{+0.04}$  \\
    (SPEMD) & $\theta_\text{PA,GH}$ [$^\circ$] & 106.8 & $110.5_{-9.2}^{+16.0}$ & 133.9 & $128.3_{-4.0}^{+4.6}$ \\
    & $r_\text{core,GH}$ ["] & 13.85 & $12.1_{-1.7}^{+1.2}$ & 8.65 & $9.0_{-1.7}^{+1.7}$ \\
    & $\theta_\text{E,GH}$ [\arcsec] & 24.13 & $25.6_{-1.3}^{+1.2}$ & 19.11 & $20.4_{-1.6}^{+1.5}$ \\
    & $\gamma_\text{GH}$ & 0.94 & $0.9_{-0.1}^{+0.01}$ & 0.79 & $0.9_{-0.1}^{+0.10}$ \\
    \hline
    & $r_\text{tr,ref}$ [\arcsec] & 5.86 & $3.8_{-1.8}^{+1.6}$ & 5.42 & $4.5_{-1.4}^{+1.1}$ \\
    & $\theta_\text{E, ref}$ [\arcsec] & 0.044 & $0.2_{-0.1}^{+0.2}$ & 0.44 & $0.6_{-0.3}^{+0.3}$ \\
    Group Members & $\theta_\text{E,1}$ [\arcsec] & 0.89 & $1.3_{-0.6}^{+0.7}$ & 1.66 & $1.8_{-0.5}^{+0.6}$ \\
    (dPIE) &$\theta_\text{E,2}$ [\arcsec] & 0.81 & $0.7_{-0.3}^{+0.4}$ & 1.54 & $1.5_{-0.3}^{+0.3}$ \\
    & $\theta_\text{E,6}$ [\arcsec] & 1.24 & $1.8_{-0.8}^{+0.9}$ & 2.89 & $2.3_{-1.0}^{+0.7}$ \\
    & $\theta_\text{E,7}$ [\arcsec] & 2.48 & $2.8_{-0.4}^{+0.6}$ & 2.09 & $2.1_{-0.4}^{+0.4}$ \\
    \hline
    S0 (PIEMD) & $\theta_\text{E,S0}$ [\arcsec] &- &- & 4.36 & $4.5_{-1.1}^{+1.2}$ \\
    \hline
    External & $\gamma_\text{ext}$ & 0.033 & $0.03_{-0.02}^{+0.03}$ & 0.063 & $0.06_{-0.004}^{+0.004}$ \\
    Shear & $\theta_\text{ext}$ [$^\circ$] & 190.72 & $182.9_{-28.6}^{+28.6}$ & $179.39$ & $179.6_{-1.4}^{+1.4}$ \\
    \hline
  \end{tabular*}
  \tablefoot{The parameters listed are the BGG truncation radius $r_\text{tr,BGG}$, the BGG Einstein radius $\theta_\text{E,BGG}$, the group halo centroid ($x_\text{GH}$, $y_\text{GH}$) with respect to the reference coordinates in Fig.~\ref{fig:CSWA31}, the group halo axis ratio $q_\text{GH}$, position angle $\theta_\text{PA,GH}$ measured counterclockwise from the x-axis, core radius $r_\text{core,GH}$, Einstein radius $\theta_\text{E,GH}$, and density slope $\gamma_\text{GH}$, the reference galaxy truncation radius $r_\text{tr,ref}$, and Einstein radius $\theta_\text{E,ref}$, the Einstein radius of group members 1, 2, 6, 7, the Einstein radius of S0 $\theta_\text{E,S0}$ for the multiplane model, the magnitude of the external shear $\gamma_\text{ext}$, and the shear angle $\theta_\text{ext}$ measured counterclockwise from the x-axis. The Einstein radii of group members 1, 2, 6, and 7 are fitted separately, while their truncation radius are scaled to the reference galaxy (see text for details). Note that all $\theta_\text{E}$ values shown in the table are scaled for sources at redshift $z = \Infinity$. To obtain the actual $\theta_{\rm E}$ for a given source at redshift $z_{\rm s}$, the displayed $\theta_{\rm E}$ needs to be multiplied with the respective $D_{\rm ds}/D_{\rm s}$.}
  \label{tab:img_chi2}
\end{table*}

Foreground galaxies in proximity to the lensed images have a significant impact on the deflection angle $\boldsymbol \alpha$ of the light rays from the corresponding background source. To account for this, we fitted separately the Einstein radius $\theta_{\rm E}$ of four group members (1, 2, 6, and 7 in Fig.~\ref{fig:input}). The truncation radius of these galaxies is comparable to their half-mass radius and typically larger than the scales where the multiple images emerge. Consequently, we do not expect good constraints on $r_{\rm tr}$ from the strong lensing configuration of CSWA\,31 and we kept their truncation radius $r_{\rm tr}$ in the scaling relations. Note that the aforementioned group members 3, 4, 5 induce significant light contamination to the main arc, but their surface brightness in the F160W band suggests lower masses and hence smaller contributions to the light deflection. For both BGG and group members, we assumed vanishing core radii since the strong lensing constraints do not cover the inner region of these individual galaxies.

We chose the SPEMD to model the mass distribution of the underlying extended dark-matter halo. This profile has more flexibility than the NFW profile thanks to its variable surface mass-density slope $\gamma$. We imposed flat priors on the parameters of the group-scale SPEMD and, in particular, we restricted its core radius to $r_{\rm core, GH}=6$\arcsec--14\arcsec. This range is motivated by recent strong lensing studies of extended dark-matter halos \citep[e.g.,][]{richard21} and using strict boundaries allowed us to exclude models with arbitrarily large SPEMD cores. In addition, other group members located out of the MUSE field-of-view are not part of the scaling relations. A constant external shear component was added to account for the tidal stretching to the multiple images induced by such neighbouring galaxies and by other mass components in CSWA\,31. 

We then ran MCMC chains to sample the maximal posterior probability $P$ (see Eq.~\ref{eq:bayes}), which is presented in terms of priors and $\chi_{\rm img}^2$ defined as
\begin{equation}
\chi_{\rm img}^2 = \sum_{i = 1}^{N_\text{set}} \sum_{j = 1}^{N_\text{img}} \frac{\left(\theta_{j,i}^\text{obs}- \theta_{j,i}^\text{pred}(\boldsymbol \eta , \boldsymbol \beta)\right)^2}{\sigma_{j,i}^2},
\label{eq:chi2_img}
\end{equation}
where $N_{\rm set}$ is the number of multiple image families (see Fig.~\ref{fig:input}), $N_{{\rm img},i}$ is the number of individual counter-images in family $i$, $\theta_{j,i}^{\rm obs}$ is the observed position of image $j$ from multiple image family $i$, $\theta_{j,i}^{\rm pred}$ is the predicted image position, and $\sigma_{j,i}$ is the position uncertainty of image $j$ from multiple image family $i$.

We introduced elliptical positional uncertainties for the multiple images with high magnification that form the extended arcs, and circular uncertainties for S0(d) and image set S4 (Table~\ref{tab:multima}). For sets S0, S1 and S2, we fixed the positional uncertainties along the elliptical minor and major axes to one and two {\it HST} image pixels, respectively, with major axes oriented along the direction of the arcs. For S3(b), we used four pixels as positional error along the major axis given the diffuse morphology of this extended arc, and for S4 identified by MUSE, we also used larger errors (e.g. three pixels) due to the larger MUSE PSF. For set S5, we used larger errors along both elliptical axes because S5 is fainter than other image sets detected with {\it HST}. After comparing the $\chi^2$ of various models, we rescaled these uncertainties by a factor of three in order to get a $\chi_{\rm img}^2/{\rm d.o.f} = 1$. This allowed us to derive realistic parameter uncertainties, to directly compare the models with different structures (e.g. single-plane versus multiplane), and to account for the simplistic dark-matter halo parametrizations, the number of possible missing group members out of the field of MUSE, the possible asymmetries in the lens mass distribution, and other perturbations along the line-of-sight. The latter include, for instance, six galaxies detected with MUSE at $z=1.357$, which are all distant from the BGG centroid and should therefore only produce a small perturbation to the lens mass modeling.

\begin{table*}
  \renewcommand\arraystretch{1.5}
  \centering
  \caption{Best-fit and marginalized values with 1-$\sigma$ uncertainties for the parameters of our lensing-only models based the extended surface brightness distribution of sources S0 and S3, and on the positions of multiple images from other sets.}
  \begin{tabular*}{\textwidth}{c@{\extracolsep{\fill}}*{6}{c}}
    \hline
    \hline
    & &\multicolumn{2}{c}{Esr2-MP (L)}
    &
    \multicolumn{2}{c}{Esr2-MP$_{\rm test}$ (L)*} \\\cmidrule(r){3-4}\cmidrule(l){5-6}
    Component & Parameter & Best-fit & Marginalised & Best-fit & Marginalised \\
    \hline
    BGG & $r_\text{tr,BGG}$ [\arcsec] & 15.010 & $15.010_{-0.005}^{+0.003}$ & 14.107 & $14.11_{-0.06}^{+0.07}$ \\
    (dPIE) & $\theta_\text{E,BGG}$ [\arcsec] & 8.558 & $8.54_{-0.05}^{+0.06}$ & 8.419 & $8.33_{-0.06}^{+0.06}$ \\
    \hline
    & $x_\text{GH}$ [\arcsec] & 72.688 & $72.687_{-0.007}^{+0.007} $ & 72.741 & $72.743_{-0.01}^{+0.01}$ \\
    & $y_\text{GH}$ [\arcsec] & 63.767 & $63.77_{-0.01}^{+0.01}$ & 63.859 & $63.87_{-0.02}^{+0.02}$ \\
    Group Halo & $q_\text{GH}$ & 0.870 & $0.870_{-0.001}^{+0.001}$ & 0.846 & $0.845_{-0.002}^{+0.001}$ \\
    (SPEMD) & $\theta_\text{PA,GH}$ [$^\circ$] & 138.882 & $138.7_{-0.3}^{+0.4}$ & 134.946&  $132.3_{-0.3}^{+0.3}$ \\
    & $r_\text{core,GH}$ & 12.232 & $12.1_{-0.2}^{+0.2}$ & 9.979 & $9.8_{-0.1}^{+0.1}$ \\
    & $\theta_\text{E,GH}$ [\arcsec] &22.646  &$22.65_{-0.05}^{+0.05}$ & 22.967 & $23.09_{-0.08}^{+0.08}$ \\
    & $\gamma_\text{GH}$ & 0.871 & $0.86_{-0.01}^{+0.02}$ & 0.713 & $0.700_{-0.007}^{+0.007}$ \\
    \hline
    & $r_\text{tr,ref}$ [\arcsec] & 5.000 & $4.997_{-0.005}^{+0.003}$ & 4.698 & $4.64_{-0.06}^{+0.07}$ \\
    & $\theta_\text{E, ref}$ [\arcsec] & 0.506 & $0.502_{-0.006}^{+0.005}$ & 0.442 & $0.434_{-0.006}^{+0.006}$ \\
    Group Members & $\theta_\text{E,1}$ [\arcsec] & 1.283 & $1.177_{-0.005}^{+0.005}$ & 1.204 & $1.205_{-0.007}^{+0.008}$ \\
    (dPIE) & $\theta_\text{E,2}$ [\arcsec] & 0.749 & $0.748_{-0.006}^{+0.007}$ & 0.777 & $0.777_{-0.008}^{+0.008}$ \\
    & $\theta_\text{E,6}$ [\arcsec] & 0.921 & $0.94_{-0.05}^{+0.05}$ & 1.345 & $1.36_{-0.05}^{+0.05}$ \\
    & $\theta_\text{E,7}$ [\arcsec] & 2.672 & $2.69_{-0.02}^{+0.02}$ & 2.942 & $2.97_{-0.03}^{+0.03}$ \\
    \hline
    S0(PIEMD) & $\theta_\text{E,S0}$ [\arcsec] & 1.756 & $1.76_{-0.02}^{+0.02}$ & 1.651 & $1.61_{-0.03}^{+0.03}$ \\
    \hline
    External & $\gamma_\text{ext}$ & 0.0905 & $0.090_{-0.001}^{+0.0001}$ & 0.0939 & $0.093_{-0.001}^{+0.001}$ \\
    Shear & $\theta_\text{ext}$ [$^\circ$] & 167.387 & $167.5_{-0.3}^{+0.3}$ & 162.814 & $162.3_{-0.2}^{+0.3}$ \\
    \hline
  \end{tabular*}
  \tablefoot{We refer to Table~\ref{tab:img_chi2} for details on the parameters.}
  \label{tab:esr_chi2}
\end{table*}

\begin{figure*}
  \centering
  \begin{subfigure}[b]{0.475\textwidth}
    \centering
    \includegraphics[width=0.9\textwidth]{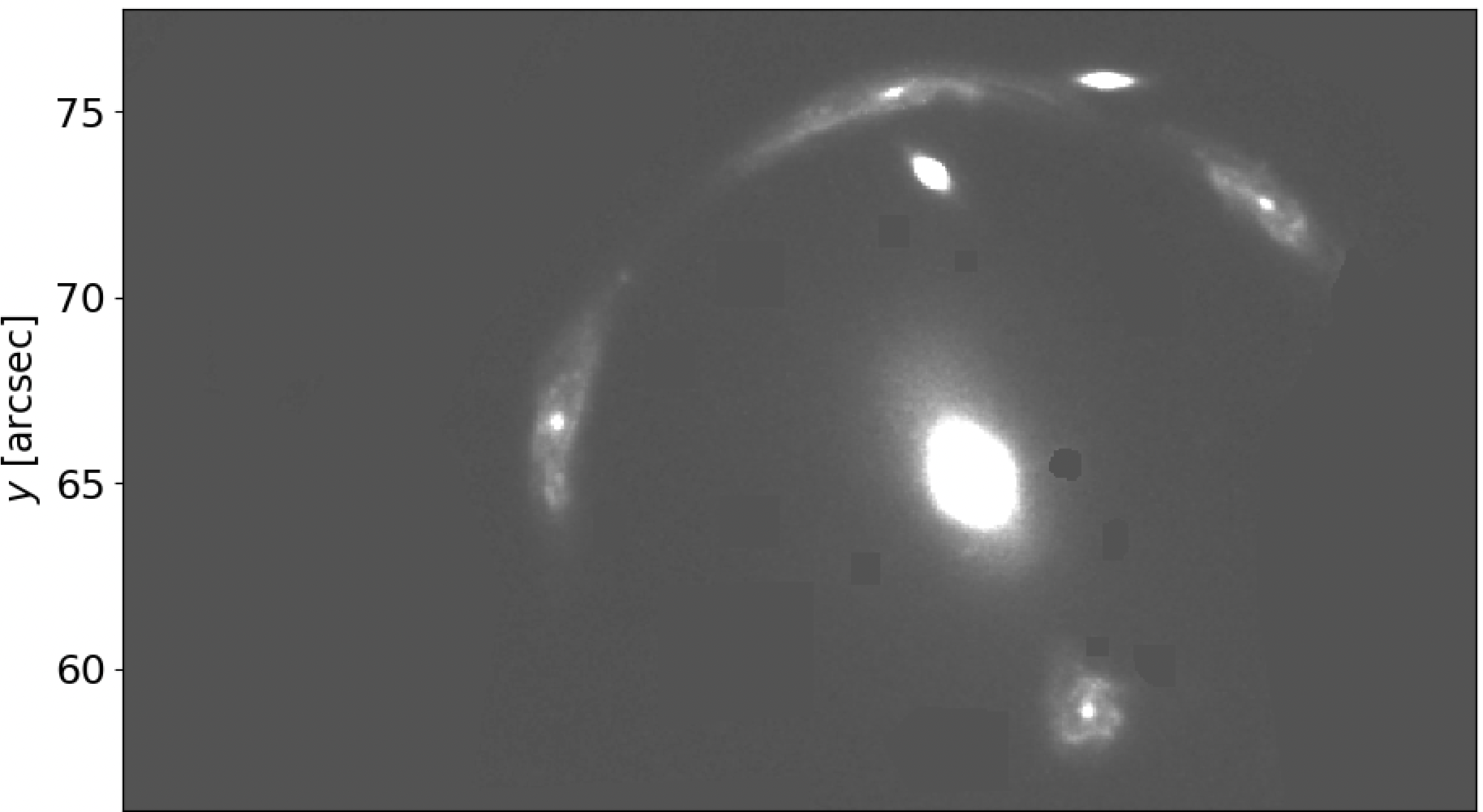}
  \end{subfigure}
  \hfill
  \begin{subfigure}[b]{0.475\textwidth}  
    \centering 
    \includegraphics[width=0.9\textwidth]{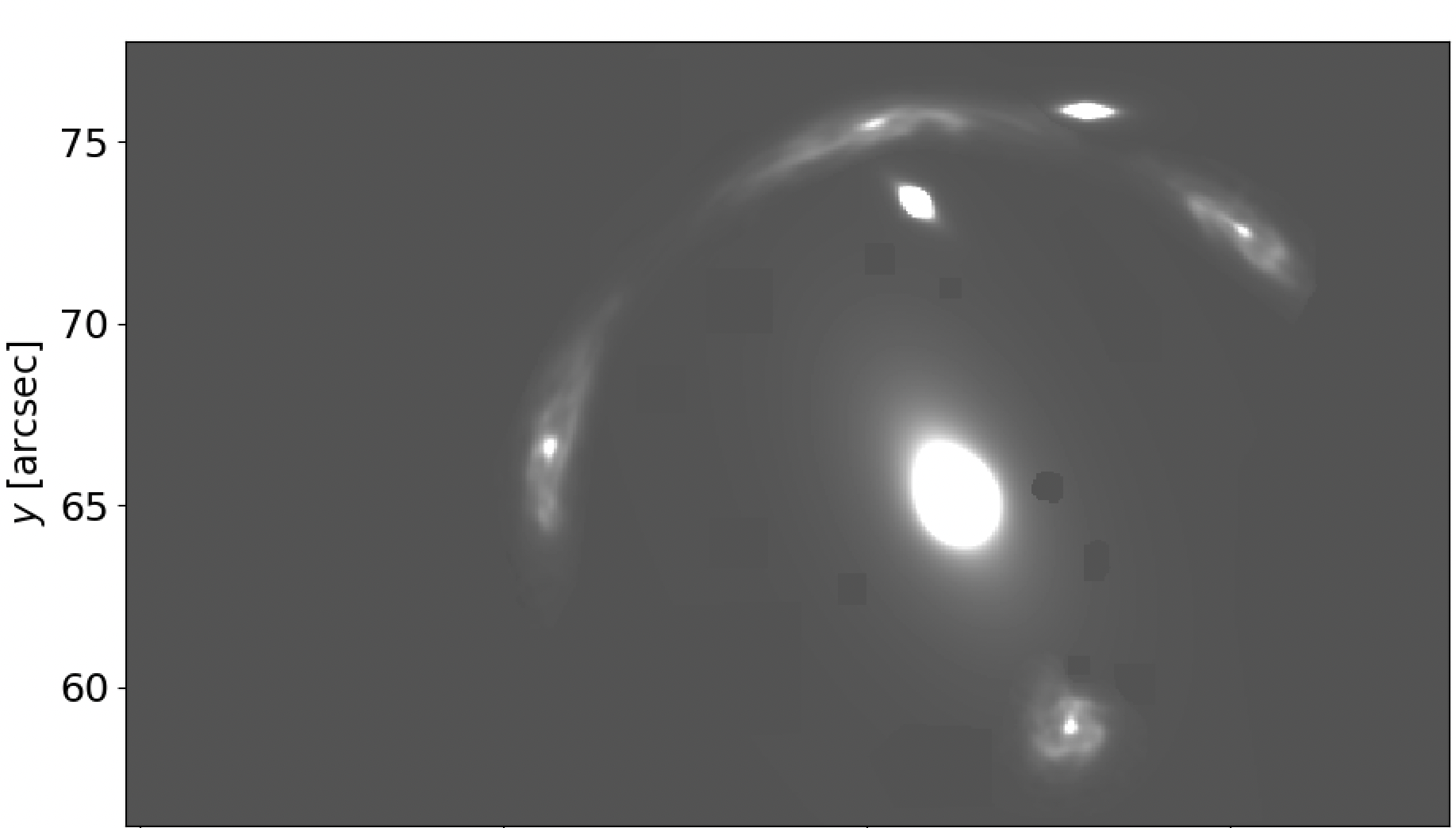}  
  \end{subfigure}
    \begin{subfigure}[b]{0.475\textwidth}
    \centering
    \includegraphics[width=0.95\textwidth]{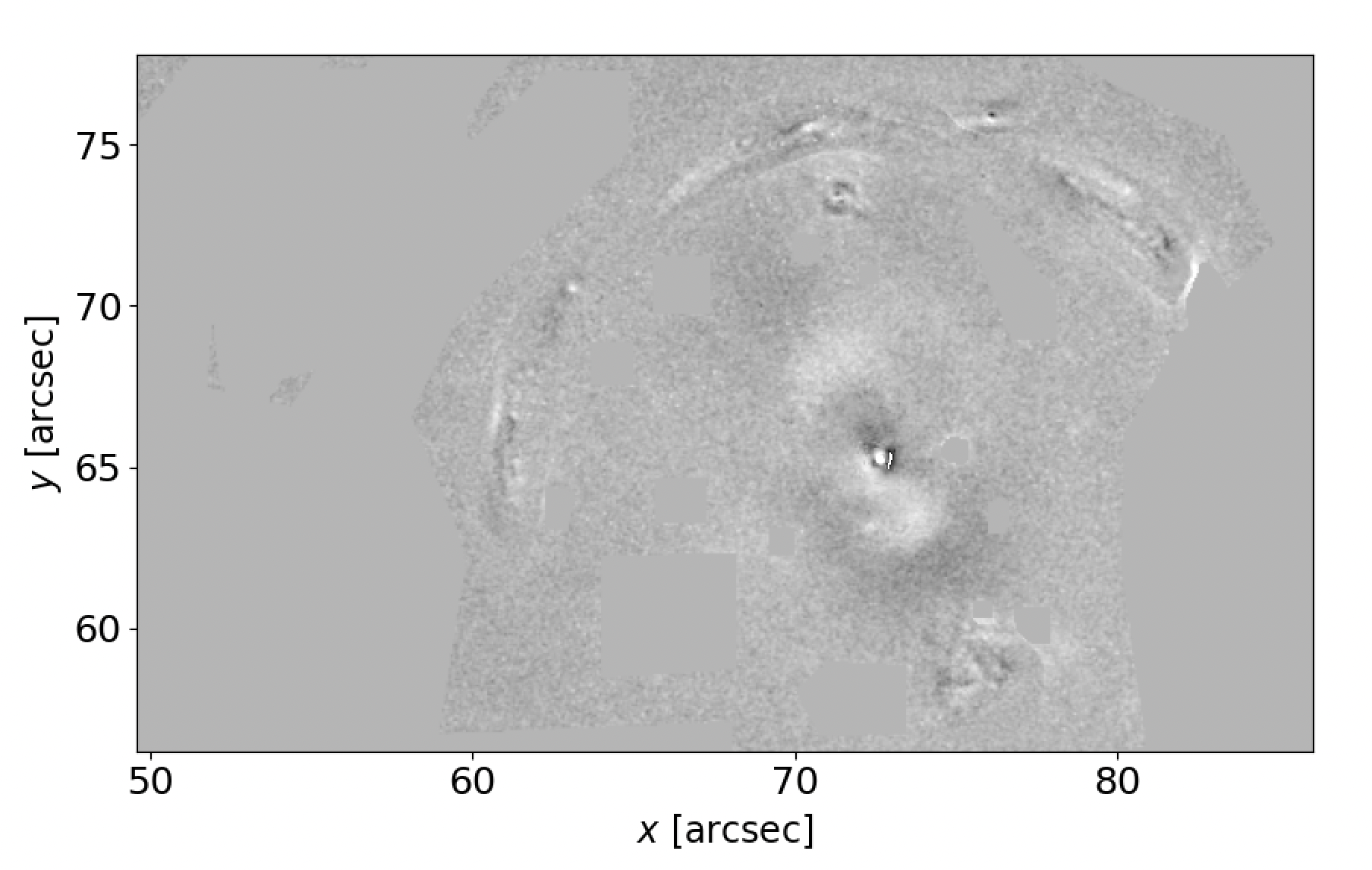}
  \end{subfigure}
  \hfill
  \begin{subfigure}[b]{0.475\textwidth}  
    \centering 
    \includegraphics[width=0.6\textwidth]{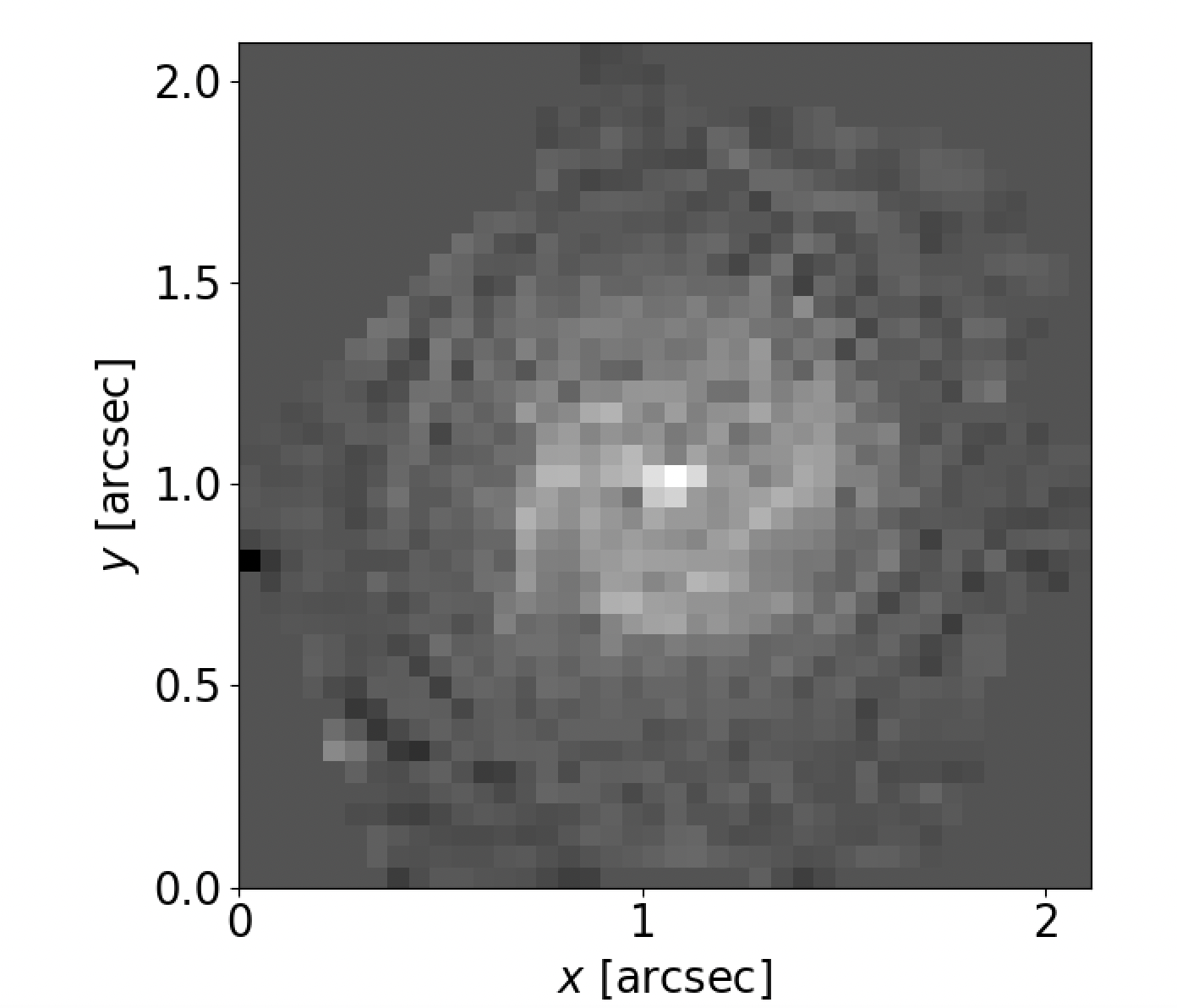}  
  \end{subfigure}  
  \caption{Surface brightness reconstruction of source S0 at redshift $z=1.4869$ from the best-fit model Esr2-MP$_{\rm test}$ (L). Two nearby group members are included in the modeling, while other objects within the grey regions are masked out. We show the observed {\it HST} F160W image over a field-of-view of 36\arcsec~$\times$~22\arcsec\ ({\it top left}), the best-fit model of the main arc, BGG and group members ({\it top right}), the normalized residual ({\it bottom left}) in a range between $-7\sigma$ to $7\sigma$, and the source-plane reconstruction of S0 over a field of 2.1\arcsec\,$\times$\,2.1\arcsec\ ({\it bottom right}). The arc is well fitted and does not show significant residuals in the image plane. S0 has a clear spiral shape in the source plane. It is well reconstructed in all our lens models, with an intrinsic morphology consistent with our reference model Esr2-MP$_{\rm test}$ (L).}
  \label{fig:arc1_mod}
\end{figure*}

\subsubsection{Results in the single and multiplane scenarios}

First, we followed the single-plane modeling approach, considering that the light rays from all background sources are deflected once by the BGG, group members and extended dark-matter halo located within a single lens plane at redshift $z=0.6828$, with negligible deflections from mass perturbations along the line-of-sight. A MCMC chain was run to minimise the $\chi^2_{\rm img}$ in Eq.~\ref{eq:chi2_img}, with the $\theta_j^\text{pred}(\eta, \beta)$ terms computed in the single-plane framework. Fig.~\ref{fig:input} shows that the observed positions of multiple images are well reproduced by this model, which we refer to as Img-SP (L), given the best-fit parameters $\eta$ of the mass profiles and the best-fit source positions $\beta$. Before rescaling the reduced $\chi_{\rm img}^2$ we obtained $\chi_{\rm img}^2/{\rm d.o.f.}=14.76$. The best-fit and marginalised parameter values with 1-$\sigma$ uncertainties are displayed in Table~\ref{tab:img_chi2}.

Second, we expanded this model to multiple lens planes, as done in strong lens modeling of galaxies \citep[e.g.,][]{gavazzi08} or galaxy clusters \citep[e.g.,][]{daloisio14,bayliss14,chirivi18} to account for the line-of-sight mass distribution and reduce the offset between observed and predicted image positions. In CSWA\,31, we observe that the weighted mean source positions of S0, S1 and S5 inferred from model Img-SP (L) have small projected separations (see Fig.~\ref{fig:input}). This means that light rays from S5 can be deflected substantially by S0 (and S1 which belongs to the same galaxy) in addition to the main lens potential at $z=0.6828$. We thus conducted the multiplane mass modeling with S0 as a secondary lens at $z=1.4869$, using a spherical isothermal mass profile fixed at the best-fit source position from Img-SP (L). Other sources forming sets S2, S3, and S4 reside in between the observer and S5, but their weighted mean positions are poorly aligned with each other and with S5, and we therefore ignored their impact on the total deflection angles. We obtained $\chi_{\rm img}^2/{\rm d.o.f}=8.87$ for this model called Img-MP (L), which is a significant improvement with respect to Img-SP (L). The results are also listed in Table~\ref{tab:img_chi2}. The best-fit value of the Einstein radius of S0, $\theta_{\rm E, S0}$, is 4.36\arcsec\ which indicates a non-negligible perturbation along the line-of-sight. 

Model Img-MP (L) has smaller rms for all sets of multiple images compared to Img-SP (L), given the best-fit parameters $\boldsymbol \eta$ of the mass profiles and the source positions $\boldsymbol \beta$. Multiple images of sources 0, 1, 2, 5 are accurately reproduced, within 0.2\arcsec\ and 0.3\arcsec\ on average in Img-MP (L) and Img-SP (L), respectively (see Fig.~\ref{fig:input}). Sets S3 and S4 show larger uncertainties and can be reproduced within 0.5\arcsec\ in Img-MP (L) and within 0.6\arcsec\ in Img-SP (L). Models Img-SP (L) and Img-MP(L) predict a third image for set S3, falling in the vicinity of the BGG centroid (see Fig.~\ref{fig:input}). This image is however demagnified by factors $\mu = 5\times 10^{-5}$ and $\mu = 0.8$ in Img-SP (L) and Img-MP (L), respectively, and it is not detected in our imaging and spectroscopic data due to strong blending with the BGG.

We conducted several tests to quantify the impact of the adopted parametrization and the selection of image constraints on the model results. Firstly, we added all group members, including those near the lensed arcs, to the scaling relations in order to model their mass exclusively via the parameters $\theta_{\rm E,ref}$ and $r_{\rm tr,ref}$. Secondly, we tested the influence of the core radius of the group-scale halo on the overall analysis by fixing this parameter to $r_{\rm core,GH}=0$ during optimisation. These two tests substantially increased the $\chi_{\rm img}^2$/d.o.f., because the output mass models were unable to recover the image positions of the most distant source galaxy (S5) at $z=4.205$. Lastly, S4 shows the largest offset between the predicted and observed image positions, and we determined the influence of S4 on the best-fit parameters by deriving new models excluding this set of multiple images. We found that the resulting best-fit parameters were within 1-$\sigma$ uncertainties compared to Img-SP (L) and Img-MP (L), and we therefore kept S4 as constraints in our models.

The marginalized parameter values given in Table~\ref{tab:img_chi2} and the probability distribution functions shown in Appendix~\ref{sec:pdf} indicate that most parameters are constrained with similar precision for our two models, and consistent within 2-$\sigma$. The values of $\theta_{\rm E}$ are listed for sources at redshift $z = \Infinity$. After rescaling to the correct source redshifts, we obtained an Einstein radius for the BGG $\theta_{\rm E,BGG}$ of about 3\arcsec\ for $z=1.4869$. This is comparable with $R_{\rm eff}$ obtained in the lens light modeling and consistent with the range of $\theta_{\rm E,BGG}$ measured for group-scale lenses in \citet{newman15}. In contrast, we observe larger angular separations approaching $\sim$10\arcsec\ for multiple images of S0 in the {\it HST} F160W frame. These separations are closer to the best-fit $\theta_{\rm E,GH}$ for $z=1.4869$ in both models and therefore likely primarily caused by the extended dark-matter halo. In addition, we find that some best-fit parameter values are more physical in model Img-MP (L), such as $\theta_{\rm E,ref} = 0.44$\arcsec\ which converges to zero in model Img-SP (L). The core of the group-scale halo is more extended in model Img-SP (L), with a best-fit value approaching the prior upper limit, and which would diverge to unrealistically large cores $r_{\rm core, GH} \gg 14$\arcsec\ for broader priors. For these reasons, and given that Img-MP (L) reproduces the image positions slightly better than Img-SP (L), we chose Img-MP (L) as the reference lens model based on image positions.

\subsection{Extended image modeling}
\label{ssec:sb_model}

Most lensed background sources detected in our {\it HST} image exhibit extended morphologies. Moreover, the bright extended arc from image set S3 at $z=2.763$ is diffuse and, contrary to source S0, the lack of structure introduces large uncertainties in our image position modeling of S3. To account for the source surface brightness distributions and to better constrain our mass models, we conducted the extended image modeling of CSWA\,31 with {\tt GLEE}. Given the surface brightness conservation $I(\boldsymbol \theta) = I(\boldsymbol \beta)$, the software reconstructs $I(\boldsymbol \beta)$ on a grid of pixels in the source plane, and then maps $I(\boldsymbol \beta)$ back to the image plane via the lens equation to obtain the predicted image morphology $I(\boldsymbol \theta)$. We optimized the lens mass parameters by minimizing the offset between the predicted light intensity $d^\text{pred}_j$ and the observed light intensity $d_j$ in each pixel $j$ of the extended arcs and foreground lens galaxies included in the fit. This is identical to minimize
\begin{equation}
  \chi_\text{esr}^2 = (\boldsymbol d - \boldsymbol d^\text{pred})^T C_\text{D}^{-1} (\boldsymbol d - \boldsymbol d^\text{pred}),
  \label{eq:chi2_esr}
\end{equation}
where $d$ includes the contributions of the lens galaxies and multiple images as $\boldsymbol d = \boldsymbol d^\text{lens} + \boldsymbol d^\text{image}$. The light intensity is written to a vector with length $N_\text{d}$, equal to the number of pixels from our F160W image included in the fit. $C_\text{D}$ is the image covariance matrix \citep{suyu06} with diagonal terms only,
\begin{equation}
   C_\text{D} = \text{diag}(\sigma_{\text{total,} 1}^2, \sigma_{\text{total,} 2}^2, ... ,\sigma_{\text{total,} N}^2)
\end{equation}
where $\sigma_{\text{total,} j}$ is derived from Eq.~\ref{Eq:noise}, $N$ is the number of pixels. $C_\text{D}$ presents the noise correlation between adjacent pixels on the image plane, induced by charge transfer and drizzling. In this work, we adopt a robust assumption that the noise is uncorrelated in observed data for simplicity.

We modeled the extended images of sources S0 and S3 which have the highest S/N in our {\it HST} images, using their surface distribution in the F160W band. Source S5 at redshift $z=4.205$ forms a fainter arc towards the outskirts of CSWA\,31 and its S/N is insufficient to perform an extended image modeling. We kept the same parametrization of the foreground lens mass potential and, following our results in Sect.~\ref{ssec:pos_model}, we focused on the multiplane scenario with a secondary lens at $z=1.4869$. We fitted the light intensity of pixels forming the extended arcs from sets S0 and S3, which corresponds to $\sim$18000 and $\sim$3000~pixels, respectively. In addition, we also included the positions of other multiple images in sets S2, S4, and S5 as constraints. Hence, we ran MCMC chains to minimize a combination of Eq.~\ref{eq:chi2_img} and \ref{eq:chi2_esr} as follows
\begin{equation}
  \chi_\text{esr,img}^2 =  \chi_\text{esr}^2 + \chi_\text{img}^2.
  \label{eq:chi2_esr_img}
\end{equation}

Our first model Esr2-MP (L) has a $\chi_{\rm esr,img}^2 = 3.1 \times 10^4$ before rescaling and fits the multiple images of S0 very well. This model however significantly overfits the northern portion of the compact image S3 (a) in set S3 (see left column of Fig.~\ref{fig:arc2_mod}). The counterimage of this region corresponds to the southern end of the extended arc S3(b), where the stellar continuum has a lower surface brightness. This faint and diffuse region extends up to image S0(c) and since it lacks a spectroscopic redshift from MUSE, the separation between the northern end of arc S0(c) and the southern end of S3(b) is ambiguous. Given the unverified redshift and the strong normalized residuals, it is also possible that this diffuse area is associated to a separate source instead of S3 at $z=2.763$. To test, we ran a second model Esr2-MP$_{\rm test}$ (L) based on an alternative mask focusing on the bright portion of the arc (see right column of Fig.~\ref{fig:arc2_mod}). To account for the uncertainties in the mask design, we boosted the uncertainties at the mask boundary where the faint and bright regions blend with each other. This second model Esr2-MP$_{\rm test}$ (L) improves the fit of the counterimage of S3 and results in a lower $\chi_{\rm esr,img} = 2.7 \times 10^4$. The centroid, clumps and spiral arms of multiple images of S0 are well reproduced by both models despite the strong tangential distortions, as highlighted by the residuals in Fig.~\ref{fig:arc1_mod} for Esr2-MP$_{\rm test}$ (L). In both models, S0 and S3 are also well reconstructed on the source plane.

\begin{figure*}
  \centering
  \includegraphics[width=.32\linewidth]{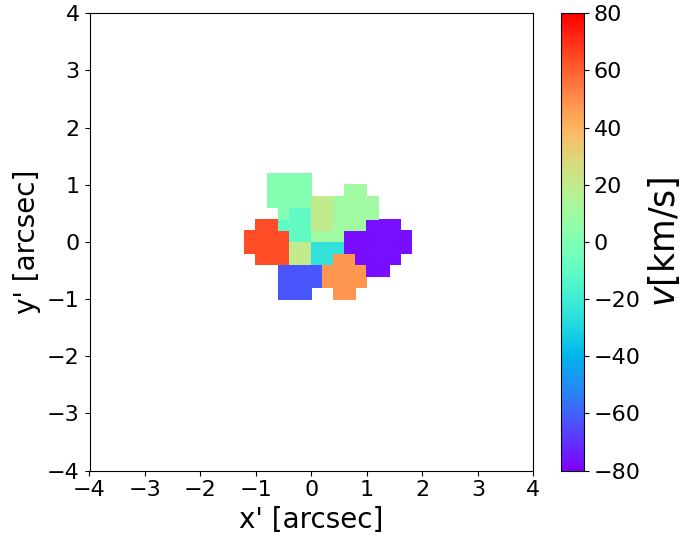}
  \includegraphics[width=.32\linewidth]{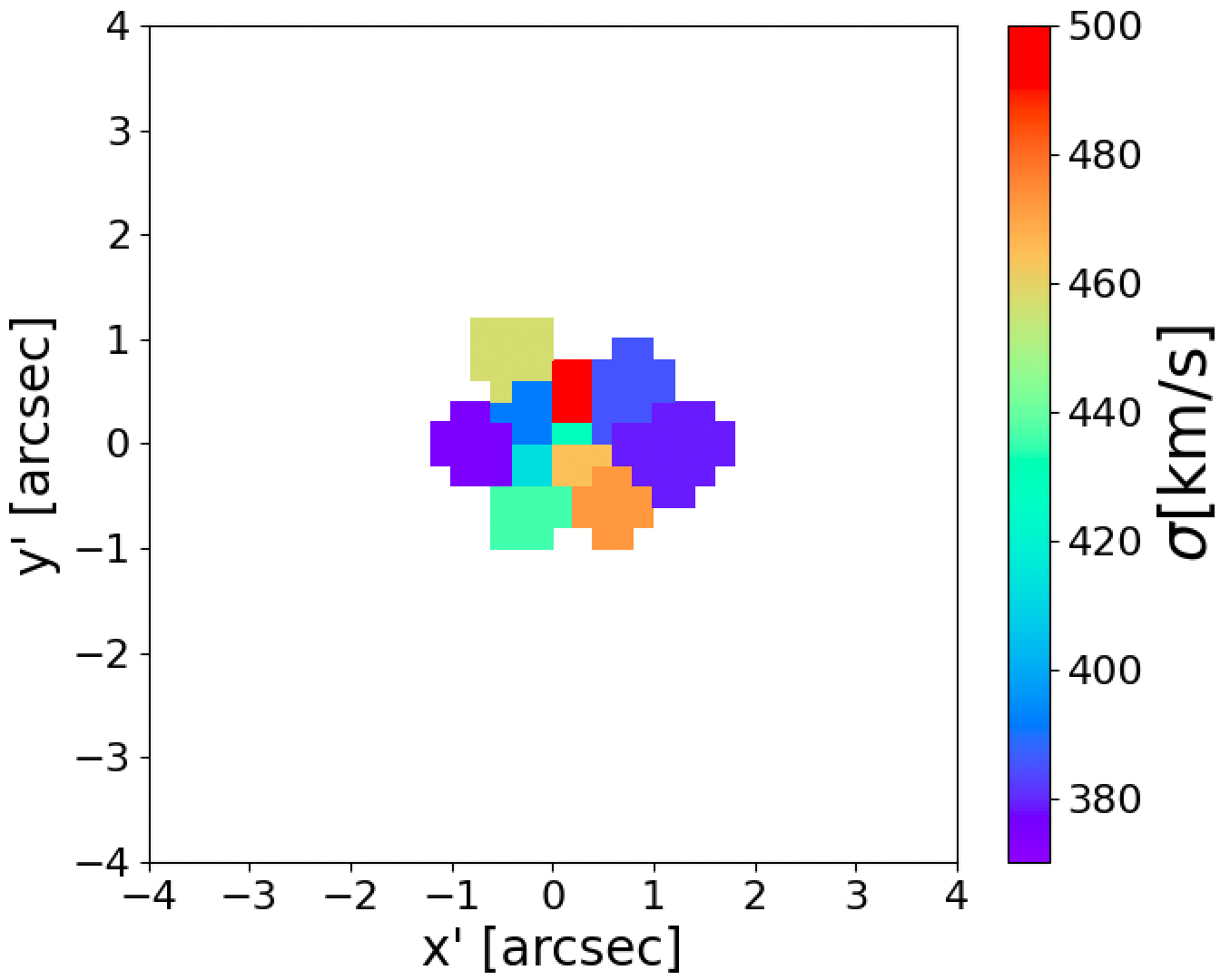}
  \includegraphics[width=.33\linewidth]{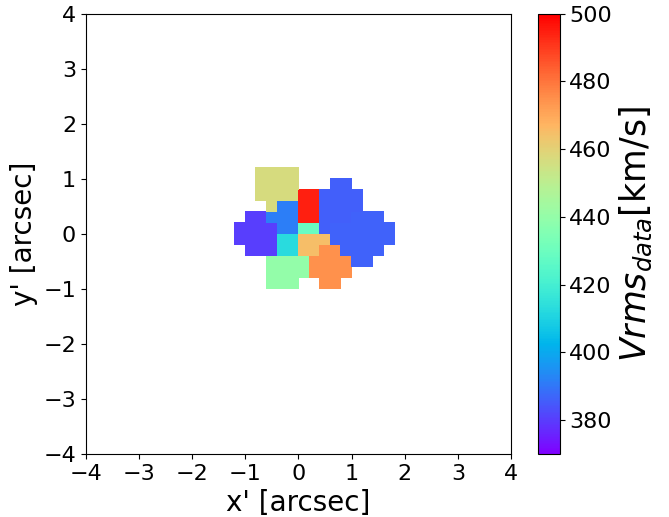}
  \caption{Spatially-resolved stellar kinematics of the BGG in 11 Voronoi bins with sufficient S/N extracted over a 8\arcsec\,$\times$\,8\arcsec\ field-of-view. The maps show the line-of-sight velocities $v$ ({\it left}), the projected velocity dispersions $\sigma$ ({\it middle}), and the second velocity moments $V_\text{rms}$ ({\it right}). The orientation of the figure differs from previous figures, as indicated in Fig.~\ref{fig:CSWA31}. The $x'$  axis points to the north and is aligned with the major axis of the BGG, while the $y'$  axis points to the west. Note that the spectrum shown in Fig.~\ref{fig:spec0} corresponds to the central resolution element.}
  \label{fig:kin_data}
\end{figure*}

The best-fit and marginalised parameter values of these two mass models are given in Table~\ref{tab:esr_chi2}. Some of the parameters, i.e., $\theta_\text{PA,GH}$, $r_{\rm core,GH}$,  $\gamma_{\rm GH}$, $r_\text{tr,BGG}$ and $\theta_\text{E,6}$ differ by $>$3$\sigma$ between Esr2-MP (L) and Esr2-MP$_{\rm test}$ (L) and others are consistent within 3-$\sigma$ uncertainties. All parameters remain consistent with the 1-$\sigma$ contours from our image position reference model Img-MP (L). The strongest variation concerns the best-fit Einstein radius of S0 in the secondary lens plane, which decreases from 4.36\arcsec\ in Img-MP (L) to $\simeq$1.7\arcsec\ in our extended image models. As mentioned above, Esr2-MP (L) and Esr2-MP$_{\rm test}$ (L) fit the image positions of other lensed sources (S2, S4, S5) together with the surface brightness distribution of S0 and S3. The constraints from these image positions are nonetheless overwhelmed by the constraints from the pixel intensities on the extended arcs. This effect explains the deviations in the best-fit values of Tables~\ref{tab:img_chi2} and \ref{tab:esr_chi2}. Due to this weighting, both extended image models also lead to larger offsets between the observed and predicted positions of images in sets S2, S4, and S5, which are on average $\simeq$0.9\arcsec. In the end, we chose the model Esr2-MP$_{\rm test}$ (L) with lower $\chi_{\rm esr,img}^2/{\rm d.o.f}$ and minor residuals as our reference lens model based on image morphologies.

\section{Mass modeling with strong lensing and stellar dynamics}
\label{sec:glad_model}

Our constraints on the mass distribution of CSWA\,31 can be significantly improved towards the lens center by jointly modeling the strong lensing observables with the lens stellar dynamics. In this section, we use the spatially-resolved kinematics of the BGG as additional information to constrain the mass components within the inner few kpc. In Sect.~\ref{ssec:kin}, we describe the extraction of stellar kinematics from the MUSE data cube. In Sect.~\ref{ssec:glee_glad}, we present a joint dynamics and strong lens modeling assuming a parametrization identical to our previous models. In Sect.~\ref{ssec:dm_bar}, we attempt to separate the total dark-matter contribution in CSWA\,31 from the baryonic mass components inferred from the near-infrared surface brightness of galaxies. 

\begin{figure}
  \centering
  \includegraphics[width=1.\linewidth]{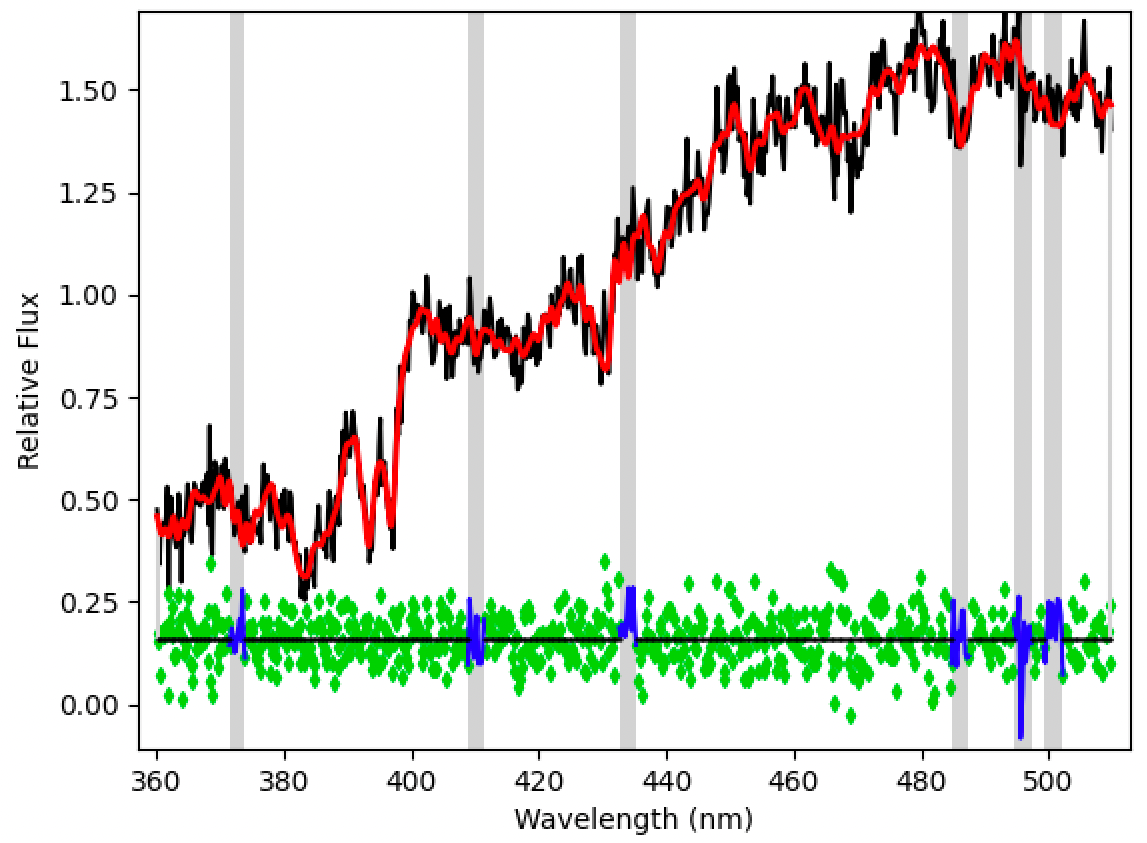}
  \caption{Example of spectral fitting with pPXF for one resolution element in our binned data cube (the central bin in Fig.~\ref{fig:kin_data}). The black and red lines show the observed spectrum normalized to unity and the best-fit stellar template, respectively. The green symbols at the bottom are the fit residuals, while the blue lines with grey-shaded regions show positions of gas emission lines masked during the fit. The most prominent spectral features used for the fit are the Ca~H and K lines at 3935~\AA\ and 3970~\AA\ in the rest-frame, the continuum break at 4000~\AA, the G-band feature at 4305~\AA, and the H$\beta$ line at 4863~\AA. Such high S/N spectra are used to extract robust stellar kinematics for the BGG out to $\simeq$0.5$R_{\rm eff}$ (see Fig.~\ref{fig:kin_data}).}
  \label{fig:spec0}
\end{figure}

\subsection{Stellar kinematics of the BGG}
\label{ssec:kin}

We used the 2D spectra over a central 8\arcsec\,$\times$\,8\arcsec\ field-of-view extracted from our MUSE data cube (see Fig.~\ref{fig:CSWA31}) in order to measure the spatially-resolved stellar kinematics of the BGG, namely the line-of-sight velocities $v$ and the projected stellar velocity dispersions $\sigma$. We extensively tested the estimation of stellar kinematics from spectra with various S/N, and found that velocity dispersions tend to be overestimated for low and moderate S/N, in particular for massive early-type galaxies with broader absorption lines (Ca\~nameras et al., in prep.). To optimize the S/N, we first binned the cube in the spectral dimension to a wavelength resolution of 3.75~\AA~pix$^{-1}$. We then conducted Voronoi tessellations \citep{cappellari03} of the MUSE data cube to obtain a binned map with adequate S/N while preserving sufficient spatial information. We used a target S/N of 35 per spectral bin over the rest-frame 3800$-$4400~\AA\ range where the main absorption lines emerge. This threshold ensures robust measurements for the BGG \citep[see also][]{yildirim21} and results in a total of 17 resolution elements.

The kinematic information was extracted with spectral fitting, using the Penalised PiXel-Fitting (pPXF) method \citep{cappellari04}. We selected a subset of the 105 stellar templates from the National Optical Astronomy Observatory library \citep{valdes04}, by focusing on G, K, M spectral classes to match the typical stellar populations of early-type galaxies. These stellar templates cover a wavelength range from 3465\,\AA\ to 9469\,\AA, with an intrinsic spectral resolution of 1.35\,\AA, sufficient to fit the MUSE data. The spectra in each bin were fitted in the rest-frame range 3600$-$5100~\AA\ with pPXF, to obtain $v$, $\sigma$, and the Gauss-Hermite moments of the line-of-sight velocity distributions $h_{\rm 3}$ and $h_{\rm 4}$. Firstly, we followed an iterative method to calibrate the BGG redshift in order to get unbiased line-of-sight velocities. We derived the best-fit redshifts for the 17 bins with pPXF, used the mean value to update the BGG redshift, and repeated the fitting procedure until the mean value stayed within $<10^{-6}$. The final BGG redshift is $z=0.6828$. Secondly, we fixed the redshift to estimate the best-fit values of $v$, $\sigma$, $h_3$, and $h_4$ per bin together with their 1$\sigma$ uncertainties. Six bins towards the galaxy outskirts do not meet our target S/N and have significantly larger parameter uncertainties. We focused on the 11 remaining bins within the central 2\arcsec\,$\times$\,2\arcsec\ to obtain reliable stellar kinematics for our joint lensing and dynamical modeling.

The maps of the best-fit $v$, $\sigma$, and second velocity moments $V_\text{rms} =\sqrt{v^2 + \sigma^2}$ are shown in Fig.~\ref{fig:kin_data} for the 11 bins. We obtained the 1-$\sigma$ uncertainties $\sigma_{V_{\rm rms}}$ of each bin $l$ using
\begin{equation}
\sigma_{V_{\text{rms},l}} = \frac{\sqrt{(v_l\times \delta v_l)^2 + (\sigma_l \times \delta \sigma_l)^2}}{v_{\text{rms},l}}   
\end{equation}
where $\delta v$ ($\simeq$25\,km s$^{-1}$) and $\delta \sigma$ ($\simeq$30\,km s$^{-1}$) are the 1-$\sigma$ uncertainties of the line-of-sight velocities and velocity dispersions, respectively, obtained from pPXF. Fig.~\ref{fig:spec0} illustrates the quality of our fits. The main absorption lines that drive the stellar kinematic estimates, namely the Ca~H and K lines at 3935\,\AA\ and 3970\,\AA, the G-band feature at 4305\,\AA, and H$\beta$ at 4863\,\AA, are correctly fitted in all bins, as well as the highly prominent 4000\,\AA\ discontinuity. The BGG does not show significant rotation, as also found for the majority of massive early-types in the local Universe \citep[e.g.,][]{emsellem07}, and $\sigma$ varies between 380 and 500~km s$^{-1}$. The $V_\text{rms}$ map shows higher values along the minor axis of the BGG (y' direction) and lower values along the major axis (x' direction) with significant scatter. Note that due to stellar orbital anisotropies, the 2D projected $V_{\rm rms}$ can vary as a function of galactocentric radius, even for intrinsically constant velocity dispersions and total density profile slopes in 3D.

We extensively checked our measurements to exclude systematic biases which would drastically affect the JAM outputs \citep[e.g.,][]{yildirim20}. Varying the bias parameter in the range from 0 to 1 does not affect the values of $v$ and $\sigma$, and we therefore fixed this parameter to the default value 0. Likewise, the polynomial order for continuum corrections has a minor influence on the results. In addition, we tested whether possible residuals in the subtraction of skylines falling at 3744.325\,\AA\ and 4305.407\,\AA\ in the rest-frame of the BGG can impact the fitting (see Sect.~\ref{sec:data}). Masking out a few spectral bins around each skyline would cut the blue tail of the G-band absorption feature and would bias the stellar kinematics, with an unphysical average velocity dispersion of 650\,km s$^{-1}$. Alternatively, excluding the entire G-band feature leads to $v$ and $\sigma$ values consistent with the results inferred from the entire wavelength range. We therefore used the full MUSE spectra to derive our final results, assuming negligible residuals from the skyline subtraction.

\begin{figure*}  
  \centering 
  \includegraphics[width=0.45\linewidth]{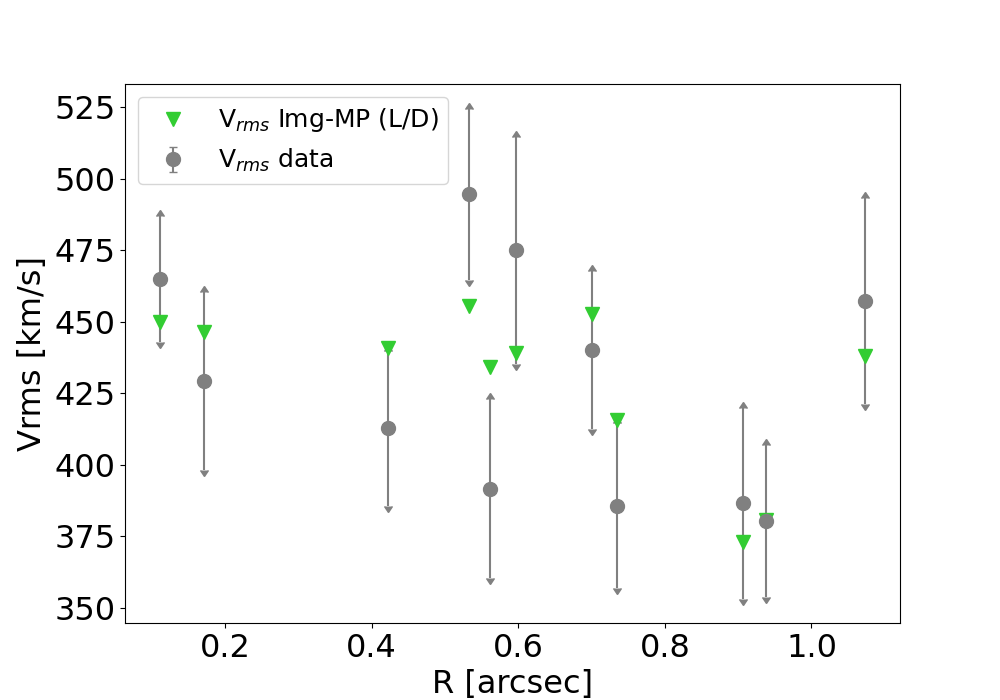} 
  \includegraphics[width=0.4\linewidth]{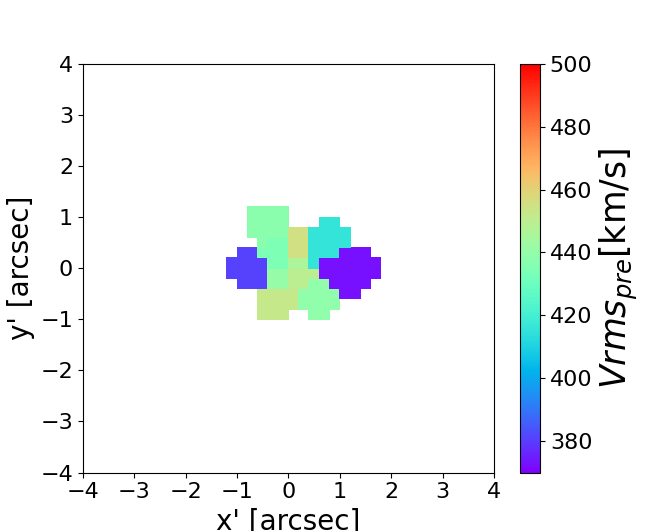}
  \caption{Modeled stellar kinematics of the BGG. \textit{Left:} The $V_{\rm rms}$ radial profile predicted by the joint lensing and dynamical modeling Img-MP (L/D) (green triangles) from the most-probable model compared to the $V_{\rm rms}$ measured over the 11 Voronoi bins and their 1-$\sigma$ uncertainties (gray points). The radial position of each resolution element is defined with respect to the BGG centroid. The observed $V_{\rm rms}$ are consistent with a flat profile and predicted values are within the 1-$\sigma$ ranges in most bins. \textit{Right:} The predicted $V_{\rm rms}$ map from model Img-MP (L/D), showing that this model does not fully recover the angular structure of $V_{\rm rms}$ in Fig~\ref{fig:kin_data}.} 
  \label{fig:onedim}
\end{figure*}

The radial profile of $V_{\rm rms}$ is plotted in Fig.~\ref{fig:onedim} and the $\sigma$ profile of the BGG is essentially similar. The variations within $\lesssim$8~kpc are consistent with a flat radial profile given the 1$\sigma$ errors. These results are in agreement with the lack of significant stellar velocity dispersion gradients in the majority of BGGs at $z<0.5$ analyzed by \citet{newman15}, and with the stellar kinematics of other individual early-type galaxies \citep[e.g., the Cosmic Horseshoe at $z=0.44$,][]{spiniello11,schuldt19}. At lower redshifts, \citet{veale18} obtained $\sigma$ radial profiles for local early-type galaxies with $M_* > 4 \times 10^{11}$~M$_{\odot}$ from the MASSIVE survey. They find a majority of negative gradients within the central few kpc and logarithmic slopes ranging between $-$0.2 and 0. In addition, \citet{veale18} detect a higher fraction of positive $\sigma$ gradients in the highest-mass bin $\simeq$10$^{12}$~M$_{\odot}$, but only for the outer 10$-$25~kpc radial range which we can not probe due to limited S/N in the MUSE data. Finally, from the best-fit stellar kinematics, we obtain a luminosity-weighted average velocity dispersion of the BGG of $\bar{\sigma} = 430 \pm 29 $\,km s$^{-1}$, consistent with the value $\bar{\sigma} = 450 \pm 80$\,km s$^{-1}$ previously measured by \citetalias{grillo13} from lower S/N SDSS spectroscopy. The elevated $\bar{\sigma}$ confirms that the BGG is an ultra-massive early-type galaxy \citep{loeb03}.

\begin{table}
  \renewcommand\arraystretch{1.5}
  \centering
  \caption{Best-fit and marginalised values with 1-$\sigma$ uncertainties for model Img-MP (L/D) based on joint constraints from strong lensing and stellar dynamics.} 
  \begin{tabular}{cccc}
    \hline
    \hline
    Component & Parameter & Best-fit & Marginalised \\
    \hline
    Kinematics & $\beta_z$ & 0.52 & $0.44_{-0.12}^{+0.10}$ \\
    & $\cos{i}$ & 0.080 & $0.11_{-0.07}^{+0.07}$ \\
    \hline
    BGG & $r_{\rm tr,BGG}$ [\arcsec] & 15.5 & $17.0_{-3.9}^{+4.8}$ \\
    (dPIE) & $\theta_{\rm E,BGG}$ [\arcsec] & 6.8 & $7.7_{-1.0}^{+0.9}$ \\
    \hline
    & $x_{\rm GH}$ [\arcsec] & 72.8 & $72.7_{-0.3}^{+0.3}$ \\
    & $y_{\rm GH}$ [\arcsec] & 63.7 & $63.5_{-0.4}^{+0.4}$ \\
    Group Halo & $q_{\rm GH}$ & 0.81 & $0.84_{-0.04}^{+0.03}$ \\
    (SPEMD) & $\theta_{\rm PA,GH}$ [$^\circ$] & 131.5& $132.7_{-4.7}^{+5.1}$ \\
    & $r_{\rm core,GH}$  [\arcsec] & 7.8 & $8.4_{-1.5}^{+1.8}$ \\
    & $\theta_{\rm E,GH}$ [\arcsec] & 20.4 & $20.0_{-1.4}^{+1.5}$ \\
    & $\gamma_{\rm GH}$ & 0.81 & $0.83_{-0.1}^{+0.1}$ \\
    \hline
    & $r_{\rm tr,ref}$ [\arcsec] & 5.1 & $5.5_{-2.6}^{+3.0}$ \\
    & $\theta_{\rm E,ref}$ [\arcsec] & 0.61 & $0.5_{-0.2}^{+0.2}$ \\
    Group Members & $\theta_{\rm E,1}$ [\arcsec] & 1.3 & $1.7_{-0.5}^{+0.5}$ \\
    (dPIE) & $\theta_{\rm E,2}$ [\arcsec] & 1.5 & $1.3_{-0.4}^{+0.3}$ \\
    & $\theta_{\rm E,6}$ [\arcsec] & 3.1 & $2.7_{-0.8}^{+0.4}$ \\
    & $\theta_{\rm E,7}$ [\arcsec] & 1.8& $2.1_{-0.4}^{+0.4}$ \\
    \hline
    S0 (PIEMD) & $\theta_{\rm E,S0}$ [\arcsec] & 4.6 & $4.2_{-1.0}^{+1.0}$ \\
    \hline
    External & $\gamma_{\rm ext}$ & 0.06 & $0.06_{-0.004}^{+0.004}$ \\
    Shear & $\theta_{\rm ext}$ [$^\circ$] & 180.6 & $179.8_{-1.3}^{+1.3}$ \\
    \hline
  \end{tabular} 
  \tablefoot{We show the anisotropy $\beta_z$ and inclination angle $i$ of the kinematic model, followed by the same list of parameters as in Tables~\ref{tab:img_chi2} and \ref{tab:esr_chi2}. Img-MP (L/D) assumes the same mass parametrization as our lensing-only models. It is constrained by the positions of multiple images from six different families and by the lens stellar kinematics over 11 Voronoi bins. Due to different implementations in the {\tt GLaD} and {\tt GLEE} softwares, we scaled the truncation and Einstein radii of group members with respect to the BGG and to the reference galaxy, respectively, optimizing the parameters $r_{\rm tr,BGG}$ and $\theta_{\rm E,BGG}$ instead of $r_{\rm tr,ref}$ and $\theta_{\rm E,ref}$. This modification does not impact the final results.}
  \label{tab:glad_chi2}
\end{table}

\subsection{Comparing joint and lensing-only models}
\label{ssec:glee_glad}

The combination of strong lensing and stellar kinematic data has been proven highly successful in constraining the total mass-density slope and dark-matter fraction of galaxy-scale lenses \citep[e.g.,][]{koopmans03,treu04,auger09,sonnenfeld15,shajib21}. In CSWA\,31, the six sets of multiple images provide direct constraints on the large-scale mass distribution, while the innermost image S4(b) falls around 20~kpc from the BGG center. The spatially-resolved MUSE kinematics covering the central 8~kpc in 11 bins are thus expected to improve our diagnostics significantly. To test whether or not the joint modeling gives consistent results while breaking parameter degeneracies, we derived a first model based on the same configuration as our image position reference model Img-MP (L). This ensures similar contributions from lensing and dynamics to the total $\chi^2$ defined as
\begin{equation}
  \chi^2 = \chi_\text{kin}^2 + \chi_\text{img}^2 = \sum_l^{N_\text{bin}} \frac{\left(V_{\text{rms}, l}-\sqrt{\overline{v^2_{{\text{LOS},l}}}}\right)^2}{\sigma^2_{V_{\text{rms},l}}} + \chi_\text{img}^2.
\label{eq:kin_chi2}
\end{equation}
Combining the stellar kinematics with extended arc surface brightness would on the other hand minimize the contribution of the kinematic data, with a $\chi^2$ dominated by the larger number of pixels on the extended arcs. For this reason, we did not attempt a comparison to Esr2-MP$_{\rm test}$ (L).

We used the {\tt GLaD} software \citep{chirivi20,yildirim20} based on the JAM formalism presented in \citet{cappellari03} and \citet{cappellari08}. This approach relies on the solutions of the Jeans equations in axisymmetric geometry, which is more flexible than assuming spherical Jeans models \citep[e.g.,][]{chen21,birrer21}. In short, the code takes the two-dimensional surface brightness and surface mass density maps from the lensing-only model, and deprojects these maps using multi-Gaussian expansions (MGEs) in order to infer analytical descriptions of the gravitational potentials, and to predict the second-order velocity moments $V_{\rm rms}$. {\tt GLaD} minimizes the $\chi^2$ in Eq.~\ref{eq:kin_chi2} by sampling the parameter space and iterating the lensing and dynamical modeling sequentially. For CSWA\,31, we used flat priors on the orbital anisotropy parameter $\beta_z$, between $-0.8$ and 0.8 and on the inclination angle $i$, between 80$^\circ$ and 90$^\circ$. $\beta_z$ presents the intersection shape of the velocity ellipsoid on the meridional plane ($v_R$, $v_z$). When $\beta_z = 0$, the ellipsoid is isotropic on ($v_R$, $v_z$), indicating the intersection shape is everywhere circle, when $\beta_z$ deviates from 0, the ellipsoid becomes anisotropic, stretching into an oblate ($\beta_z > 0$) or prolate ($\beta_z < 0$) \citep{Cap_aniso}. The angle $i$ presents the way how real galaxies are projected on to the observed plane, i.e., a galaxy observed edge-on will have $i = 90^\circ$. The multi-Gaussian expansion only included the dPIE describing the BGG given the negligible contributions of the group-scale halo to the inner $\kappa$. We ran an MCMC chain to infer $\beta_z$, $i$, and lens-mass parameters.

The best-fit model Img-MP (L/D) has a $\chi^2/{\rm d.o.f.} = 5.22$ before rescaling, lower than our lens models based on image positions, and it can reproduce the positions of multiple images of sets S0, S1, S2, and S5 within in average 0.2\arcsec, similarly to model Img-MP (L). The modeled positions of images in sets S4 and S5 are within 0.5\arcsec. The best-fit and marginalized parameter values of Img-MP (L/D) are shown in Table~\ref{tab:glad_chi2}. The deprojection of the surface brightness via the inclination angle $i$ to obtain the intrinsic luminosity density is not unique, implying that $i$ is usually poorly constrained and degenerate with anisotropy $\beta_z$. This effect explains the lack of constraints on the dynamical parameters of the BGG. On the other hand, parameters describing the mass profiles are consistent with the values in Table~\ref{tab:img_chi2}.

Model Img-MP (L/D) has a $\chi_{\rm kin}^2/ \rm d.o.f.= 0.7$ and fits the MUSE stellar kinematics relatively well. Fig.~\ref{fig:onedim} shows that the predicted $V_{\rm rms}$ are within the measured 1-$\sigma$ ranges in most bins. However, the predictions do not fully recover the angular structure in the 2D map and underestimate $V_{\rm rms}$ along the BGG minor axis. We postulate that this discrepancy could be due to small asymmetries in the mass distribution of the BGG in CSWA\,31, which would contradict the axisymmetry assumptions in ${\tt GLaD}$. We indeed note that the near-infrared light emission of the BGG shows a south/north asymmetry, which could also be present in the underlying mass distribution. In fact, our reference lensing-only models suggest that the group-halo mass center is offset from the BGG mass center by $\simeq 16.2$~kpc, and the best-fit group-halo axis ratio is $q \sim 0.8$. Given the small, but non-negligible contribution of the group-halo SPEMD within $R_{\rm kin}$< 8~kpc, these properties induce a small asymmetry in the modeled mass distribution at the inner region covered by the stellar kinematic data. For these reasons, the properties of the BGG could approach the intrinsic limitations of JAM. Nonetheless, the overall good agreement between the predicted and observed radial profile of $V_\text{rms}$ (Fig.~\ref{fig:onedim} left panel) suggests adequate reconstruction of the lens mass distribution.

\subsection{Separating the baryonic and dark-matter components}
\label{ssec:dm_bar}

While keeping these limitations in mind, we attempted to disentangle the baryonic and dark-matter components in the BGG to constrain the overall dark-matter mass distribution in CSWA\,31 (from the group-scale halo and BGG). \citetalias{grillo13} analysed the ROSAT X-ray emission towards CSWA31, excluding significant hot gas emission from the intergalactic medium over 150~kpc around the BGG. For this reason, we assumed that stellar mass dominates the total baryonic mass budget within the radial ranges under consideration, and we modeled this component by scaling the near-infrared surface brightness distribution with a stellar mass-to-light ratio $\Gamma$. We reproduced the light modeling of the BGG with two Chameleon profiles \citep[e.g.,][]{dutton11,maller00,suyu14} which can be related to PIEMDs to ease the calculation of lensing quantities, while mimicking the two Sérsic profiles in Table~\ref{tab:light_BGG}. These Chameleon profiles fit the BGG surface brighteness equally well as in Fig.~\ref{fig:light_mod_six}, with a $\chi_{\rm BGG, Chameleon}^2/\text{d.o.f} = 2.4$, only slightly higher than with Sérsic profiles ($\chi_{\rm BGG, Sersic}^2/\text{d.o.f} = 2.3$). The analytical structure of each Chameleon profile $I(x,y)$ given in Appendix~\ref{sec:chameleon} is equivalent to two PIEMDs with different core radii, such that the baryonic surface mass density can be defined as,
\begin{equation}
\kappa_\text{baryon} (x, y)  \mid_{z_s = \infty}  = \sum_{i=1}^2 \Gamma I_{i}(x, y).
\end{equation}

We kept the same configuration as Img-MP (L/D), except that we modeled the baryonic and dark-matter components of the BGG with four PIEMDs and a non-cored SPEMD, respectively, instead of a single dPIE. For simplicity, we used a single, spatially-constant stellar mass-to-light ratio $\Gamma$ to scale these PIEMDs, despite the mass-to-light radial gradients detected in samples of massive galaxies with independent constraints on the dark-matter density profiles \citep[e.g.,][]{sonnenfeld18}. We fixed the SPEMD profile of the BGG to the centroid position of the best-fit Chameleon profiles, and used the same parametrizations and priors as before for the group halo, group members, external shear, and S0 at $z=1.487$. 

The best-fit model Img-MP (L\&D) can recover the measured $V_{\rm rms}$ within 1-$\sigma$ uncertainties similarly to model Img-MP (L/D), and can reproduce the positions of multiple images of S0, S1, S2, and S3 in average within 0.2\arcsec, and within 0.6\arcsec\ for sets S4 and S5. The parameter $\Gamma$ is relatively well constrained, to $0.81_{-0.12}^{+0.15}$, while the anisotropy $\beta_z$, and inclination angle $i$ are poorly constrained for the reasons given in Sect.~\ref{ssec:glee_glad}. Moreover, other mass components are not strongly affected by the new parametrization for the BGG, and the best-fit profiles remain stable with respect to our previous models.

\begin{figure*}
  \centering
  \includegraphics[scale=0.3]{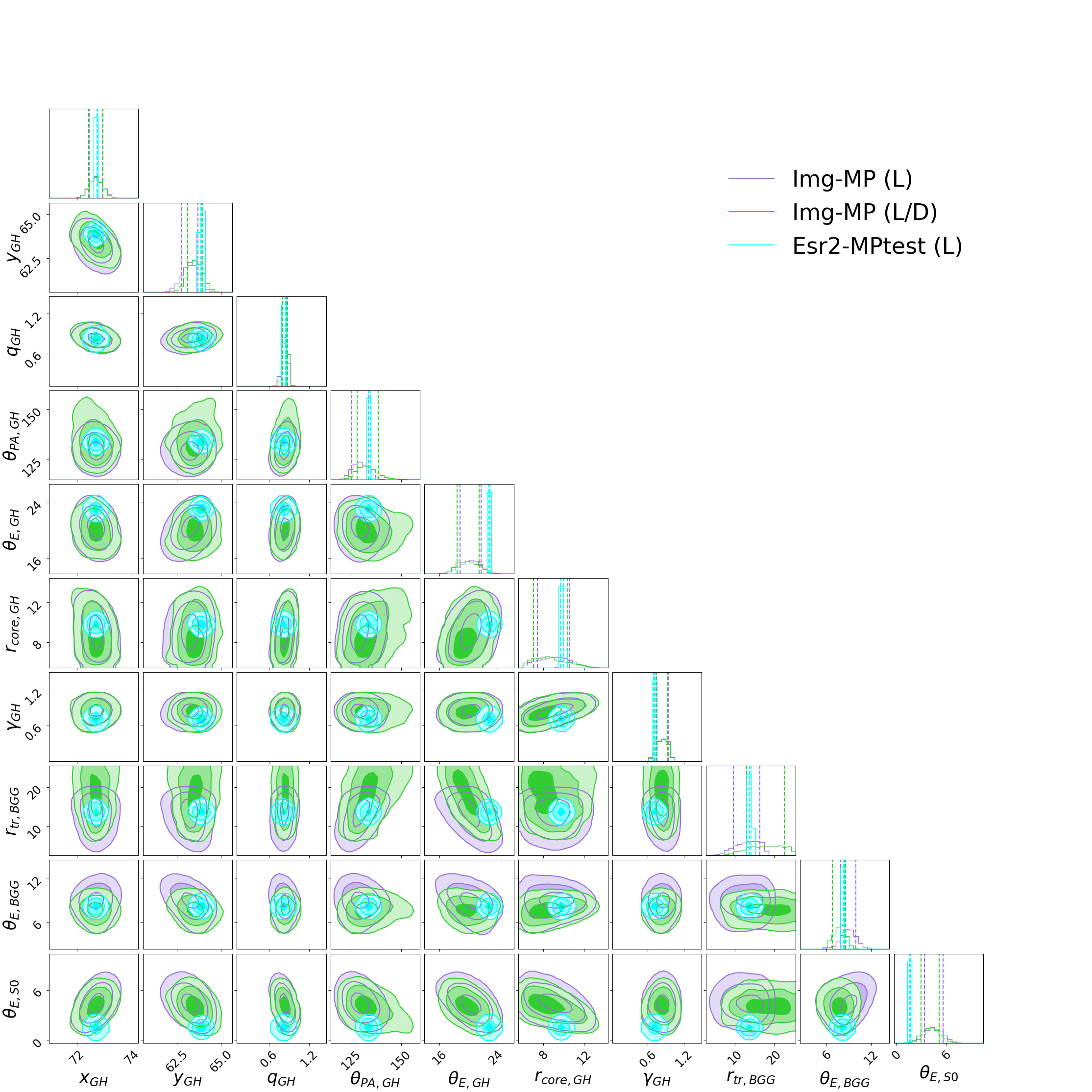}
  \caption{Joint posterior probability distribution functions for our three reference models, corresponding to the lensing-only models based on image positions (Img-MP (L), purple contours), and on extended surface brightness distributions (Esr2-MP$_{\rm test}$ (L), cyan contours), and to the joint modeling with strong lensing and dynamics (Img-MP (L/D), green contours). We focus on the most important mass parameters describing the BGG and the extended group-scale halo. The three shaded areas on the joint PDFs show the $68.3\%$, $95.4\%$, and $99.7\%$ credible regions. The 1-D histograms show the marginalized PDFs for the selected mass parameters and the vertical lines mark the 1-$\sigma$ confidence intervals. The model based on extended image fitting has narrower contours due to the large number of constraints from the lensed source morphologies. }
  \label{fig:deg_chi2all}
\end{figure*}

\section{Model comparison and discussion}
\label{sec:model_summary}

In Sect.~\ref{ssec:constraints}, we compare the results from our lensing-only reference models based on image position (Img-MP (L)), and extended light (Esr2-MP$_{\rm test}$(L)) reconstructions, and from our joint strong lensing and dynamical modeling (Img-MP (L/D)). In Sect.~\ref{ssec:contrib}, we present the robust separation between the total mass of BGG and extended group halo and, in Sect.~\ref{ssec:slope}, we compute the slope of the total mass-density profile and we compare with the literature. Then, in Sect.~\ref{ssec:fbaryon}, we disentangle the baryonic component within BGG from dark matter using model Img-MP (L$\&$D), and we compare the resulting baryonic fraction with the SED analysis.

\subsection{Constraints on the mass parameters}
\label{ssec:constraints}

To test the consistency between models and probe parameter degeneracies, we plot in Fig.~\ref{fig:deg_chi2all} the posterior PDFs for the parameters describing the BGG, extended group-scale halo, and secondary lens S0 at $z=1.487$. The marginalized PDFs indicate that mass parameters are much better constrained for Esr2-MP$_{\rm test}$ (L), due to the large number of constraints from the lensed source morphologies. While the best-fit values of $\theta_{\rm E,GH}$, $r_{\rm core, GH}$, and $\theta_{\rm E,S0}$ differ from the other two models, the joint PDFs remain consistent with Img-MP (L) and Img-MP (L/D). 

For the lensing-only models, we notice that the Einstein radius of the group-scale dark-matter halo $\theta_{\rm E,GH}$, and $r_{\rm tr, BGG}$ and $r_{\rm E,S0}$ are less constrained in Img-MP (L) due to the joint contributions from the BGG, group-scale halo, and mass perturbations along the line-of-sight. $\theta_{\rm E,GH}$ is larger for smaller BGG truncation radii $r_{\rm tr, BGG}$, and for smaller $\theta_{\rm E,S0}$. The slope of the group-halo SPEMD, $\gamma_{\rm GH}$, is strongly degenerate with its core radius $r_{\rm core, GH}$. While such parameter degeneracies usually complicate the separation of the central galaxy and host dark-matter halo in group-scale lenses \citep[e.g.,][]{more12}, they have a lower impact for CSWA\,31 (see Sect.~\ref{ssec:contrib}). In particular, including extended arcs helps break these degeneracies since the additional constraints from S0 and S3 cover the same ranges as the 1D marginalized PDFs for $\theta_{\rm E,GH}$ and $r_{\rm tr, BGG}$ in Img-MP (L). We nonetheless note that the best-fit parameters of Esr2-MP$_{\rm test}$ (L) are likely slightly biased to the values optimizing the reconstruction of S0, which is more extended in the image plane. For the joint model Img-MP (L/D), the additional constraints from the BGG stellar kinematics only make the marginalized PDF of $\theta_\text{E,BGG}$ in Fig.~\ref{fig:deg_chi2all} slightly narrower than Img-MP (L), others are comparable with Img-MP (L). The best fit and marginalized parameter values do not vary significantly between both models, and the parameter degeneracies are not completely broken in Img-MP (L/D), likely due to the large uncertainties in our stellar kinematics measurements (see Fig.~\ref{fig:onedim}). We expect that increasing the number of resolution elements with sufficient S/N, and using broader spectral coverage to decrease the errors on $v$ and $\sigma$ would improve the constraints from the joint modeling.

These results do not motivate the inclusion of a secondary extended mass component, as done for merging galaxy clusters \citep[e.g.,][]{lagattuta17,mahler18}, further suggesting that CSWA\,31 is an isolated galaxy group. We note that the reference models discussed in this section are resulting from an extensive exploration of the parametrization of the foreground gravitational potential. Other choices of mass profiles, such as a NFW for the group-scale halo, significantly degrade the fit.

\begin{figure}
  \centering
  \includegraphics[width=1.1\linewidth]{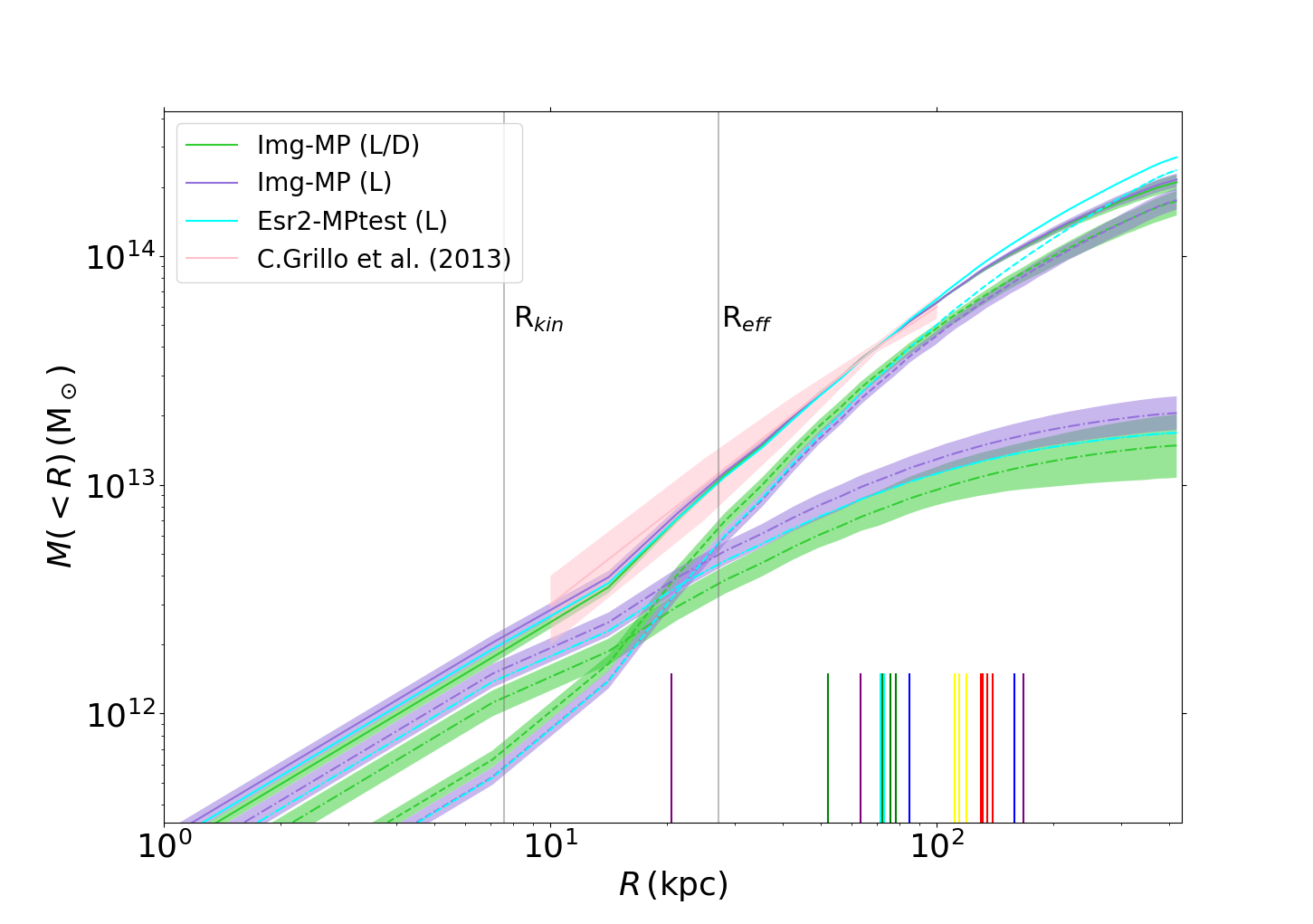}
  \caption{Cumulative mass profiles as a function of radial separation from the BGG center, for the BGG (dot-dashed lines), group-scale (dashed lines), and total (solid lines) mass components of CSWA\,31. The shaded regions show the 1-$\sigma$ uncertainties on the mass distributions estimated from the 16th and 84th percentiles of the posterior PDFs. The reference models Img-MP (L), Esr2-MP$_{\rm test}$ (L), and Img-MP (L/D) are compared using the same colors as in Fig.~\ref{fig:deg_chi2all}. Vertical lines in the bottom indicate the positions of multiple images used as constraints. The pink solid line with shaded regions shows the total mass of CSWA 31 with 1-$\sigma$ uncertainties estimated by \citetalias{grillo13}, using only image positions of set S0 (green vertical lines) as constraints. The gray line at $R_{\rm kin} = 7.6$~kpc shows the coverage of the MUSE spatially-resolved stellar kinematics. We also mark $R_{\rm eff}$ to ease comparison with other lenses in the literature.}
  \label{fig:totalmass}
\end{figure}

\subsection{Contributions from the BGG and extended group-halo}
\label{ssec:contrib}

We computed the cumulative projected mass profiles as a function of radial separation from the BGG light center for the main components in CSWA\,31, namely the BGG and extended group-scale halo, and the results are shown in Fig.~\ref{fig:totalmass} for our three reference models. Other mass components with minor contributions to the cumulative mass profiles are ignored. The radial distributions are broadly consistent with each other in the range 20$-$150~kpc covered by multiple images of background sources. In particular, over this radial range, the total mass distributions are in excellent agreement with the mass model of \citetalias{grillo13} constrained exclusively from S0 marked by pink solid line with 1$\sigma$ uncertainties in Fig.~\ref{fig:totalmass}, but with much smaller 1$\sigma$ uncertainties due to the additional constraints in our analysis. The group halo, and total masses are consistent with each other in our three mass models in the range 20$-$100~kpc where the brightest lensed images of S0 and S3 emerge, despite the small differences seen in Fig.~\ref{fig:deg_chi2all} for $\gamma_{\rm GH}$, the power-law slope of the group-halo SPEMD. The BGG mass from model Img-MP (L/D) is overall slightly lower but still comparable with other two reference models within 1$\sigma$ uncertainties, which is induced by the smaller fitting values of $\theta_\text{E,BGG}$ in Img-MP (L/D). The total enclosed masses are broadly similar to the virial masses inferred by \citet{munoz13} for strong-lensing galaxy groups at similar redshifts as CSWA\,31. The masses enclosed within $R_{\rm Ein,S0} \sim 70$~kpc are $0.96_{-0.01}^{+0.01}\times10^{13}$~M$_{\odot}$, $2.93_{-0.01}^{+0.01}\times10^{13}$~M$_{\odot}$, and $4.05_{-0.01}^{+0.01}\times10^{13}$~M$_{\odot}$ for the BGG, group-halo, and all components, respectively, in model Esr2-MP$_{\rm test}$ (L).

In the outer regions $>$100~kpc, the best-fitting group-halo and total mass profiles in Esr2-MP$_{\rm test}$ (L) differ from the other two models by more than 1$\sigma$. While this model includes the external arcs located between 100 and 150~kpc, these constraints have lower weights than the extended arcs closer to the BGG center which are reconstructed with {\tt GLEE} (see Sect.~\ref{ssec:sb_model}). The difference at $>$100~kpc is also related to the degeneracy between $\theta_{\rm E,GH}$ and $\theta_{\rm E,S0}$ parameters. In model Img-MP (L), this leads to a very large best-fit $\theta_\text{E,S0} = 4.36$\arcsec\, while the best-fit $\theta_\text{E,GH}$ is the lowest among all mass models. This degeneracy is drastically decreased in Esr2-MP$_{\rm test}$ (L), which increases the mass in the outskirts and decreases the best-fit $\theta_\text{E,S0} = 1.65$\arcsec\ to a realistic value. In the inner regions $<$15~kpc, the slope of BGG and group-halo mass profiles are slightly steeper for Img-MP (L/D) than for lensing-only models. This joint model provides the best constraints on the relative contributions from the BGG and group-halo within $R_{\rm kin} = 7.6$~kpc covered by the spatially-resolved stellar kinematics. It is nonetheless important to note that extrapolating the lensing-only models constrained at $>$20~kpc towards the inner regions $<R_{\rm kin}$ leads to total mass distributions consistent with Img-MP (L/D). The masses enclosed within $R_{\rm kin}$ are $1.18_{-0.14}^{+0.15}\times10^{12}$~M$_{\odot}$, $6.94_{-0.50}^{+0.58}\times10^{11}$~M$_{\odot}$, and $1.89_{-0.12}^{+0.10}\times10^{12}$~M$_{\odot}$ for the BGG, group-halo, and all components, respectively, in model Img-MP (L/D). \footnote{The total mass also accounts for group members such that the sum of BGG and group-halo only is lower than the total mass.}

Overall, despite these small variations related to the different sets of constraints, the relative contributions from the BGG and group-scale halo are remarkably consistent at all radii. Fig.~\ref{fig:totalmass} shows that the ultra-massive BGG dominates at projected separations $<$20~kpc independent of the modeling assumptions\footnote{We obtain the same results by excluding set S4 that has a counter-image at $\sim$20~kpc from the BGG center.}. In contrast to most group-scale lenses with smaller image separations \citep[e.g.,][]{auger08,limousin09b,newman15}, the peculiar configuration of CSWA\,31 is the main ingredient to get a robust decomposition between the BGG and extended components over multiple scales. 

\begin{figure}
  \centering
  \includegraphics[width=1.\linewidth]{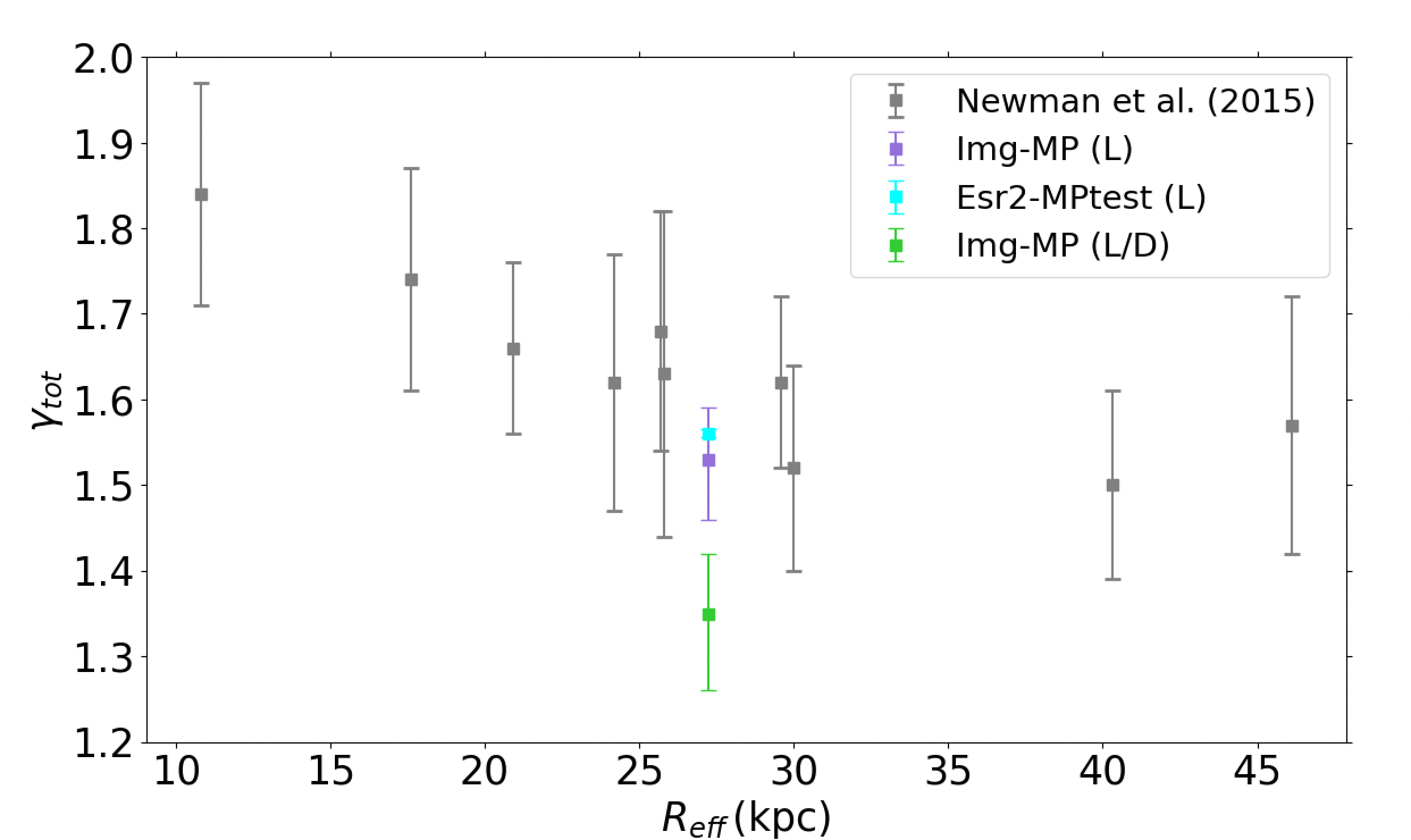}
  \caption{Comparison between the total density slope $\gamma_{\rm tot}$ at the effective radius for CSWA\,31 (colored squares), and the sample of group-scale lenses from \citet{newman15} (grey squares). The total density slopes inferred from our three reference models are shallower than the average values of other group-scale lenses. The discrepancy between lensing-only models and model Img-MP (L/D) is primarily caused by differences in the best-fit group halo component.}
  \label{fig:newmanslope}
\end{figure}

\subsection{Slope of the total mass-density profile}
\label{ssec:slope}

The radial slope of the total matter-density profile $\gamma_{\rm tot}$ within the effective radius, $R_{\rm eff}$, can be defined following \citet{dutton14} as,
\begin{equation}
\gamma_\text{tot}(r) = \frac{1}{M(r)} \int^{r}_{0}-\gamma(x) 4 \pi x^2 \rho(x) \text{d}x = 3 - \frac{4\pi r^3 \rho(r)}{M(r)},
\end{equation}
which can also be expressed as in terms of the local logarithmic slope,
\begin{equation}
\gamma_\text{tot}(r) = 3 - \frac{\text{d log}M}{\text{d log}r}. 
\label{eq:slope}
\end{equation}

The evolutionary trend of $\gamma_{\rm tot}$ with redshift has been mostly characterized for early-type galaxies in the field and remains debated. On the theoretical side, \citet{wang19} have recently studied the evolution of the total mass density profiles of $\gtrsim$10$^{11}$~M$_{\odot}$ early-type galaxies in the IllustrisTNG cosmological hydrodynamical simulations. They show that the slopes decrease from to $z \sim 2$ to become nearly isothermal by $z \sim 1$, and that the passive evolution at $z<1$, likely primarily affected by dry minor mergers, do not significantly affect the slopes. Recently, the Jeans dynamical analysis of early-type galaxies at $0.29 < z < 0.55$ by \citet{derkenne21}, based on deep {\it HST} Frontier-Fields imaging and integral-field-unit stellar kinematics from MUSE, further confirmed the lack of evolution in the average mass-density slopes over the last few Gyr. However, this contradicts the continuous mild increase of $\gamma_{\rm tot}$ from $z \sim 2$ suggested by strong lensing studies \citep[e.g.,][]{treu04,koopmans06,bolton12,sonnenfeld13,li18}. 

\begin{figure}
  \centering
  \includegraphics[width=1.1\linewidth]{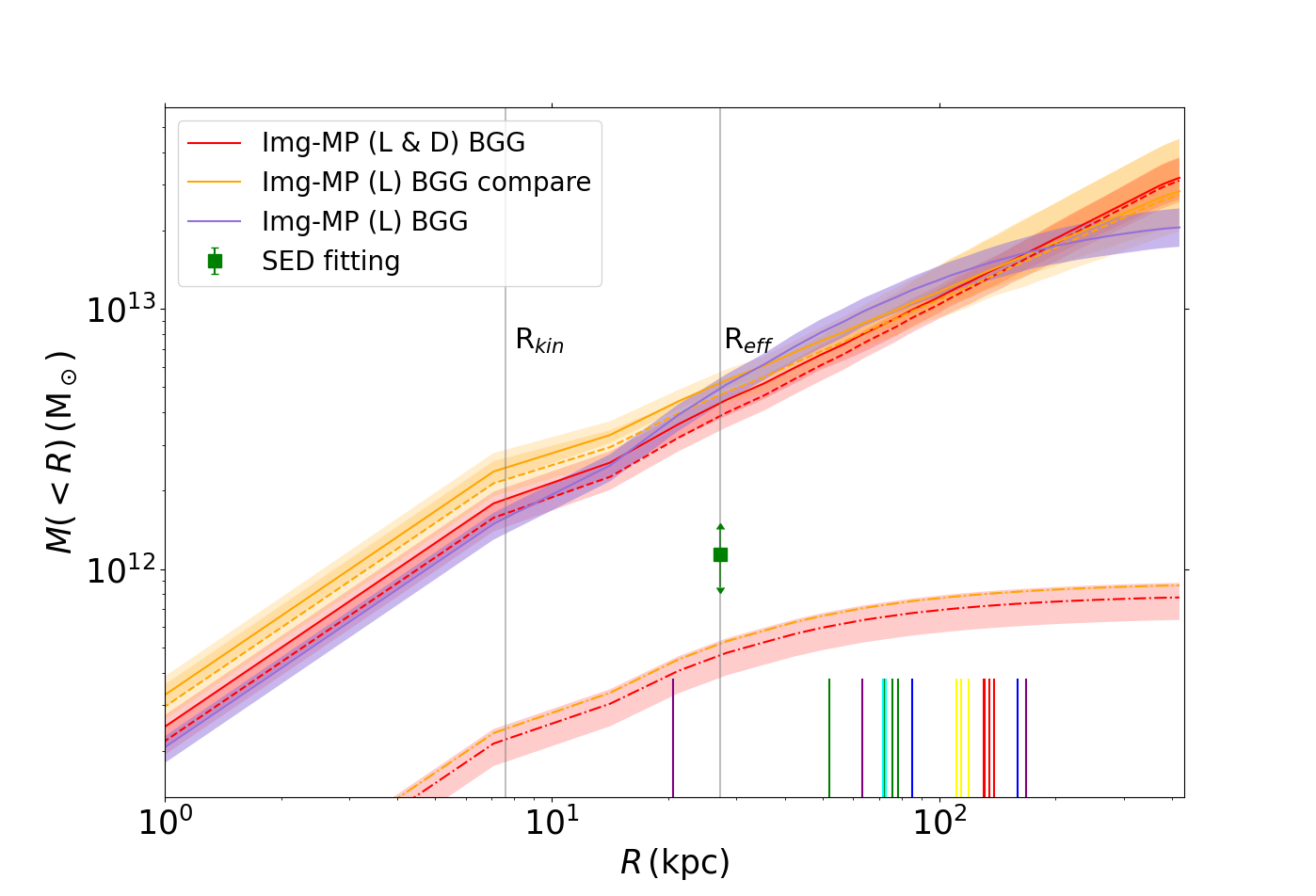}
  \caption{Decomposition of dark-matter and baryonic mass components within the BGG from model Img-MP (L\&D). We plot the cumulative profiles for the dark-matter (red dashed line) and baryonic (red dot-dashed line) components, and for the total mass of the BGG (red solid line). The purple curve shows the total mass profile of the BGG obtained in model Img-MP (L) using a single dPIE. The orange lines show the decomposition of dark-matter (orange dashed line) and baryonic (orange dot-dashed line) mass components within the BGG, from an additional lensing-only model with the same mass configuration as Img-MP (L\&D). The shaded regions show the 1-$\sigma$ uncertainties on the mass distributions estimated from the 16th and 84th percentiles of the posterior PDFs. The baryonic mass profile from Img-MP (L\&D) is lower than the stellar mass at $R_{\rm eff}$ inferred independently from SED fitting (green square), and this discrepancy is driven by lensing constraints rather than JAM modeling. Vertical lines at the bottom indicate the positions of the multiple images used as lensing constraints. }
  \label{fig:BGG_decomp}
\end{figure}

In the local universe, the total density slope is well known to be nearly isothermal ($\gamma_\text{tot} \simeq 2$) for isolated early-type galaxies \citep[e.g.,][]{koopmans06,sonnenfeld13,cappellari15,li18}. Galaxy-to-galaxy variations in the outer $\gamma_\text{tot}$ at 1--2~$R_{\rm eff}$ have been identified from stellar dynamics \citep[e.g.,][]{veale18,wang21}, but the slopes of the total mass density profiles within $R_{\rm eff}$ have much smaller scatter. Importantly, \citet{newman13b} have shown that the average slope within $R_{\rm eff}$ decreases to $\gamma_{\rm tot} = 1.16 \pm 0.05$ for bright ellipticals residing in the center of $M_{\rm 200} \simeq 10^{15}$~M$_{\odot}$ galaxy clusters. Subsequently, \citet{newman15} probed the regime of intermediate-mass dark-matter halos with a sample of 10 group-scale lenses at $z \sim 0.2$--0.45, and measured $\gamma_{\rm tot} = 1.64 \pm 0.05~({\rm stat.}) \pm 0.07~({\rm sys.})$, suggesting a smooth evolution of $\gamma_{\rm tot}$ over a broad range in host halo mass. 

To determine the $\gamma_\text{tot}$ of CSWA\,31,\footnote{We determine the $\gamma_\text{tot}$ for the total mass of BGG and group halo. We exclude the group members from the total mass in the Sec.~\ref{ssec:slope} to simplify the calculation because the mass of the group members are negligible given the results from Sec.~\ref{ssec:contrib} and most of them are located outside of the region enclosed by the effective radius of the BGG as shown in \it{HST} imaging.} we used MGE to fit the surface mass density of the BGG and group halo (as mentioned in Sec.~\ref{ssec:glee_glad}) and deprojected them into the 3D galaxy coordinates $(x, y, z)$ respectively via:
\begin{equation}
    \rho~(x, y, z) = \sum_{k=1}^N \frac{M'_k}{\sqrt{2\pi} {{\sigma'_k}^2 q_k}}~\text{exp} \left[-\frac{1}{2{{\sigma'_k}^2}}\left({x}^2+{y}^2+\frac{{z}^2}{q_{k}^2}\right)\right]
 \label{eq:mge-3D}, 
\end{equation}
where $N$ is the number of the adopted Gaussian components, $q_k$ is the intrinsic flattening, $\sigma'_{k}$ is the dispersion along the projected minor axis of the observed galaxy, $M'_{k}$ is the amplitude of each Gaussian \citep[see e.g.,][]{cappellari08}. We summed up the 3D mass density of BGG and group halo with an assumed separation between their centroids which is the observed offset in the 2D plane obtained from our lens models. Their separation in 3D is not unique after the deprojection. We probed to increase the distance between their centroids to twice larger than the adopted separation, obtaining minor variations of the slopes within the 1-$\sigma$ uncertainties for models Img-MP (L) and Img-MP (L/D), and slightly larger than the 1-$\sigma$ uncertainty in model Esr2-$\text{MP}_\text{test}$ (L). The total 3D density slope is not sensitive to the small separation between BGG and group halo because the group halo in CSWA\,31 has a large core radius and thus a flat distribution in the inner region. Then we determined the enclosed 3D mass in terms of the radial distance from the BGG centroid using the 3D density and we estimated $\gamma_\text{tot}$ using Eq.~\ref{eq:slope}.

We obtained an average mass-density slope of $\gamma_{\rm tot} = 1.48$ at the effective radius of the BGG based on the three reference models, shallower than the average value of the group-scale lenses estimated by \citet{newman15} (also measured at $R_{\rm eff}$). As shown in Fig.~\ref{fig:newmanslope}, the $\gamma_{\rm tot}$ from lensing-only models are in the lower limit of the slope range for group-scale lenses.  When compared to cluster-scale lenses instead of group-scale lenses, the $\gamma_{\rm tot} = 1.35_{-0.09}^{+0.07}$ from the model Img-MP (L/D) is near the upper limit of the cluster-scale lenses with $\gamma_\text{tot} = 1.34$ from \citet{newman15}. The difference in $\gamma_{\rm tot}$ between the lensing-only and lensing-and-dynamics models is caused by differences in the group halo distribution because the BGG, modeled by a dPIE profile, is isothermal with $\gamma = 2.0$ at $R_{\rm eff}$ in all models. The scatter of $\gamma_\text{tot}$, except for model Esr2-$\text{MP}_\text{test}$ (L) with extended arc as constraints, is not significantly smaller than the results from \citet{newman15}, since our uncertainties also account for the imperfect fitting of the surface mass density from the MGE. Our measurements extend the diagnostics obtained in \citet{newman15} from the range $\theta_{\rm E}$ to $R_{\rm eff}$, and add to the growing evidence that BGGs have shallower total mass-density slopes than isolated ellipticals.

We also compared the mass distribution in the extended dark-matter halo of CSWA\,31 with the literature. Few strong-lensing groups have sets of multiple images covering beyond 100~kpc from the lens center, which results in less accurate lensing models. We thus rather compared with higher-mass ($\gtrsim$10$^{14}$~M$_{\odot}$) cluster-scale halos, using the high-precision lens models from \citet{caminha19}. Since we can not reliably estimate $M_{\rm 200}$ and $R_{\rm 200}$ without weak lensing, we rescaled the mass distribution of CSWA\,31 to the cluster-scale lens RX J2129 with the closest $\theta_\text{E}$. After rescaling, we obtained similar projected total mass profiles for both systems, suggesting that the self-similarity between high-mass dark-matter halos obtained by \citet{caminha19} extends to intermediate-mass, group-scale halos.

\begin{figure}
  \centering
  \includegraphics[width=1.\linewidth]{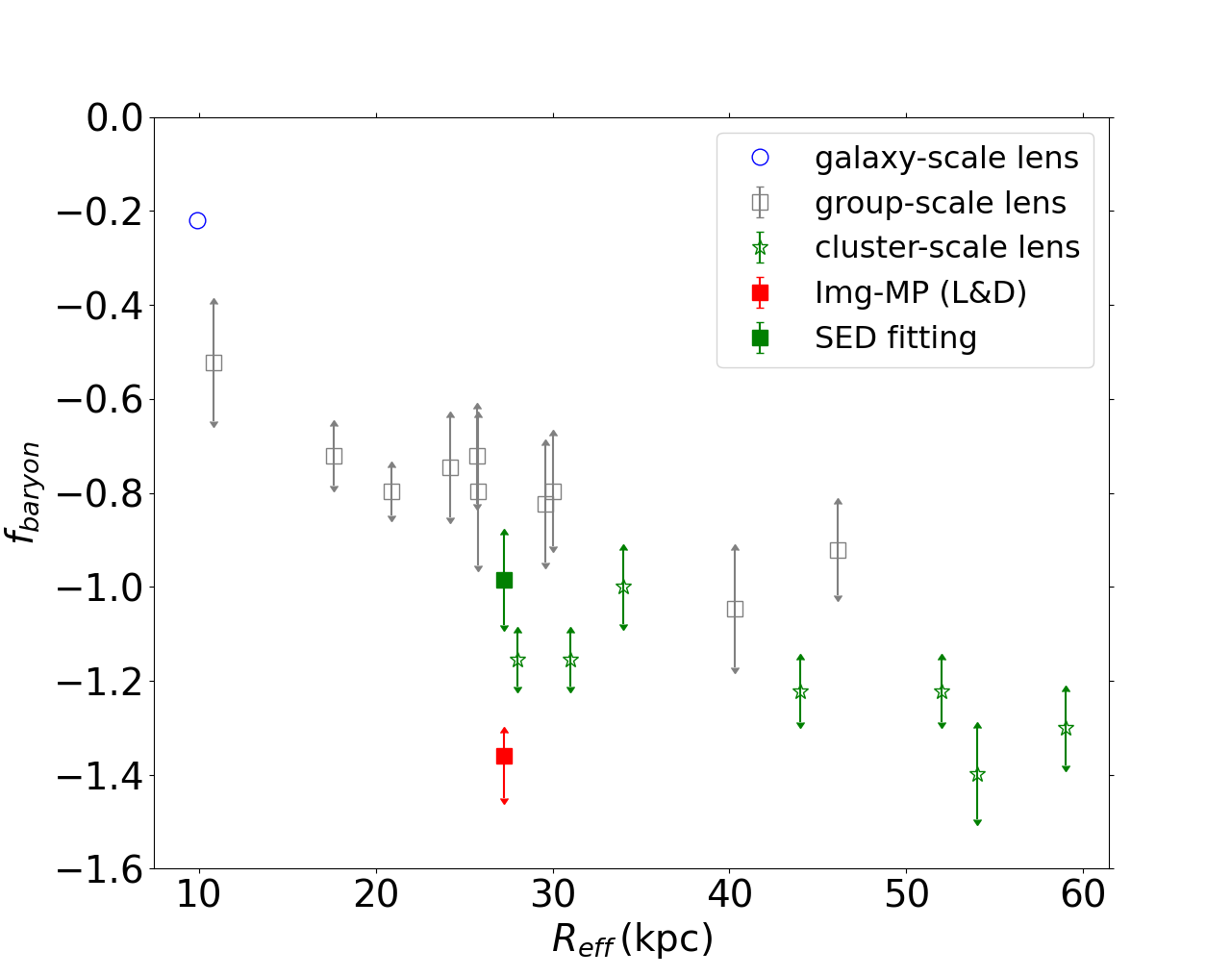}
  \caption{Baryonic mass fraction within $R_{\rm eff}$ for CSWA\,31 compared to galaxy-, group- and cluster-scale in the literature \citep[samples combined by][]{newman15}. The $f_{\rm baryon} = -1.40_{-0.09}^{+0.05}$ of CSWA\,31 at $R_{\rm eff} = 27.2$\,kpc inferred from model Img-MP (L\&D) (red square) is comparable to massive ellipticals in the center of cluster-scale lenses, but much smaller than for BGGs. Nevertheless, the $f_{\rm baryon} = -0.98_{-0.10}^{+0.10}$ obtained using the stellar mass from SED fitting (green square) is closer to group-scale lenses.}
  \label{fig:f_bar}
\end{figure}

\subsection{Constraints of the baryonic mass fraction}
\label{ssec:fbaryon}

The baryonic and dark-matter mass fractions in CSWA\,31 are distinguished in model Img-MP (L\&D). In this section, we discuss the robustness of this separation and characterize the combined dark-matter distribution. Fig.~\ref{fig:BGG_decomp} shows the cumulative projected mass profiles separately for each component in the BGG, indicating that dark-matter strongly dominates the mass distribution up to $r_{\rm tr,BGG}$.\footnote{Note that Fig.~\ref{fig:BGG_decomp} shows the decomposition of the BGG into dark-matter and baryonic component, while Fig.~\ref{fig:totalmass} shows the decomposition of the whole system CSWA\,31 into BGG and group halo. Both figures show the total mass of the BGG, however with different line styles, in Fig.~\ref{fig:BGG_decomp} with solid lines, in Fig.~\ref{fig:totalmass} with dot-dashed lines.}

The total mass profile of the BGG from Img-MP (L\&D) is consistent with the results obtained in Img-MP (L) from a single dPIE. The baryonic mass within the BGG remains well below 10$^{12}$~M$_{\odot}$ at all radii, and corresponds to an integrated stellar mass significantly lower than our independent estimate from SED fitting of $ (1.6 \pm 0.4) \times 10^{12}$~M$_{\odot}$. To compare the stellar masses obtained from these two independent methods at a common radius, we rescaled the result from SED fitting to the mass enclosed within $R_{\rm eff}$, using the baryonic mass profile from Img-MP (L\&D) and assuming no radial gradient in the stellar mass-to-light ratio.

Modeling results from Img-MP (L\&D) are used to infer the baryonic and dark-matter mass fractions in CSWA\,31 from the joint contributions of the BGG, group-halo, and other perturbers. We define the baryon to total mass fraction within $R_{\rm eff}$ as,
\begin{equation}
  f_{\rm baryon} = \log_{\rm 10}\frac{M_{\rm baryon} (R_{\rm eff}) }{M_{\rm tot}(R_{\rm eff})},
  \label{eq:bgg_frac}
\end{equation}
where $M_{\rm baryon}$ is determined from the four best-fit PIEMDs in model Img-MP (L\&D). We obtain $f_{\rm baryon} = -1.40_{-0.09}^{+0.05}$ for CSWA\,31 and compare in Fig.~\ref{fig:f_bar} with the average value for the SLACS sample of galaxy-scale lenses, and with group- and cluster-scale lenses compiled by \citet{newman15}. In the comparison samples, $f_{\rm baryon}$ is defined as the baryon mass fraction of each lens systems, measured within the effective radius of the brightest central galaxy. $f_{\rm baryon}$ decreases progressively from galaxy- to cluster-scale lenses, while $R_{\rm eff}$ increases for denser environments. The effective radius of the BGG in CSWA\,31 falls near the median value of group-scale lenses, and near the lower envelope of cluster-scale lenses. However, $f_{\rm baryon}$ is much lower than the average value $f_{\rm baryon} = -0.78$ for group-scale lenses analyzed in \citet{newman15}. If we would assume our independent stellar mass estimate from SED fitting, we would obtain $f_{\rm baryon}= -0.98_{-0.10}^{+0.10}$, placing CSWA\,31 closer to group-scale lenses. For comparison, the joint lensing and stellar dynamical analysis of the Cosmic Horeshoe of \citet{schuldt19} leads to $f_{\rm baryon}$ in the range between $-$0.40 and $-$0.52, close to the typical value of $-$0.60 for isolated early-type galaxies with masses $\gtrsim$10$^{12}$~M$_{\odot}$ \citep[e.g.][]{cappellari13}. The baryonic mass fraction in CSWA\,31 is rather comparable to massive ellipticals with $R_{\rm eff} \simeq 30$~kpc in the center of large-scale halos.

The difference between the stellar masses estimated from Img-MP (L\&D) and from SED fitting might be driven by systematic errors in either of the two methods. Since the systematics in SED fitting have been extensively discussed in the literature \citep[e.g.,][]{conroy13}, we focus on quantifying the systematic uncertainties resulting from our novel modeling approach in Img-MP (L$\&$D), testing our assumptions on lens mass profiles and on the dynamical modeling. First, we remodeled dark-matter within the BGG using a NFW profile instead of the SPEMD. This change degrades the fit mildly and further decreases $f_{\rm baryon}$, suggesting that the mass parametrization does not contribute much to the systematic error budget of Img-MP (L\&D). Second, the results from Img-MP (L\&D) might also be affected by the assumptions involved in the theoretical dynamics modeling framework (see Sect.~\ref{ssec:glee_glad}). We remodeled the separation between the baryonic and dark-matter components using the same mass configuration as Img-MP (L\&D), but using only multiple image positions as constraints and discarding the BGG stellar kinematics. The cumulative mass profiles, the 1-$\sigma$ ranges, and the integrated stellar mass obtained from this new model are closely matching those from Img-MP (L\&D), with only slightly larger error bars on the total mass profile (Fig.~\ref{fig:BGG_decomp}). This shows that the low $f_{\rm baryon}$ is driven by the lensing constraints rather than the JAM modeling. Consequently, while CSWA 31 is certainly dark-matter dominated towards the center, the actual baryonic mass fraction needs to be further studied to solve the discrepancy between these two independent methods.

\section{Summary}
\label{sec:summary}

In this work, we studied the inner structure of the group-scale lens CSWA\,31 at $z=0.683$ using {\it HST} near-infrared imaging and integral-field-unit spectroscopy from MUSE. Based on the spectroscopic confirmation of five sets of multiple images covering various projected separations from the lens center, we conducted a detailed analysis of the multi-scale mass distribution using various modeling approaches. First, we performed a lensing-only modeling, adopting a composite mass model to account for the central BGG, group members and extended group-scale halo, and we compared the single and multiplane scenarios. We used image positions and extended arc morphologies as constraints. Secondly, we measured the spatially-resolved stellar kinematics of the BGG to derive a joint lensing and dynamics model, in order to improve the constraints towards the lens center and to attempt a separation between baryonic and dark-matter mass components in the BGG. For the dynamical modeling, we used the Jeans equations in cylindrical coordinates and assumed an axisymmetric underlying mass distribution. We compared the stellar mass of the BGG estimated independently from the joint lensing and dynamical analysis, and from SED fitting.

Despite small variations related to the different sets of input constraints, the relative contributions from the BGG and group-scale halo are remarkably consistent in our three reference models, demonstrating the self-consistency between strong lensing analyses based on image position and extended image modeling. We find that the ultra-massive BGG dominates the projected total mass profiles within 20~kpc, while the group-scale halo dominates at higher radii. The BGG represents $62.4\%$ of the total mass enclosed within $R_{\rm kin} = 7.6$~kpc, and the group-scale halo represents 72.3\% of the total mass enclosed within $R_{\rm E,S0} = 70$~kpc, the position of the brightest lensed arcs. Overall, CSWA\,31 is a peculiar fossil group, strongly dark-matter dominated towards the central regions, and with a projected total mass profile similar to higher-mass cluster-scale halos. The total mass-density slope within the effective radius is shallower than isothermal, consistent with results obtained for lower-mass early-type galaxies in overdense environements ranging from galaxy groups to galaxy clusters.

In a future study, we will put the properties of the central ultra-massive galaxy into context with the evolutionary trends of ellipticals in various environments to test predictions of galaxy evolution models. Furthermore, increasing the number of galaxy groups with such reliable mass decompositions will help constrain the processes driving galaxy evolution in group-scale environments. Multiband {\it HST} imaging with higher S/N and broader stellar kinematic maps from JWST/NIRSpec would significantly increase the accuracy of the baryon and dark-matter separations towards the central regions (the present map from MUSE probes only up to $\simeq$25\% of the BGG effective radius). Finally, the small asymmetries in the best-fit mass distribution of CSWA\,31 highlight the limitations of JAM modeling, and further progress would benefit from more sophisticated dynamical modeling frameworks.

\section*{Acknowledgements}

We thank A. Halkola for software support on {\tt GLEE} and {\tt GLaD}. G.~B.~Caminha, S.~H.~Suyu, A.~Y{\i}ld{\i}r{\i}m, G. Chiriv\`i and S.~Schuldt acknowledge the Max Planck Society for support through the Max Planck Research Group for S.~H.~Suyu. GBC also thanks the academic support from the German Centre for Cosmological Lensing. This research is supported in part by the Excellence Cluster ORIGINS which is funded by the Deutsche Forschungsgemeinschaft (DFG, German Research Foundation) under Germany's Excellence Strategy -- EXC-2094 -- 390783311. This research is based on observations made with the NASA/ESA Hubble Space Telescope obtained from the Space Telescope Science Institute, which is operated by the Association of Universities for Research in Astronomy, Inc., under NASA contract NAS 5–26555. These observations are associated with program GO-15253. This work is also based on observations collected at the European Organisation for Astronomical Research in the Southern Hemisphere under ESO program 0104.A-0830(A).

\bibliographystyle{aa} 
\bibliography{reference} 

\begin{appendix}

\section{Light profiles and spectral energy distribution}
\label{sec:chameleon}

The Chameleon profile used to model the BGG light distribution in Sect.~\ref{sec:glad_model} consists of two isothermal profiles with different core radii that mimic a S\'ersic profile. It is defined in Cartesian coordinates $(x, y)$ as
\begin{equation}
\begin{split}
I (x, y) = \frac{I_0}{1 + q} \bigg( \frac{1}{\sqrt{{x}^2 + {y}^2/q^2 + 4{\omega_c}^2/(1+q)^2}} - \\
\frac{1}{\sqrt{{x}^2 + {y}^2/q^2 + 4{\omega_t}^2/(1+q)^2}}  \bigg) ,
\end{split}
\label{eq:chameleon}
\end{equation}
where $q$ is the axis ratio, $\omega_t$ and $\omega_c$ are the different core radii, $I_0$ is the amplitude. To keep the $ I (x,y)>0 $ we imposed $\omega_t > \omega_c$. The Chameleon profile can be rotated by the position angle $\theta_{\rm PA}$. It can be directly linked to isothermal mass profiles using a mass-to-light ratio.

\begin{figure}
  \centering
  \includegraphics[width=0.50\textwidth]{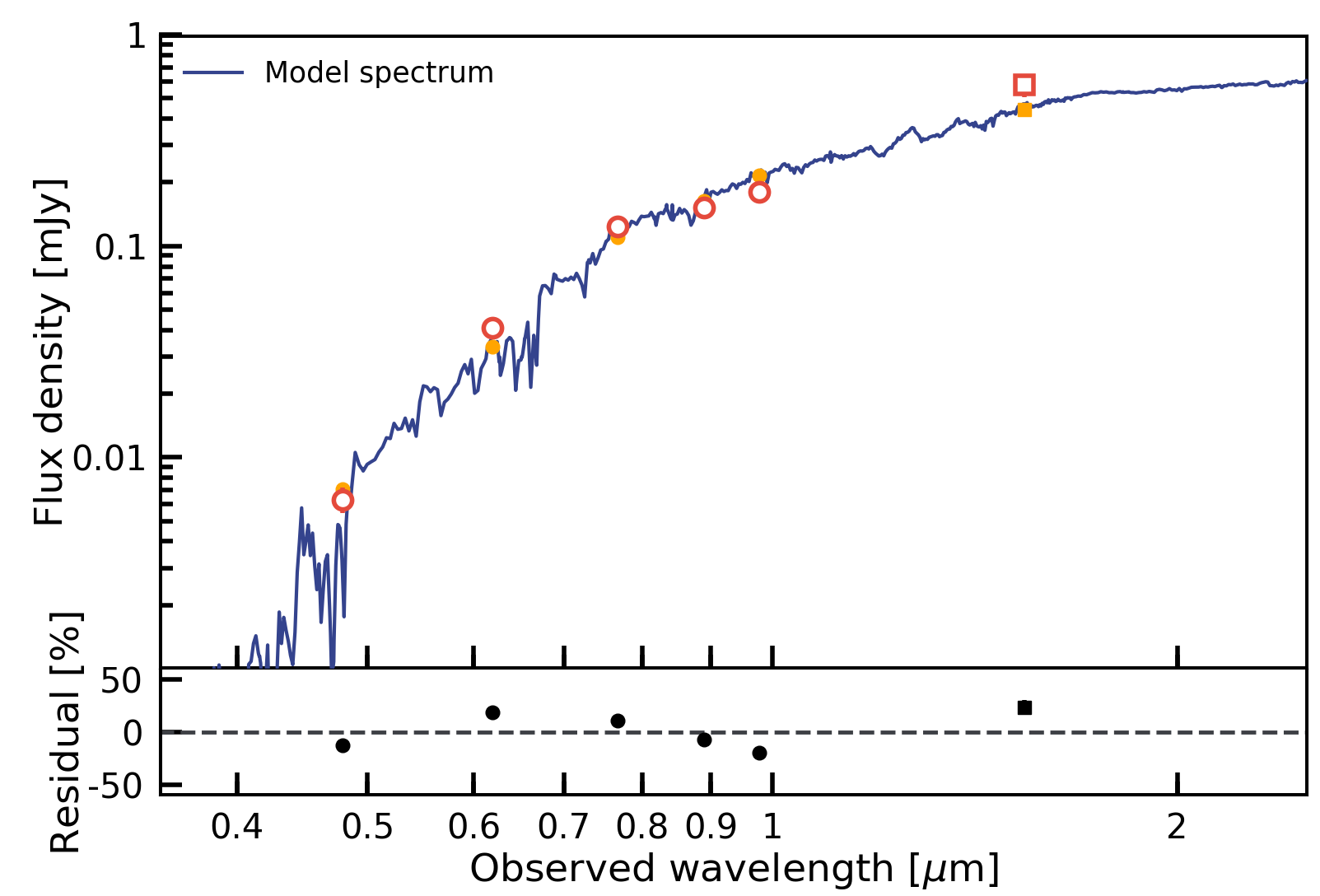}
  \caption{Spectral energy distribution of \textbf{the BGG in} CSWA\,31. The observed PanSTARRS (red circles) and {\it HST} (red square) fluxes are plotted, with uncertainties smaller than the symbols, together with the best-fit SED obtained with CIGALE (blue curve), and the corresponding model fluxes (orange markers). The bottom panel shows the relative residuals of the fit.}
  \label{fig:cigale}
\end{figure}

\section{MUSE redshift catalog and spectra of multiple images}
\label{sec:muse_multim}

\begin{figure*}
  \centering
  \includegraphics[width=0.33\textwidth]{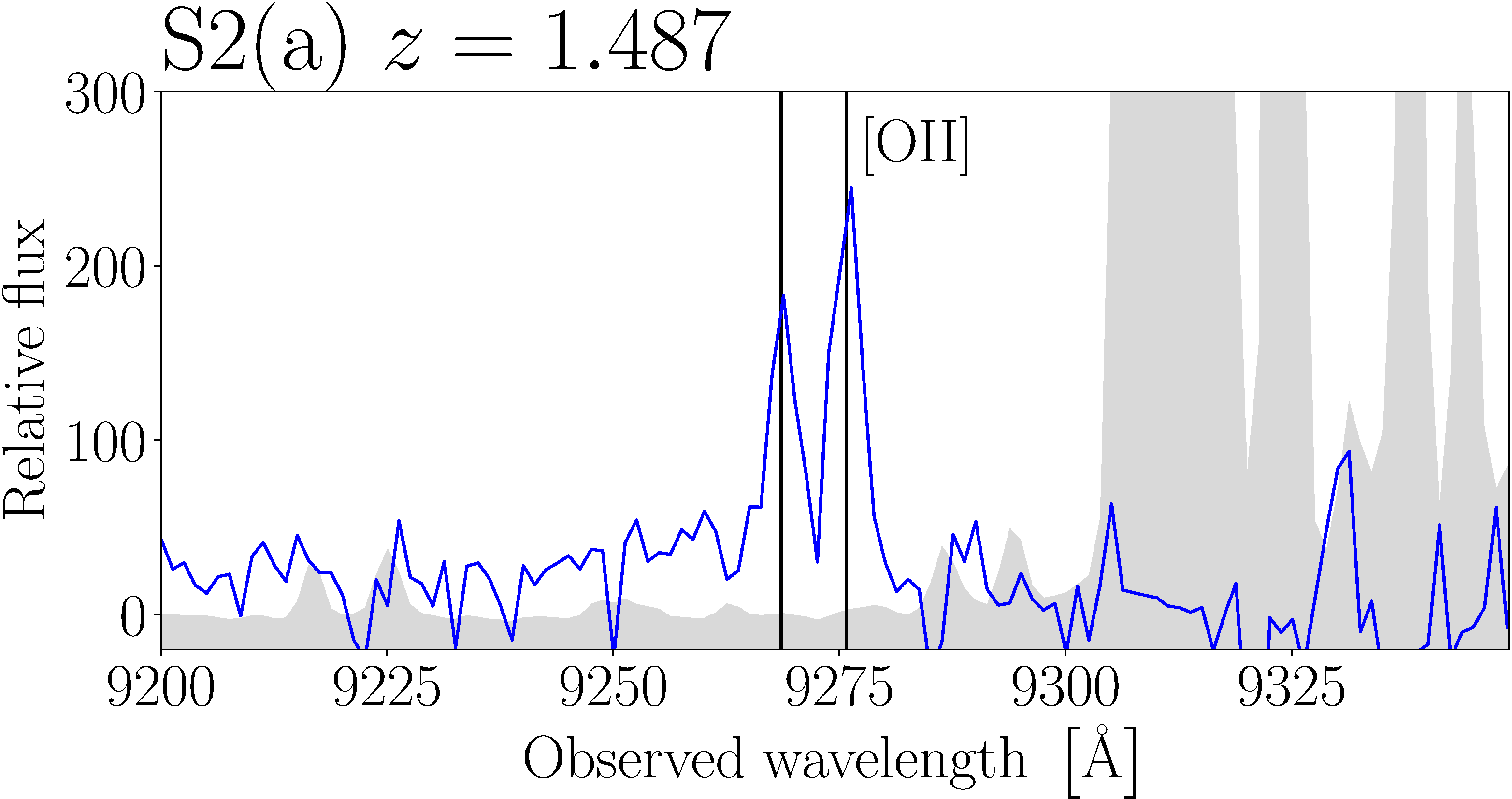}
  \includegraphics[width=0.33\textwidth]{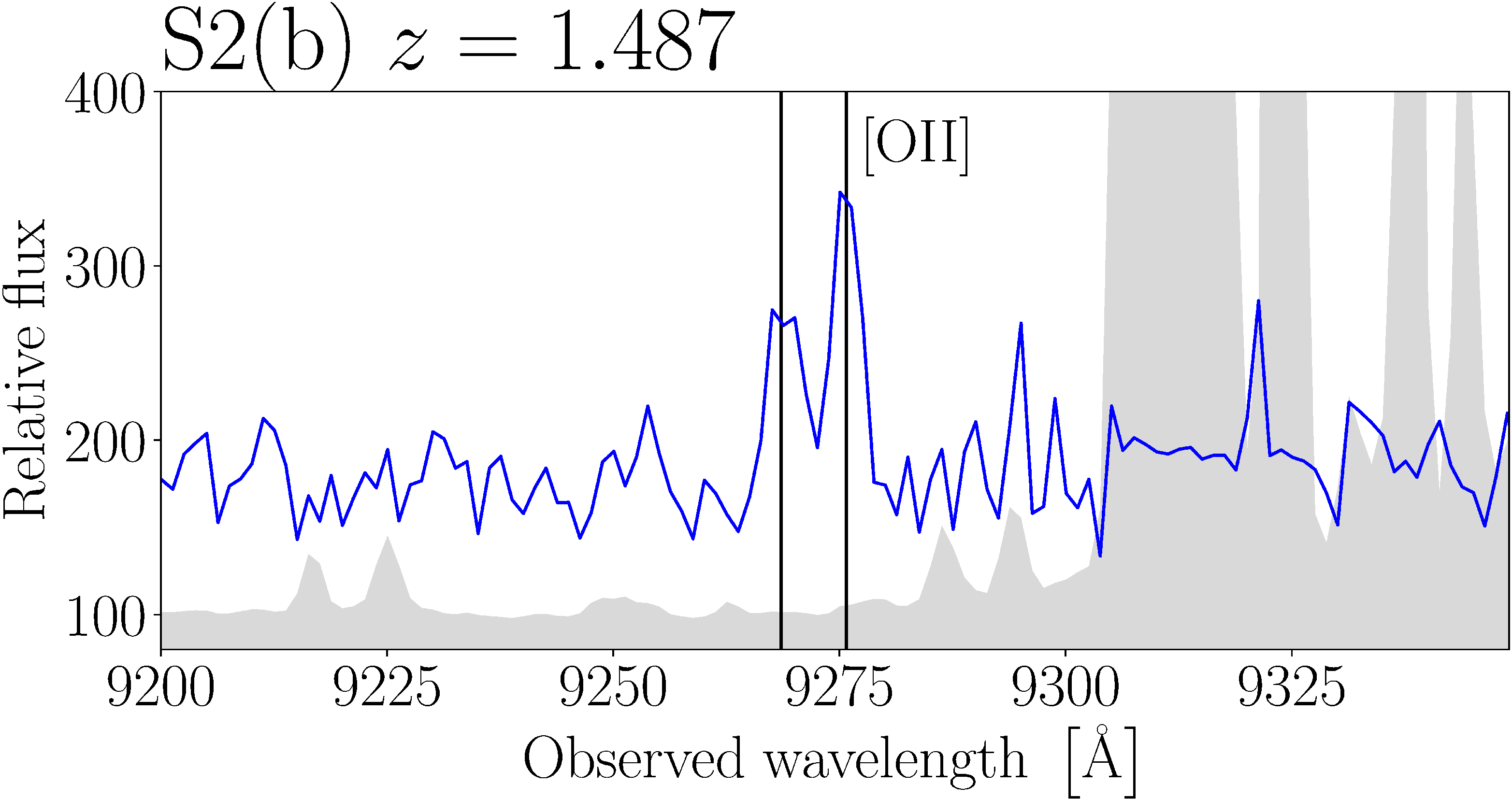}
  \includegraphics[width=0.33\textwidth]{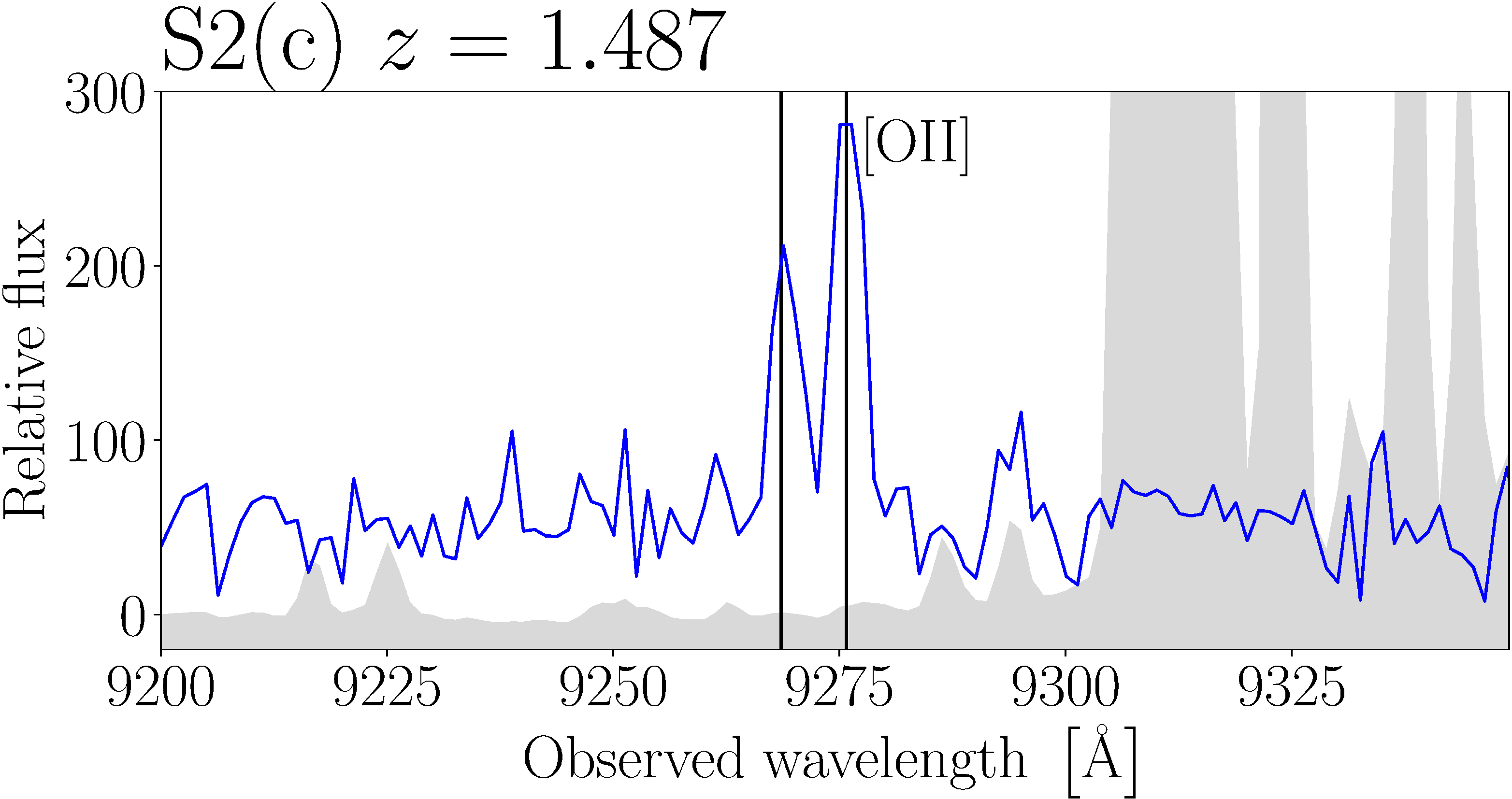}

  \includegraphics[width=0.33\textwidth]{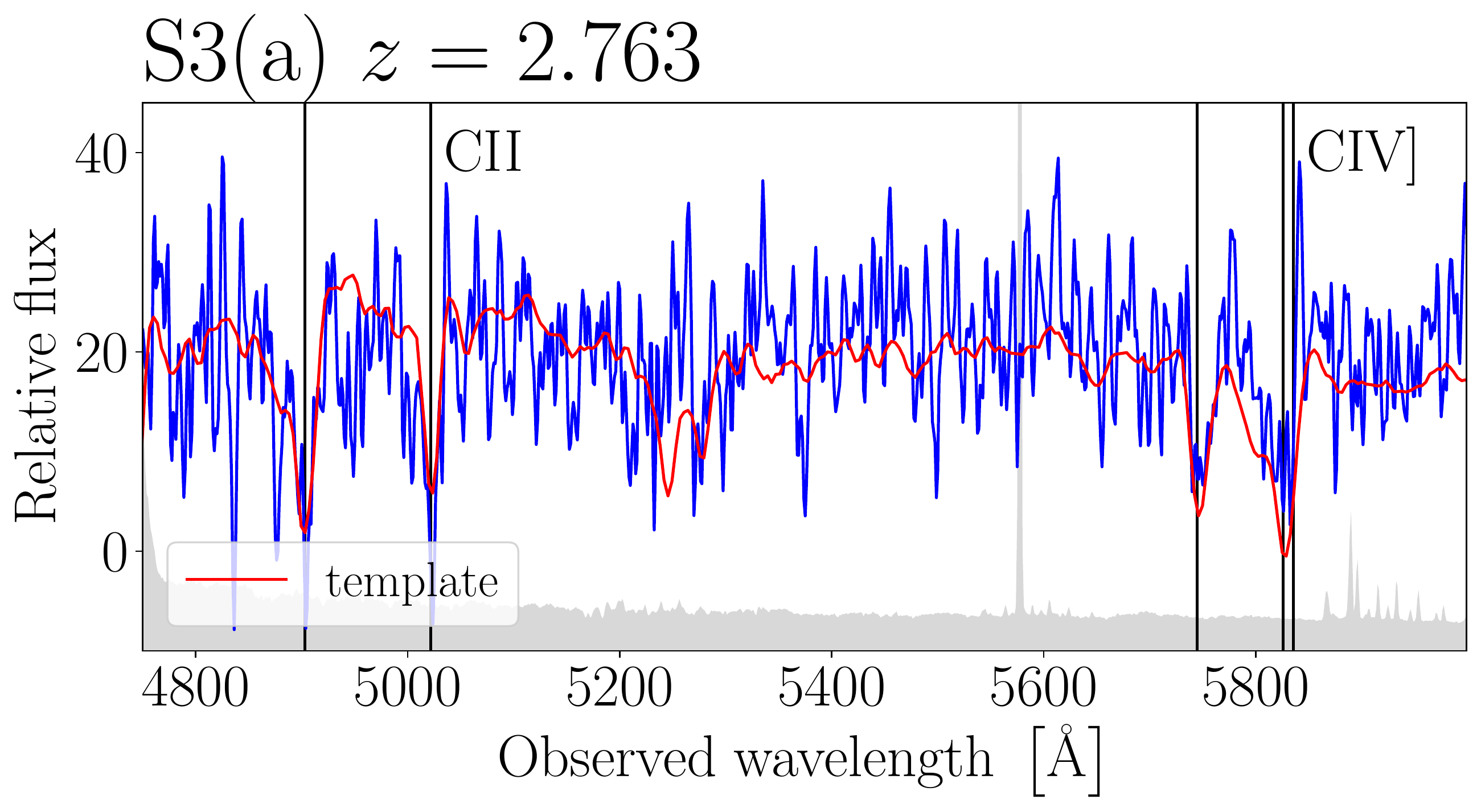}
  \includegraphics[width=0.33\textwidth]{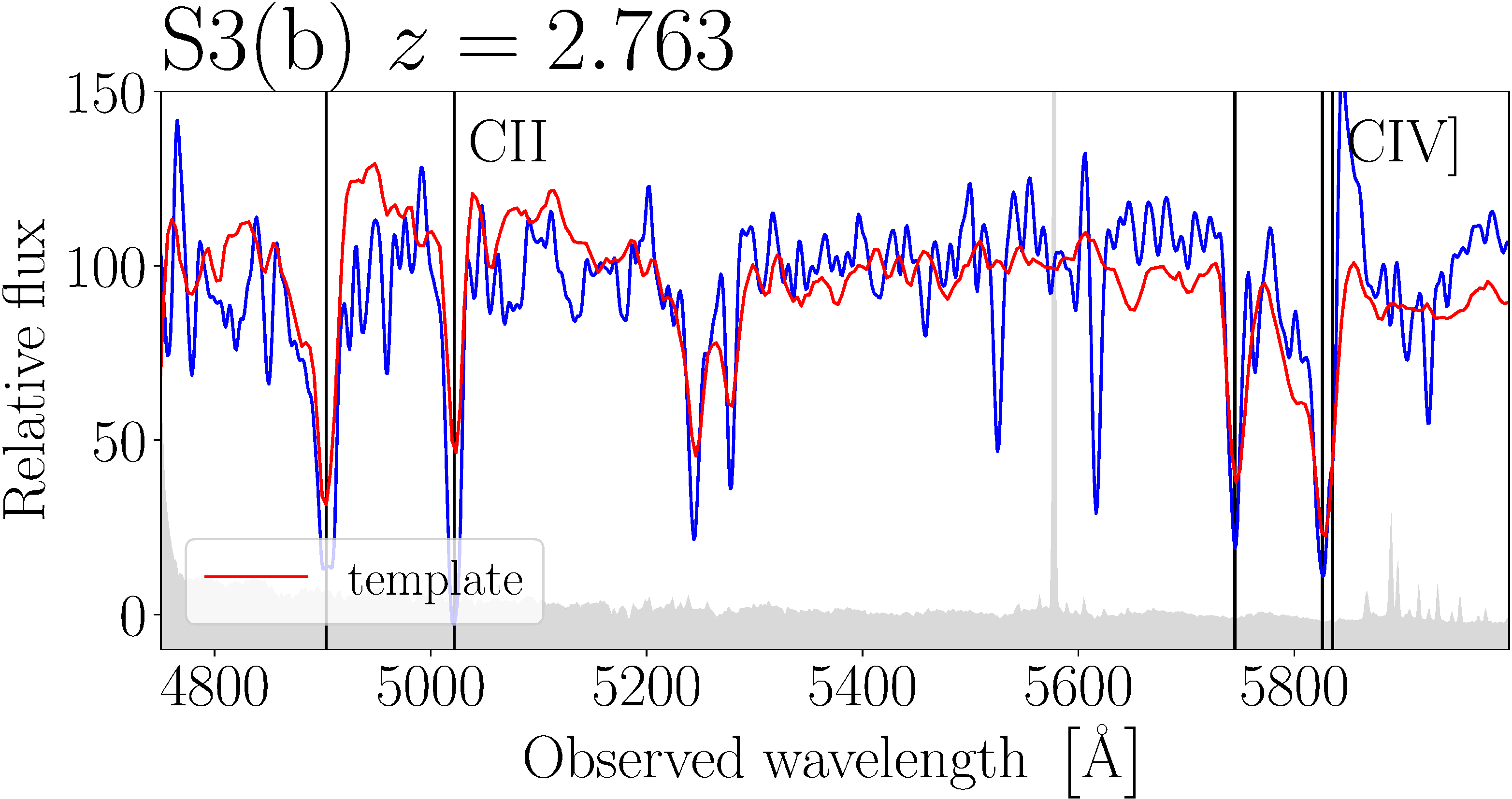}
  \includegraphics[width=0.33\textwidth]{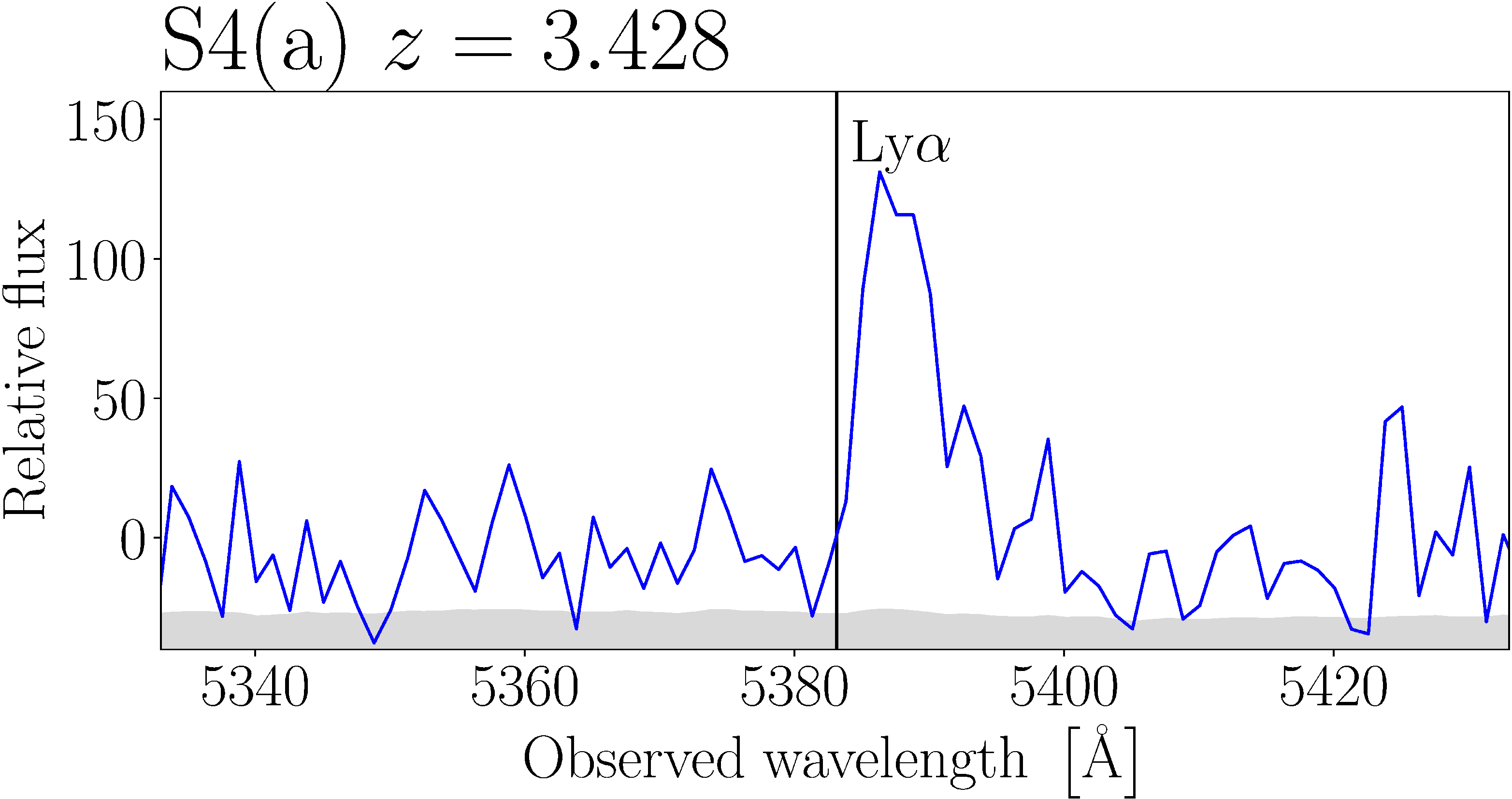}

  \includegraphics[width=0.33\textwidth]{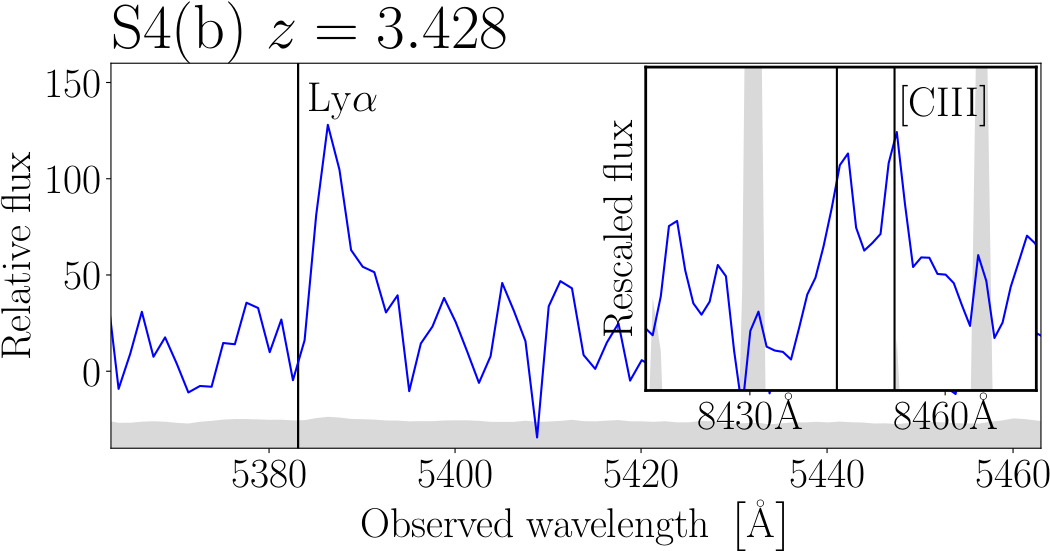}
  \includegraphics[width=0.33\textwidth]{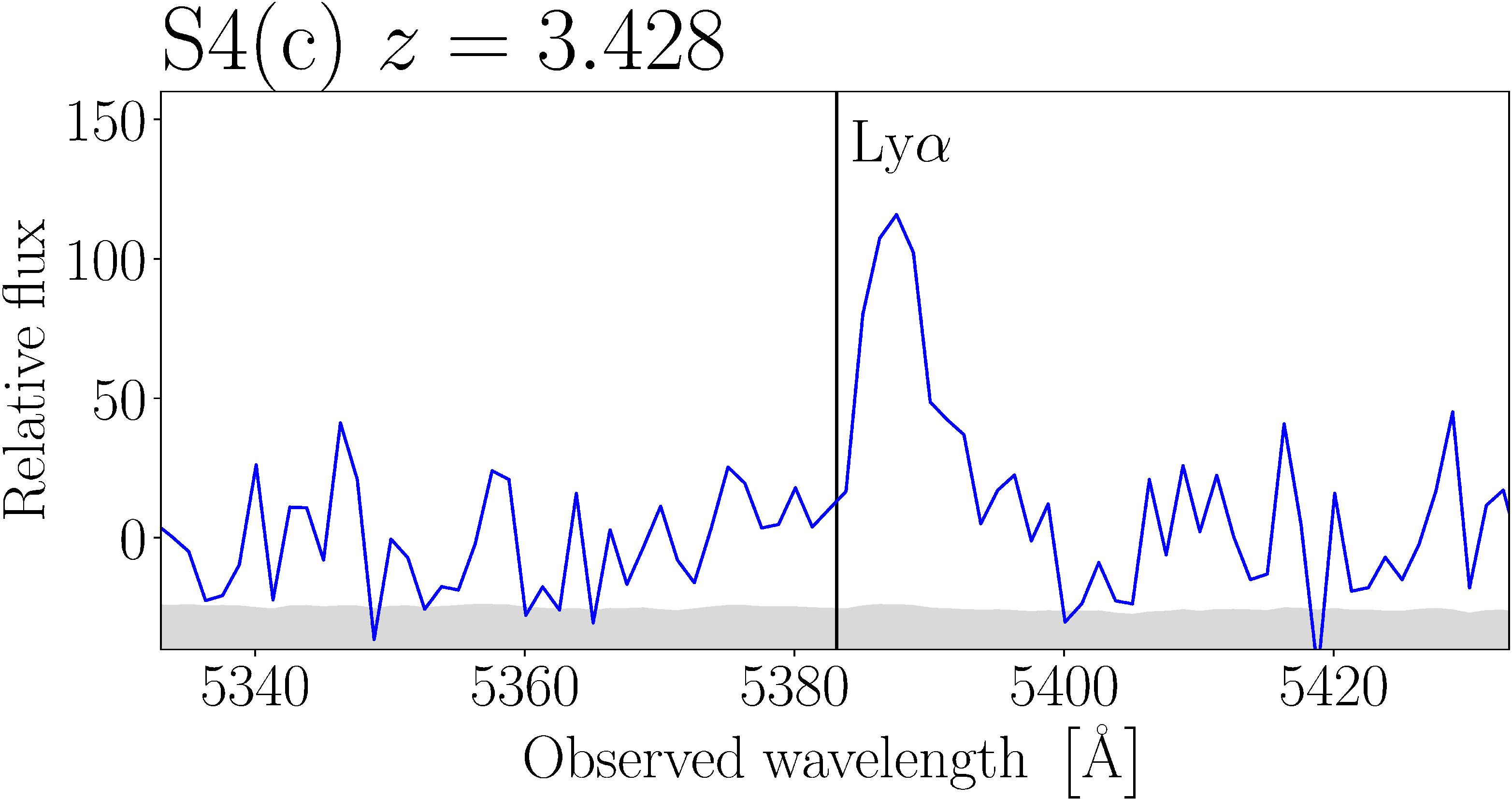}
  \includegraphics[width=0.33\textwidth]{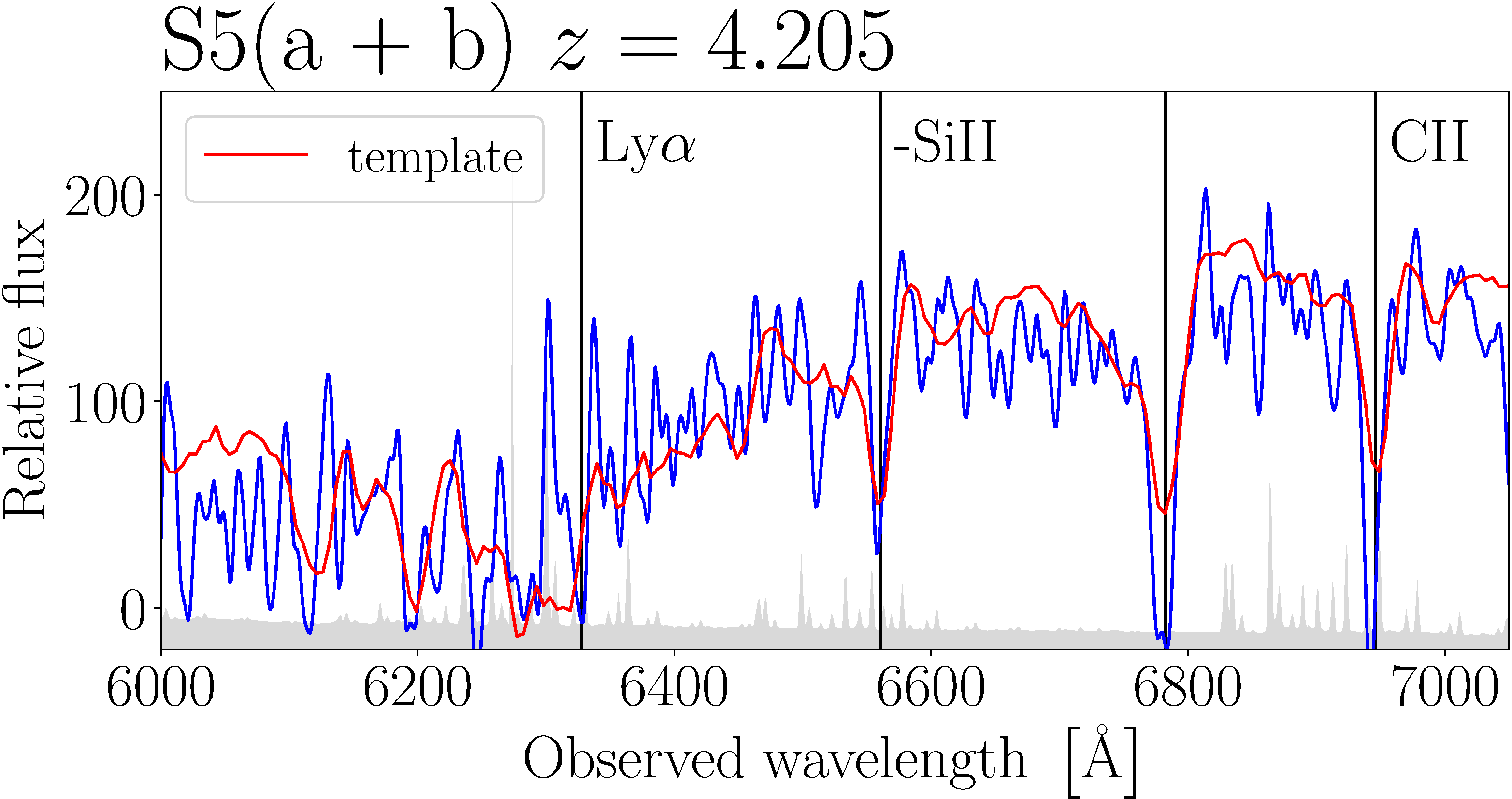}
  
  \includegraphics[width=0.33\textwidth]{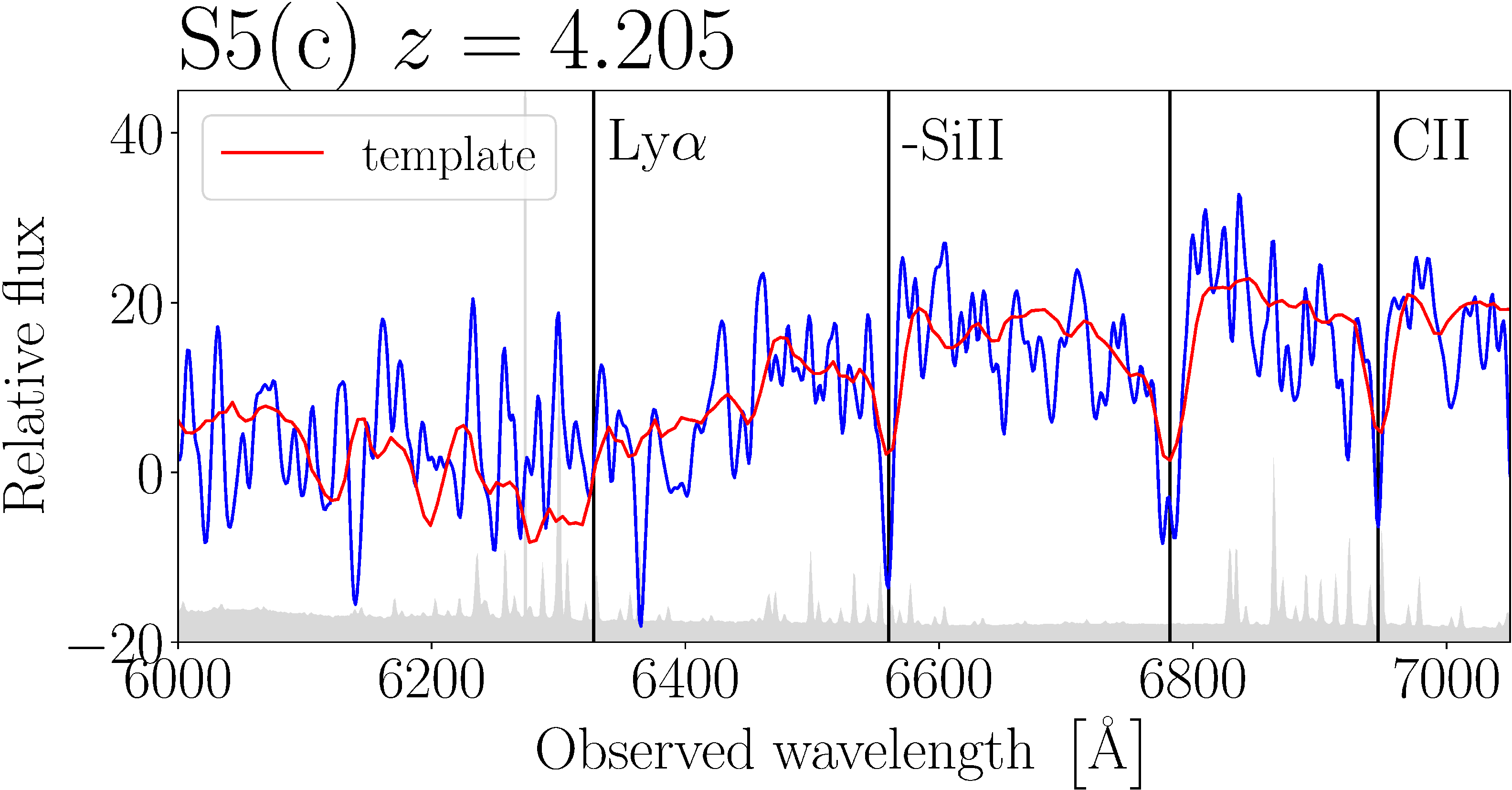}
  \includegraphics[width=0.33\textwidth]{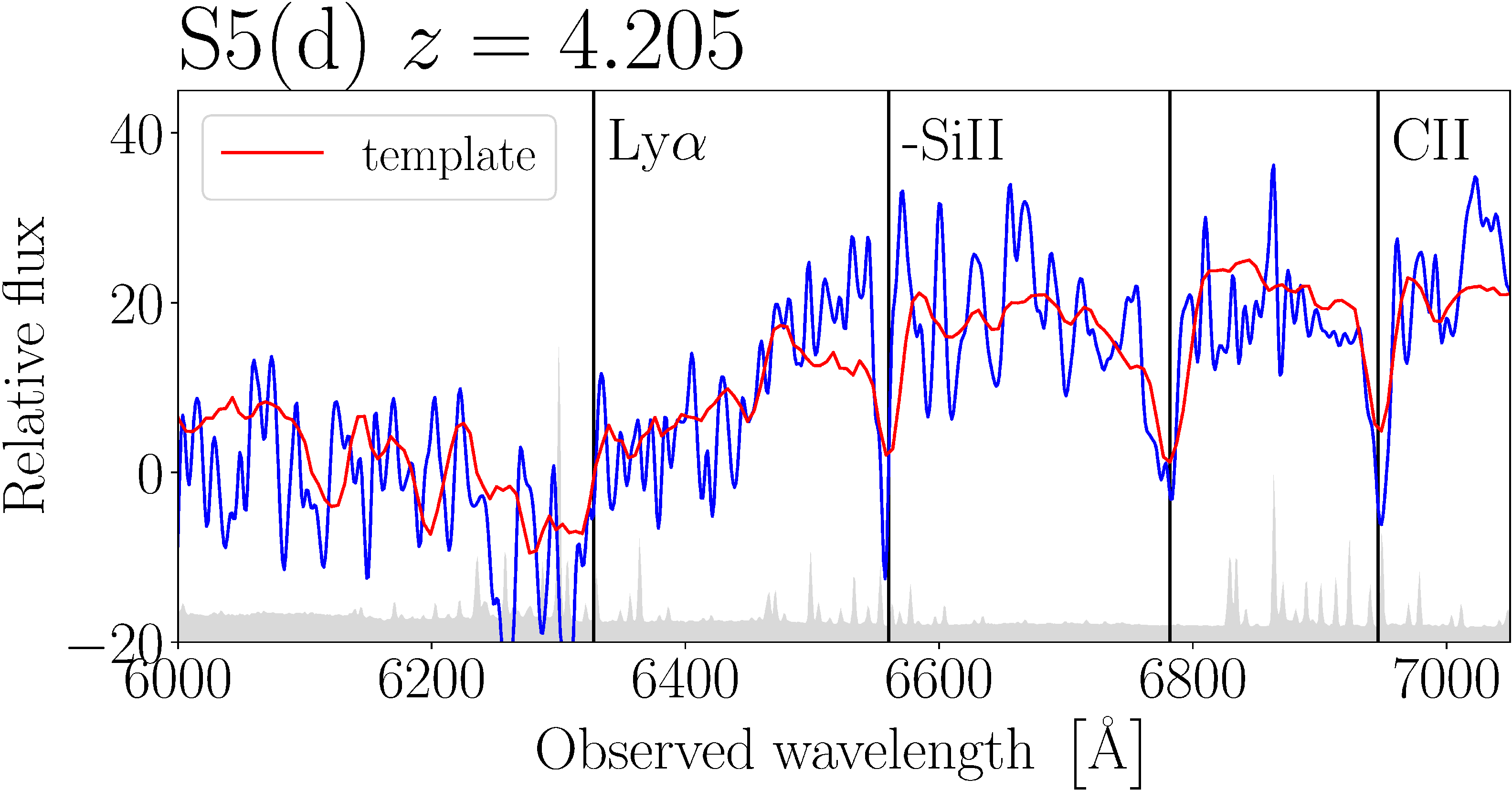}
  \caption{MUSE 1D spectra of the multiple images in sets S2, S3, S4 and S5, which are newly confirmed. The blue lines are the observed spectra in units of 10$^{-20}$\,erg\,s$^{-1}$\,cm$^{-2}$\,\AA$^{-1}$. They are extracted using circular apertures of 0.8\arcsec\ radii, except for the faint images S5(a) and S5(b) which are jointly extracted along the extended arc. The grey regions indicate the data variance. The vertical lines mark the main spectral features used to measure redshifts, and the red curves show the best-fit templates to estimate the redshifts of images in sets S3 and S5. The best-fit systemic redshift reported for set S4 comes from the [CIII] emission line detected in image S4(b). Ly$\alpha$ lines are detected in all three images of S4 and are slightly redshifted \citep[see also, e.g.,][]{verhamme18}.}
  \label{fig:spec}
\end{figure*}

\begin{table}
  \centering
  \caption{The spectroscopic redshift catalog from MUSE.}
  \begin{tabular}{ccccc}
    \hline
    \hline
    ID & RA & dec & $z_{\rm spec}$ & QF \\
    \hline
    354 & 140.3572 & 18.1715 & 0.6828 & 3 \\
    470 & 140.3604 & 18.1726 & 2.763  & 3 \\
    493 & 140.3502 & 18.1798 & 5.0547 & 3 \\
    524 & 140.3542 & 18.1798 & 0.6835 & 3 \\
    628 & 140.3558 & 18.1751 & 0.6808 & 3 \\
    636 & 140.3598 & 18.1781 & 0.6867 & 3 \\
    677 & 140.3552 & 18.1750 & 1.3576 & 3 \\
    731 & 140.3620 & 18.1784 & 0.6858 & 3 \\
    740 & 140.3574 & 18.1735 & 1.4869 & 3 \\
    760 & 140.3555 & 18.1762 & 4.2012 & 2 \\  
    \dots & \dots & \dots & \dots & \dots \\
    \hline
  \end{tabular}
  \tablefoot{The columns are: the source ID; the RA and dec coordinates; the measured spectroscopic redshift and its quality flag. The complete redshift catalogue is available in the electronic version of the paper.}
  \label{tab:redshiftOfAllProjects from MUSE}
\end{table}

\section{Image position models in the single and multiplane scenarios}
\label{sec:pdf}

\begin{figure*}
  \centering
  \includegraphics[scale=0.3]{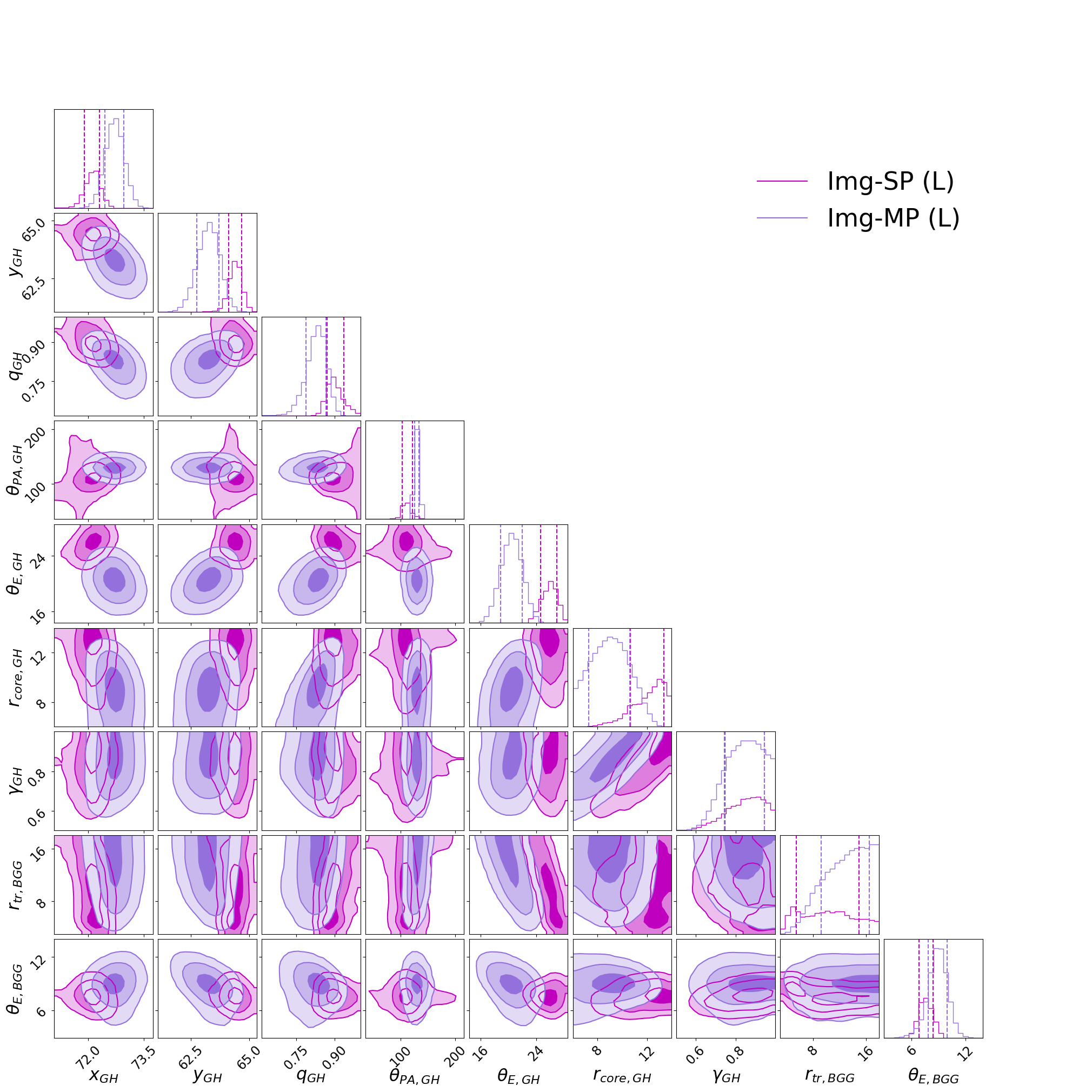}
  \caption{Joint posterior PDFs for lensing-only models Img-SP (L) and Img-MP (L) based on the centroid positions of multiple images. The three shaded areas show the $68.3\%$, $95.4\%$, $99.7\%$ credible regions. The 1-D histograms show the marginalized PDFs of each mass parameter, and vertical lines highlight the 1-$\sigma$ ranges.}
\end{figure*}

\section{Extended image modeling of source S3}

\begin{figure*}
       \centering
        \begin{subfigure}[b]{0.475\textwidth}
            \centering
           \includegraphics[width=\textwidth]{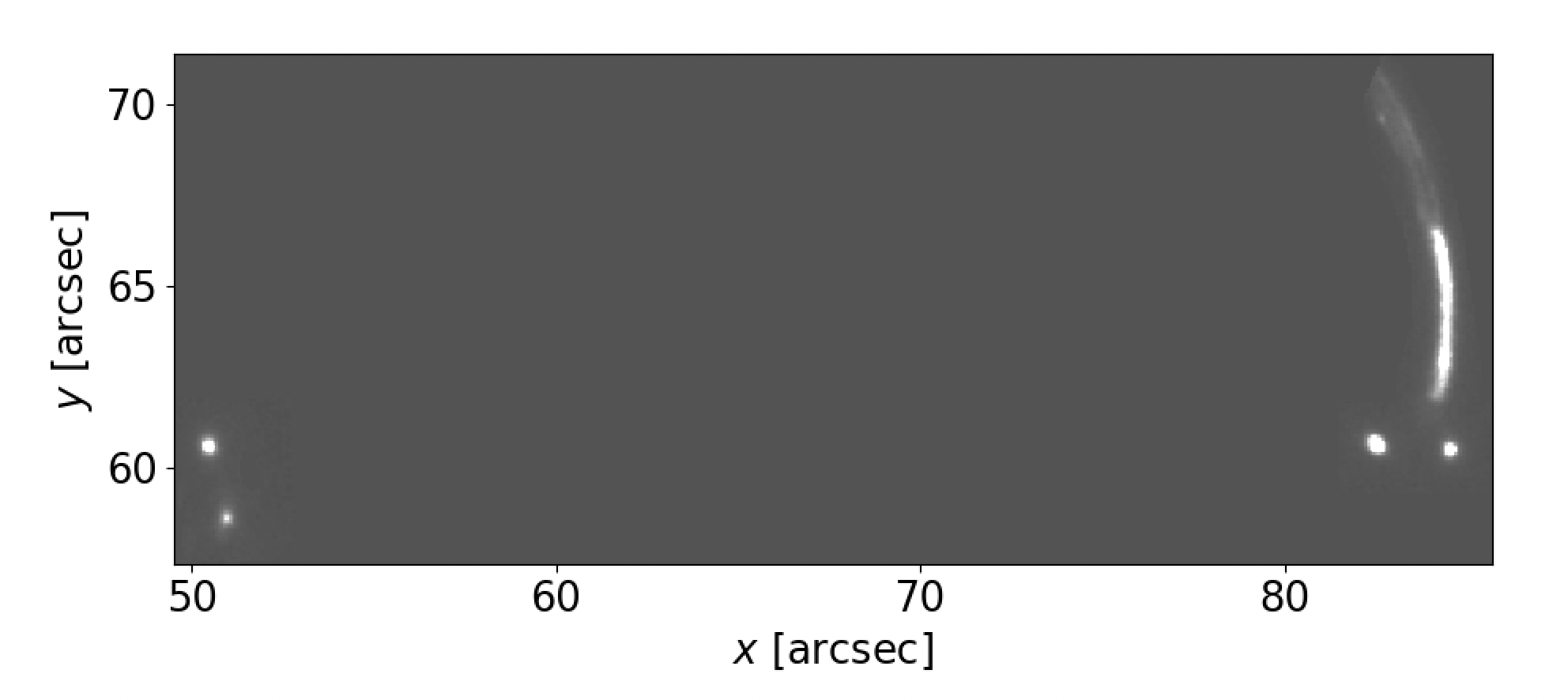}
            \caption[]%
            {{\small observed image of source 3}}    
        \end{subfigure}
        \hfill
        \begin{subfigure}[b]{0.475\textwidth}  
            \centering 
            \includegraphics[width=\textwidth]{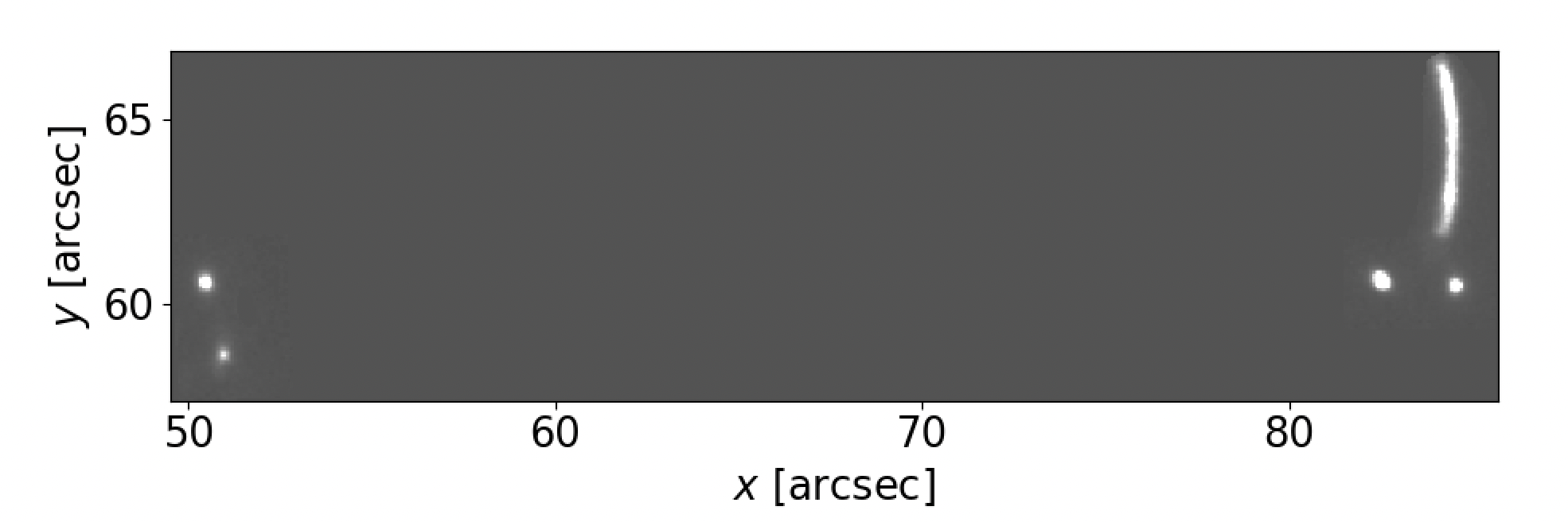}
            \caption[]%
            {{\small observed image of source 3}}    
        \end{subfigure}
        \vskip\baselineskip
        \begin{subfigure}[b]{0.475\textwidth}   
            \centering 
            \includegraphics[width=\textwidth]{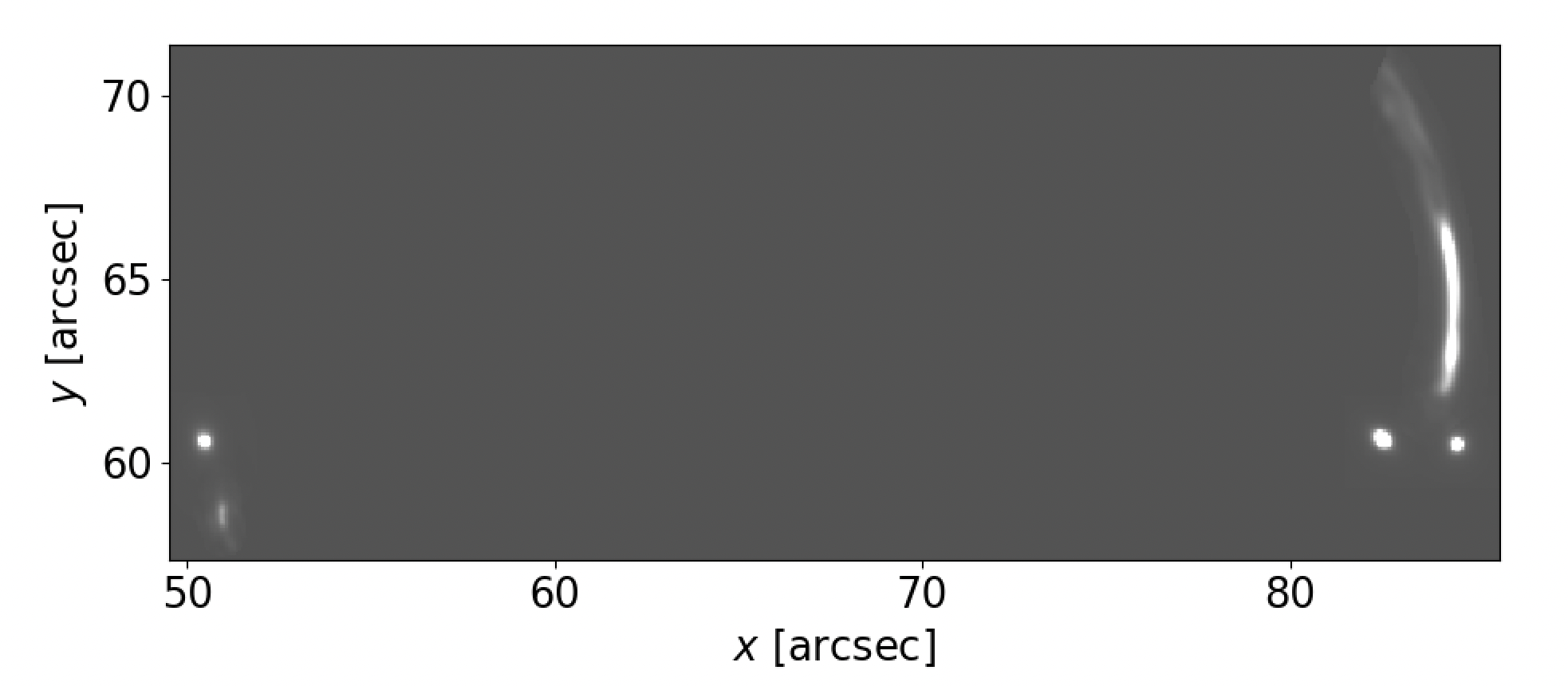}
            \caption[]%
            {{\small predicted image from model Esr2-MP  (L)}}    
        \end{subfigure}
        \quad
        \begin{subfigure}[b]{0.475\textwidth}   
            \centering 
            \includegraphics[width=\textwidth]{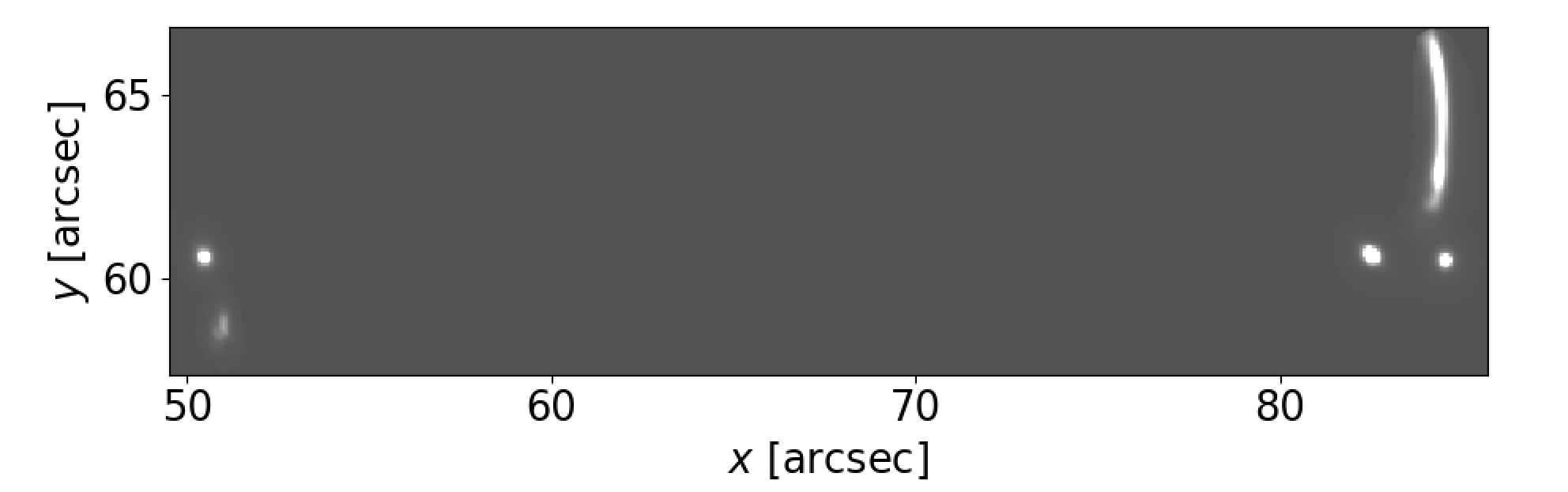}
            \caption[]%
            {{\small predicted image from model Esr2-$\text{MP}_\text{test}$ (L)}}    
        \end{subfigure}
        
         \begin{subfigure}[b]{0.475\textwidth}   
            \centering 
            \includegraphics[width=\textwidth]{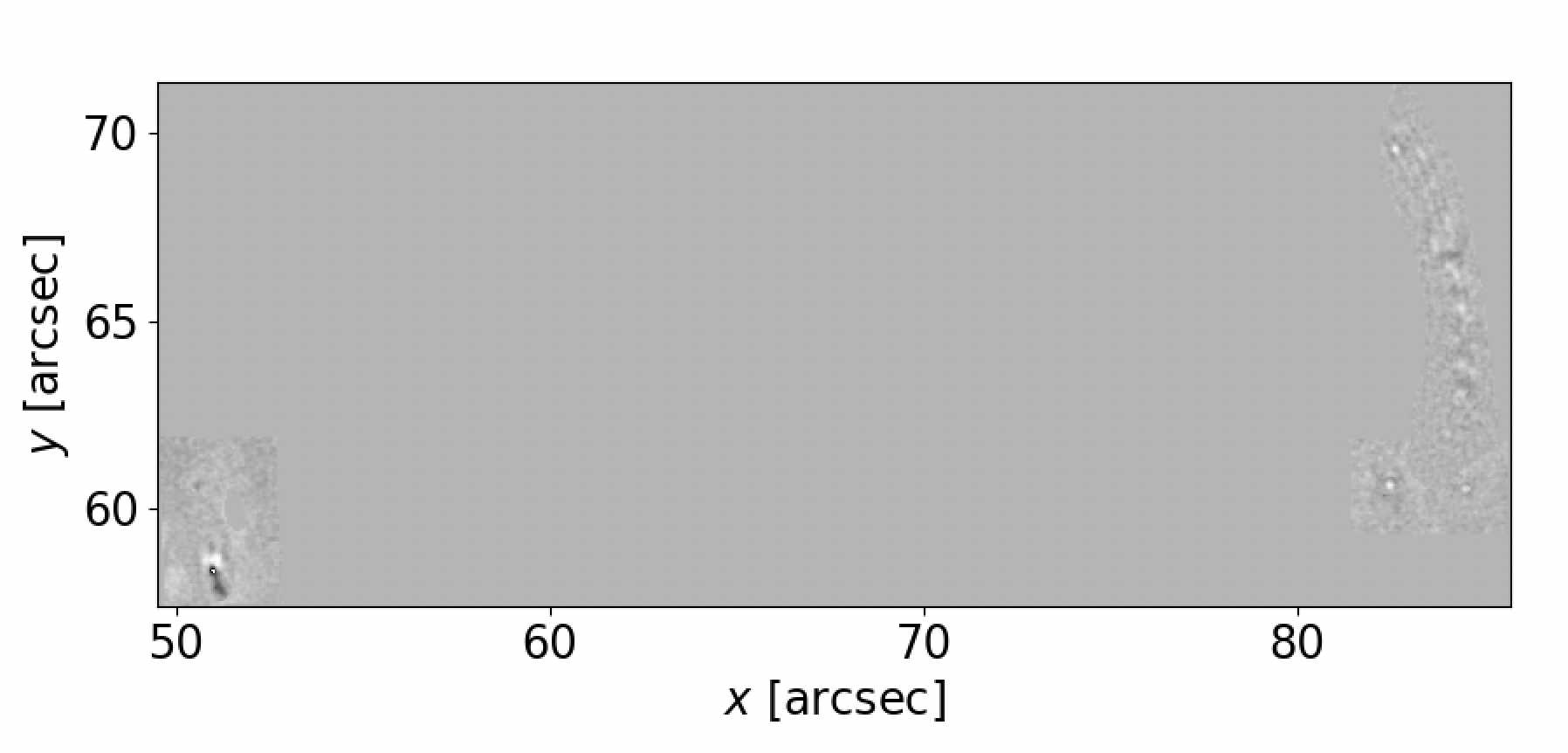}
            \caption[]%
            {{\small normalized residuals from model Esr2-MP (L)  }}    
           \label{fig:normalized residuals from model Esr2-MP (L)}
      \end{subfigure}  
        \quad
        \begin{subfigure}[b]{0.475\textwidth}   
            \centering 
            \includegraphics[width=\textwidth]{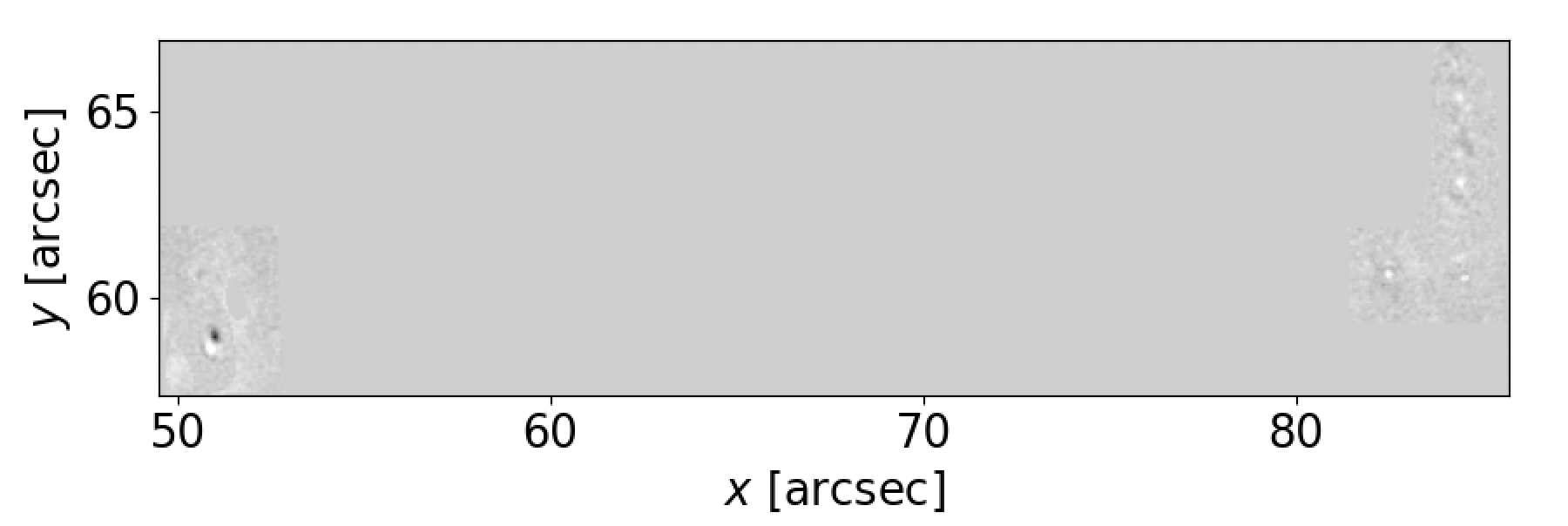}
            \caption[]%
            {{\small normalized residuals from model Esr2-$\text{MP}_\text{test}$ (L) }}    
          \end{subfigure}

         \begin{subfigure}[b]{0.375\textwidth}   
            \centering 
            \includegraphics[width=\textwidth]{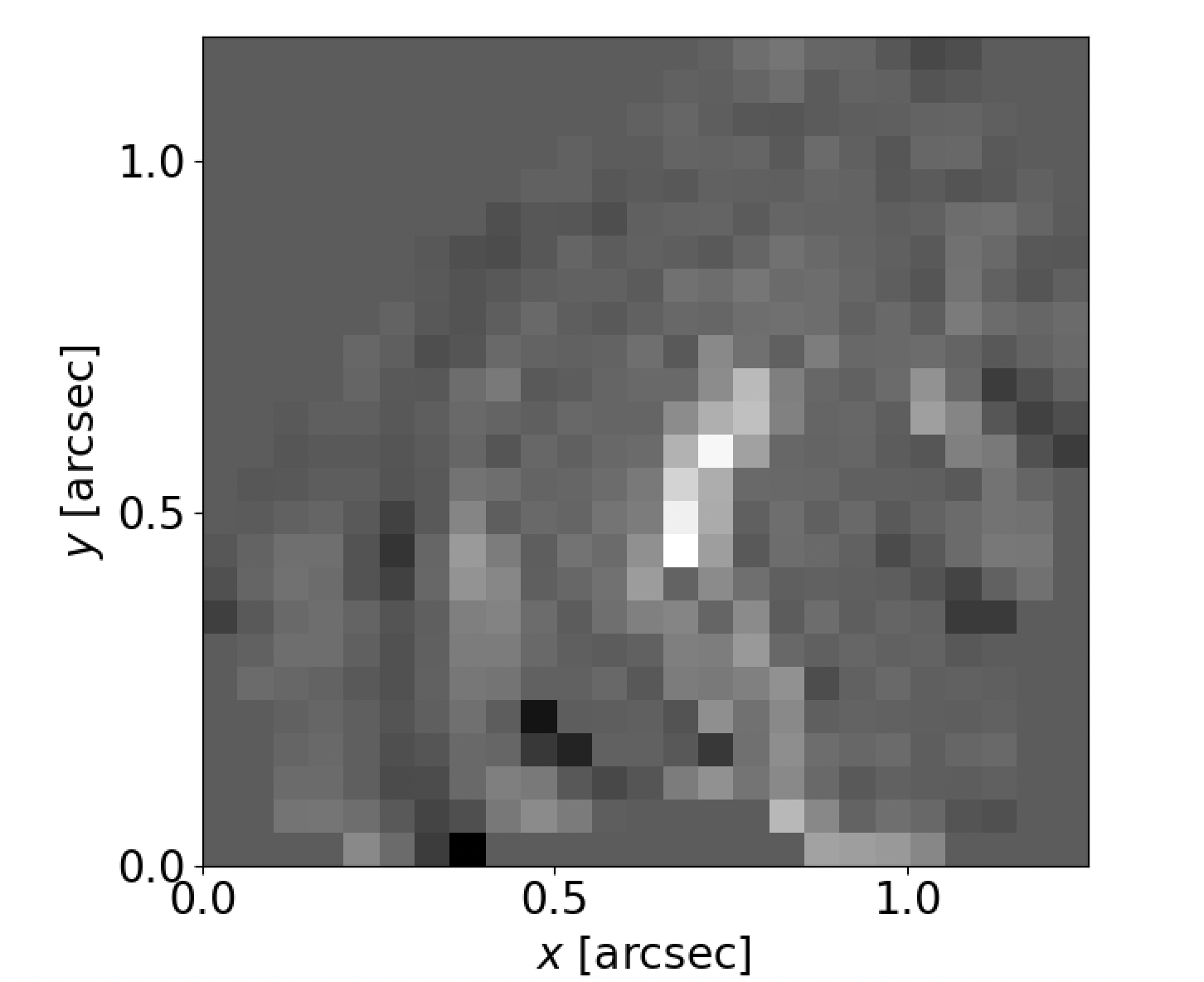}
            \caption[]%
            {{\small source reconstruction from model Esr2-MP (L)  }}    
           \label{fig:normalized residuals from model Esr2-MP (L)}
      \end{subfigure}  
        \quad
        \begin{subfigure}[b]{0.37\textwidth}   
            \centering 
            \includegraphics[width=\textwidth]{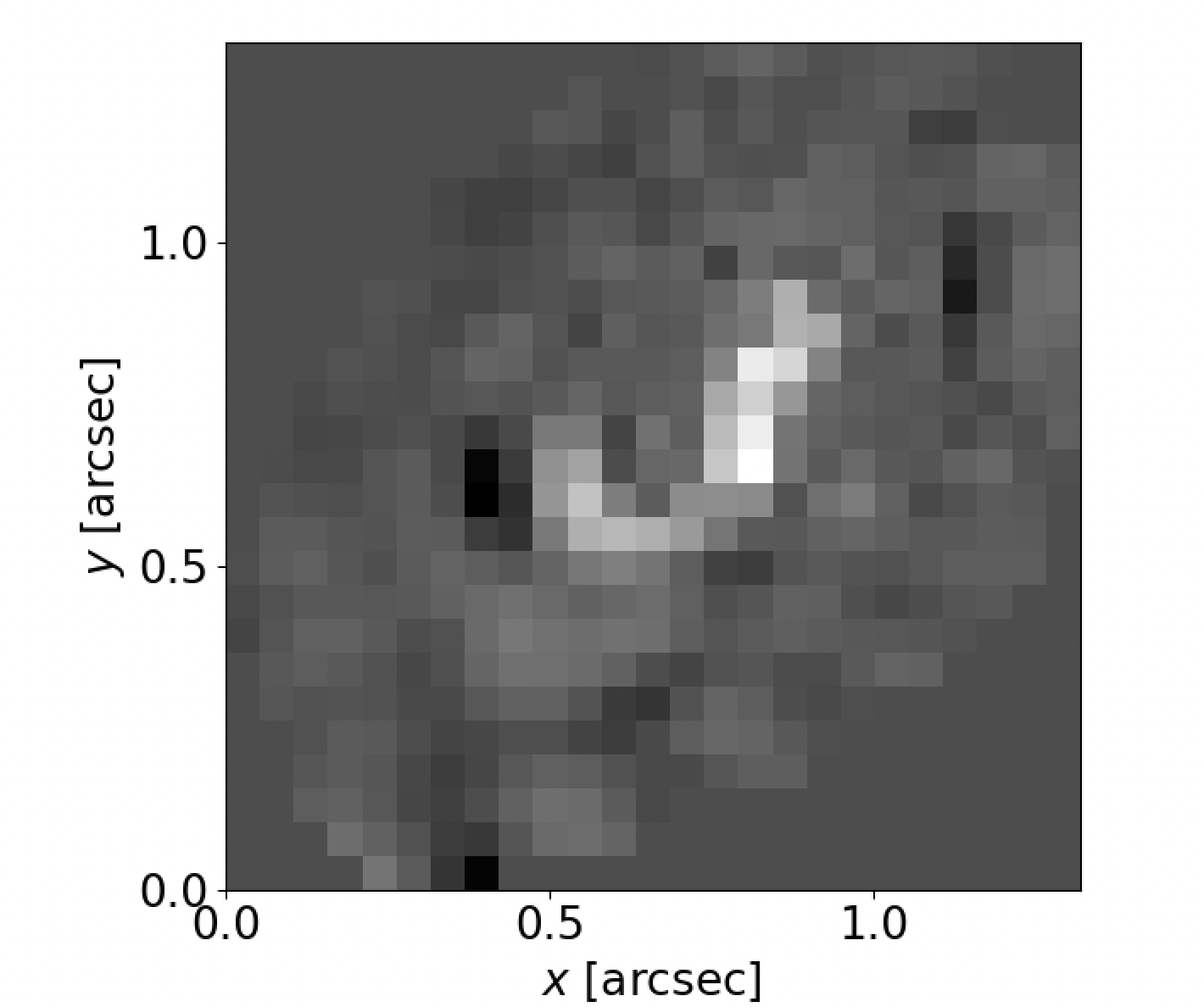}
            \caption[]%
            {{\small source reconstruction from model Esr2-$\text{MP}_\text{test}$ (L) }}    
          \end{subfigure}  
         
  \caption{Surface brightness reconstruction for lensed source S3 at redshift z = 3.4280 from two different extended image models. We show the nearby group members included in the light modeling, while other objects within the grey regions are masked out. {\it From top to bottom:} the observed {\it HST} F160W images, the best-fit models, the normalized residuals in a range between $-9.5\sigma$ to $5\sigma$, and the reconstructed source-plane morphology. {\it Left column:} Reconstruction from model Esr2-MP (L) showing excellent residuals for the rightmost arc, but significant overfitting of the light emission from the compact counterimage on the left. {\it Right column:} Same for model Esr2-$\text{MP}_\text{test}$ (L) based on a new mask excluding the upper faint, diffuse region of the arc. This model improves the fit of the counterimage of S3.} 
  \label{fig:arc2_mod}
\end{figure*}

\section{Critical curves of three reference lens models}
\label{sec:Critical curves of three reference lens models}
\begin{figure*}  
  \centering 
  \includegraphics[width=0.51\linewidth]{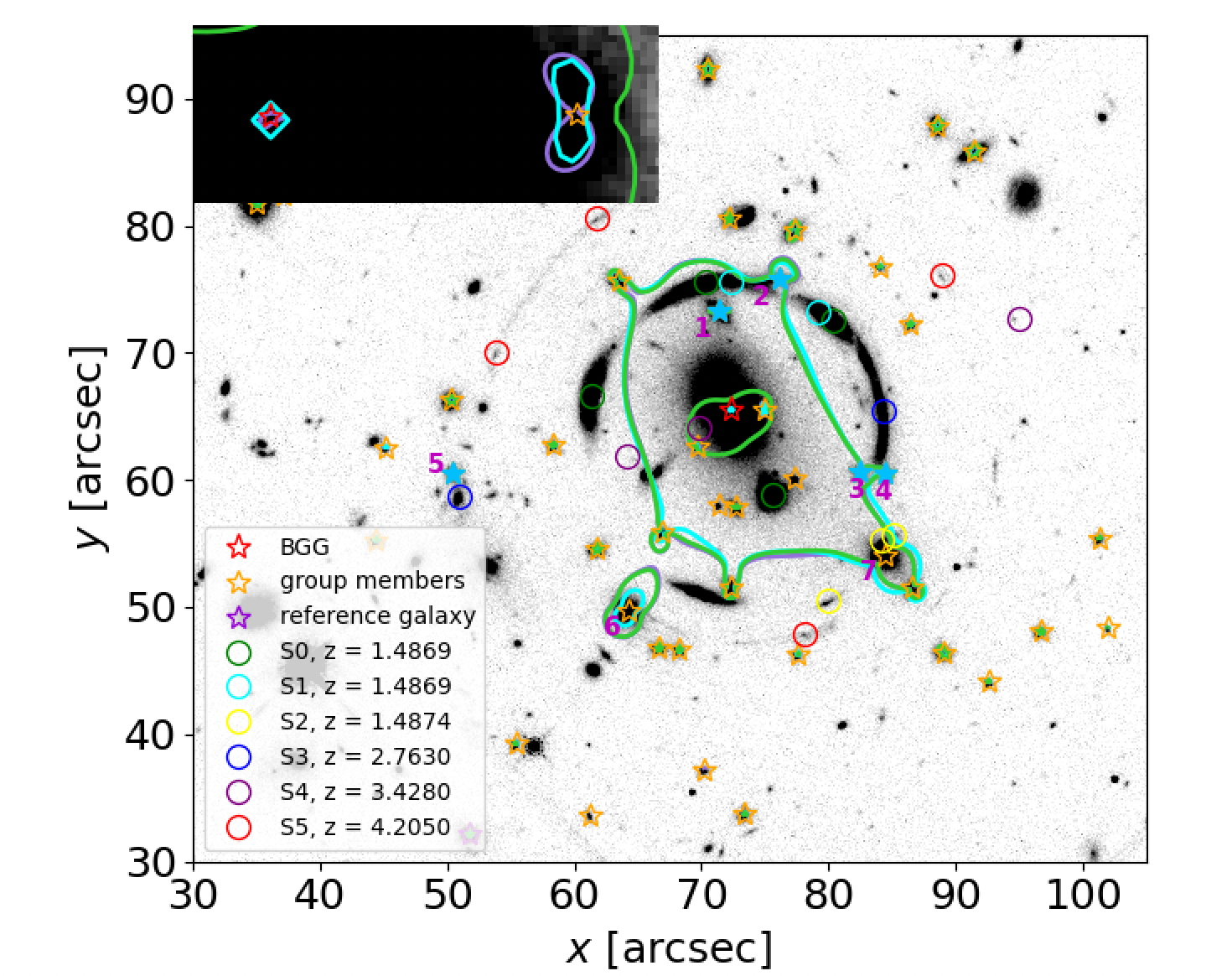} 
  \includegraphics[width=0.48\linewidth]{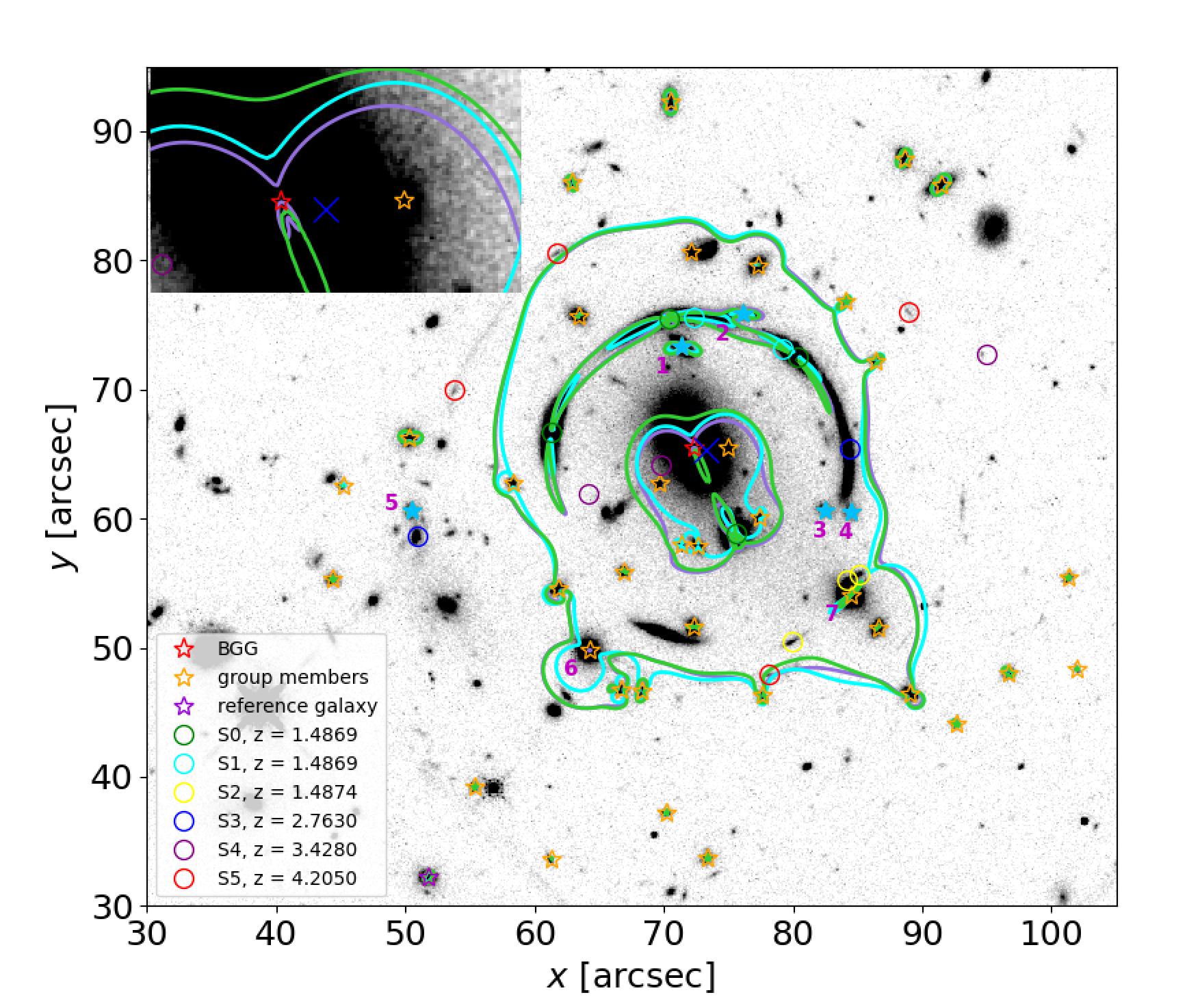}
  \caption{Critical curves of three reference lens models. \textit{Left:} Critical curves for models Img-MP (L) (purple solid line), Esr2-$\text{MP}_\text{test}$ (L) (cyan solid line) and Img-MP (L/D) (green solid line), and for the redshift of S0. The upper-left inset zooms in on a 5\arcsec\ $\times$ 3\arcsec\
  rectangle to show more details of critical curves in the central region. \textit{Right:} Same critical curves but for the redshift of S3. The upper-left inset zooms in on a 8\arcsec\ $\times$ 5\arcsec\ rectangle around the central region where the third image of S3 falls (blue 'cross'). No critical curve pass through S3(b), the extended arc of S3, indicating that it comprises a single distorted image. }
  \label{fig:critical_curves}
\end{figure*}

\end{appendix}

\end{document}